\input harvmac
\input amssym.def
\input amssym.tex
\parskip=4pt
\baselineskip=12pt
\hfuzz=20pt
\parindent 10pt

\font\iti=cmti9

\font\male=cmr9

\font\tfont=cmbx12 scaled\magstep1 

\def\md{\vskip 5mm}
\def\han{{\textstyle{1\over2}}}

\def\trh{{\textstyle{3\over2}}}
\def\trq{{\textstyle{3\over4}}}

\def\frh{{\textstyle{5\over2}}}
\def\srh{{\textstyle{7\over2}}}
\def\nrh{{\textstyle{9\over2}}}
\def\erh{{\textstyle{11\over2}}}
\def\vth{{\vartheta}}
\def\bvh{{\bar\vartheta}}
\def\tth{{\theta}}
\def\bth{{\bar\theta}}%

\def\chh{{ch}} 
\def\nt{\noindent}
\def\nl{\hfill\break}

\def\np{\vfill\eject}

\global\secno=0

\def\wt{\widetilde{|\L\rg}}

\def\half{{\textstyle{1\over2}}}
\def\ha{{\textstyle{1\over2}}}
\def\trh{{\textstyle{3\over2}}}
\def\frh{{\textstyle{5\over2}}}
\def\srh{{\textstyle{7\over2}}}

\def\halfn{{\textstyle{2\over N}}}

\def\third{{\textstyle{1\over3}}}
\def\quarter{{\textstyle{1\over4}}}

\def\twelvth{{\textstyle{1\over12}}}

\def\rg{\rangle} 

\font\fat=cmsy10 scaled\magstep5

\def\Bbullet{\raise-3pt\hbox{\fat\char"0F}}

\def\tV{\tilde V} 
\def\tv{\tilde v}
\def\hV{\hat V} \def\hL{\hat L}

\font\verysmall=cmr5

\def\righa{{\longrightarrow\kern-14pt
\raise5pt\hbox{\verysmall 12}}}

\def\sigha{{\longrightarrow\kern-14pt
\raise5pt\hbox{\verysmall 34}}}

\def\upot{{\uparrow\kern-3pt
\hbox{\verysmall 12}}}

\def\upone{{\uparrow\kern-3pt
\hbox{\verysmall 1}}}

\def\uptwo{{\uparrow\kern-3pt
\hbox{\verysmall 2}}}

\def\dthree{{\downarrow\kern-3pt
\raise2pt\hbox{\verysmall 3}}}

\def\dfour{{\downarrow\kern-3pt
\raise2pt\hbox{\verysmall 4}}}

\def\dtf{{\downarrow\kern-3pt
\raise2pt\hbox{\verysmall 34}}}

\def\nwone{{\nwarrow\kern-3pt
\raise2pt\hbox{\verysmall 1}}}

\def\nwot{{\nwarrow\kern-5pt
\raise3pt\hbox{\verysmall 12}}}

\def\swthree{{\swarrow\kern-3pt
\raise2pt\hbox{\verysmall 3}}}

\def\swtf{{\swarrow\kern-5pt
\raise1pt\hbox{\verysmall 34}}}

\def\lone{{\leftarrow\kern-8pt
\raise5pt\hbox{\verysmall 1}}}

\def\ltwo{{\leftarrow\kern-8pt
\raise5pt\hbox{\verysmall 2}}}

\def\rone{{\rightarrow\kern-11pt
\raise5pt\hbox{\verysmall 1}}~~}

\def\rtwo{{\rightarrow\kern-11pt
\raise5pt\hbox{\verysmall 2}}~~}

\def\rthree{{\rightarrow\kern-11pt
\raise5pt\hbox{\verysmall 3}}~~}

\def\rfour{{\rightarrow\kern-11pt
\raise5pt\hbox{\verysmall 4}}~~}

\def\rtf{{\longrightarrow\kern-15pt
\raise5pt\hbox{\verysmall 34}}~~}

\def\lot{{\longleftarrow\kern-14pt
\raise5pt\hbox{\verysmall 12}}}

\def\dia{~~$\diamondsuit$}

\def\bu{$\bullet$}

\def\cg{{\cal G}} \def\ch{{\cal H}}

\def\cp{{\cal P}} \def\cq{{\cal Q}} \def\car{{\cal R}}
\def\cs{{\cal S}}

\def\hcg{\hat{\cal G}} \def\hch{\hat{\cal H}}

\def\white#1{\mathop{\bigcirc}\limits_{#1}}
\def\gray#1{\mathop{\bigotimes}\limits_{#1}}
\def\riga{-\kern-4pt - \kern-4pt -}

\def\bbz{Z\!\!\!Z}
\def\bbc{C\kern-6pt I}
\def\bac{{C\kern-5.5pt I}}

\def\bbn{I\!\!N}  \def\o{{\bar 0}} \def\I{{\bar 1}}

\def\a{\alpha} \def\b{\beta} \def\g{\gamma} \def\d{\delta}

\def\eps{{\epsilon}}
\def\l{\lambda} 
\def\r{\rho}

\def\G{\Gamma} 

\def\D{\Delta} \def\L{\Lambda}
\def\hD{\hat{\Delta}} \def\hpi{\hat{\pi}}

\def\ve{\varepsilon}

\def\lra{\longleftrightarrow}

\def\mt{\mapsto}

\def\({\left(}
\def\){\right)}

\def\a{\alpha}
\def\b{\beta}
\def\d{\delta}
\def\g{\gamma}

\def\r{\rho}
\def\l{\lambda}
\def\D{\Delta}
\def\L{\Lambda}



\lref\Konishi{K. Konishi,  Phys. Lett. {\bf B135} (1984) 439-444.}

\lref\Sie{W. Siegel, Nucl. Phys. {\bf B177}, 325-332 (1981).}

\lref\HST{P.S. Howe, K.S. Stelle and P.K. Townsend,
Nucl. Phys. {\bf B192}, 332 (1981).}

\lref\GuMa{M. Gunaydin and N. Marcus, Class. Quant. Grav. {\bf 2}
(1985) L11-L18 \& L19-L23.}

\lref\FGPW{D.Z. Freedman, S.S. Gubser, K. Pilch and N.P. Warner,
Adv. Theor. Math. Phys. {\bf 3}, 363 (1999), hep-th/9904017.}

\lref\Dobsin{V.K. Dobrev,
Lett. Math. Phys. {\bf 22} (1991) 251-266.}

\lref\GIKOS{A. Galperin, E. Ivanov, S. Kalitzin, V. Ogievetsky and E.
Sokatchev, Class. Quant. Grav. {\bf 1} (1984) 469; {\bf 2} (1985) 155.}

\lref\GIOS{A.~Galperin, E.~Ivanov, V.~Ogievetsky and E.~Sokatchev,
      Annals Phys.\  {\bf 185} (1988) p. 1 \& p. 22;~
  {\it Harmonic superspace}, 306 pages, (Cambridge Univ. Press, 2001).}

\lref\AFSZ{L. Andrianopoli, S. Ferrara, E. Sokatchev and B. Zupnik,
Adv. Theor. Math. Phys. {\bf 3}, 1149 (1999) hep-th/9912007.}

\lref\FS{S. Ferrara and E. Sokatchev,
Lett. Math. Phys. {\bf 52}, 247-262 (2000),
hep-th/9912168;\
J. Math. Phys. {\bf 42}, 3015 (2001) hep-th/0010117.}

\lref\FSa{S. Ferrara and E. Sokatchev,
Int. J. Theor. Phys. {\bf 40}, 935-984 (2001).} 

\lref\CDDF{A. Ceresole, G. Dall' Agata, R. D'Auria and S. Ferrara,
 Spectrum of Type IIB Supergravity on $AdS_5 \times T^11$,
 hep-th/9905226, Phys. Rev. D61:066001  (2000).}

\lref\AES{G. Arutyunov, B. Eden \& E. Sokatchev,
Nucl. Phys. {\bf B619}, 359-372 (2001);\ 
G. Arutyunov, B. Eden, A.C. Petkou \& E. Sokatchev,
Nucl. Phys. {\bf B620}, 380-404 (2002).} 

\lref\BKRSa{M. Bianchi, S. Kovacs, G. Rossi \&  Y.S. Stanev,
JHEP 0105:042   (2001).} 

\lref\HeHoa{P.J. Heslop and P.S. Howe,
Phys. Lett. {\bf 516B}, 367 (2001);\
Nucl. Phys. {\bf B626}, 265-286 (2002);\  
    JHEP 0401 (2004) 058.} 

\lref\EdSo{B. Eden and E. Sokatchev, Nucl. Phys. {\bf B618}, 259-276 (2001).}

\lref\Koba{T. Kobayashi, Publ. RIMS, 47 (2011) 585-611;
    arXiv:1001.0224 [math.RT].}

\lref\Nahm{W. Nahm, Nucl. Phys. {\bf B135}, 149-166 (1978).}

\lref\HLS{R. Haag, J.T. Lopuszanski and M. Sohnius,
Nucl. Phys. {\bf B88}, 257-274 (1975).}

\lref\FF{M. Flato and C. Fronsdal, Lett. Math. Phys. {\bf 8}, 159-162
(1984).}

\lref\DPm{V.K. Dobrev and V.B. Petkova,
Lett. Math. Phys. {\bf 9} (1985) 287-298.}

\lref\DPf{V.K. Dobrev and V.B. Petkova, 
Fortschr. d. Phys. {\bf 35} (1987) 537-572; first as ICTP Trieste preprint
IC/85/29 (March 1985).}

\lref\DPu{V.K. Dobrev and V.B. Petkova, 
Phys. Lett. {\bf 162B} (1985) 127-132 .}

\lref\DPp{V.K. Dobrev and V.B. Petkova, 
Proceedings, eds. A.O. Barut and H.D. Doebner,
Lecture Notes in Physics, Vol. 261 (Springer-Verlag, Berlin, 1986)
p. 291 
and p. 300.} 

\lref\Min{S. Minwalla, Adv. Theor. Math. Phys. {\bf 2}, 781-846
(1998).}

\lref\Dosix{V.K. Dobrev,
J. Phys. A {\bf A35} (2002) 7079-7100; hep-th/0201076.}

\lref\GRW{M. Gunaydin, L.J. Romans \& N.P. Warner,
Nucl. Phys. {\bf B272}, 598-646 (1986).}

\lref\Mack{G. Mack, Comm. Math. Phys. {\bf 55}, 1 (1977).}

\lref\Dobsy{V.K. Dobrev, Kazhdan-Lusztig polynomials, subsingular
vectors, and conditionally invariant (q-deformed) equations, Invited
talk at Symposium "Symmetries in Science IX", Bregenz, Austria,
(1996), Proceedings, eds. B. Gruber and M. Ramek, (Plenum
Press, New York and London, 1997) pp. 47-80.}

\lref\Dobsu{V.K. Dobrev,
J. Math. Phys. {\bf 26} (1985) 235-251.}

\lref\Dobch{V.K. Dobrev,
Phys. Part. Nucl.  {\bf 38} (2007) 564-609, hep-th/0406154.}

\lref\Dobbps{V.K. Dobrev, 
Nucl. Phys. {\bf B854} (2012) 878-893; arXiv:1012.3685 [hep-th];~
Phys. Part. Nucl. {\bf 43} (2012) 616–620,
arXiv:1112.3498 [hep-th].}

\lref\KL{D. Kazhdan and G. Lusztig,~ Inv. Math. {\bf 53}, 165-184
(1979).}

\lref\Di{J. Dixmier, {\it Enveloping Algebras}, (North Holland,
New York, 1977).}

\lref\Kac{V.G. Kac, {\it Infinite--Dimensional Lie Algebras},
(Birkh\"auser, Boston, 1983).}

\lref\Kacs{V.G. Kac, Adv. Math. {\bf 26}, 8-96 (1977);\
Comm. Math. Phys. {\bf 53}, 31-64 (1977);\
the second paper
is an adaptation for physicists of the first paper.}

\lref\Kacl{V.G. Kac, Lect. Notes in Math. {\bf 676}
(Springer-Verlag, Berlin, 1978) pp. 597-626.}

\lref\Kacc{V.G. Kac, 
Comm. Algebra {\bf 5}, 889-897 (1977).}

\lref\BBMS{N. Beisert, M. Bianchi, J.F. Morales \& H. Samtleben,
    JHEP 0407:058  (2004).}

\lref\BGHR{A. Barabanschikov, L. Grant, L.L. Huang, S. Raju,
 JHEP   0601:160  (2006).} 

 \lref\HJS{J. Henn, C. Jarczak \& E. Sokatchev, 
Nucl. Phys. {\bf B730} (2005) 191-209.}

\lref\KMMR{J. Kinney, J. Maldacena, S. Minwalla, S. Raju,
Comm. Math. Phys. {\bf 275} (2007) 209.}

\lref\Dolb{F.A. Dolan,
Nucl. Phys. {\bf B790} (2008) 432-464.}

\lref\BBMR{J. Bhattacharya, S. Bhattacharyya, S. Minwalla and S. Raju,
JHEP 0802:064 (2008).} 

\lref\GPR{A. Gadde, E. Pomoni \& L. Rastelli,
YITP-SB-10-20, arXiv:1006.0015 [hep-th] (2010).}

 \lref\Dola{F.A. Dolan, 
J. Math. Phys. {\bf 47} (2006) 062303.} 

\lref\BDHO{M. Bianchi, F.A. Dolan, P.J. Heslop \& H. Osborn,
Nucl. Phys. {\bf B767} (2007) 163-226.}

\lref\Doo{V.K. Dobrev, 
Lett. Math. Phys. {\bf 9} (1985) 205-211;
Talk at the I National Congress of Bulgarian Physicists (Sofia,
1983) and INRNE Sofia preprint (1983).}

\lref\Dor{V.K. Dobrev, Rept. Math. Phys. {\bf 25} (1988) 159-181.}

\lref\Sha{N.N. Shapovalov, Funkts. Anal. Prilozh. {\bf 6} (4) 65
(1972);\ English translation: Funct. Anal. Appl. {\bf 6}, 307
(1972).}

\lref\Dobcond{V.K. Dobrev,
J. Phys. A {\bf A28} (1995) 7135-7155.}

\lref\Serg{V.V. Serganova, Appendix to the paper:
D.A. Leites, M.V. Saveliev and V.V. Serganova, in: Proc. of
Group Theoretical Methods in Physics, Yurmala, 1985
(Nauka, Moscow, 1985, in Russian) p. 377; English translation in VNU
(Sci. Press, Dordrecht, 1987).} 


\lref\BL{I.N. Bernstein and D.A. Leites,
C.R. Acad. Bulg. Sci. {\bf 33}, 1049 (1980).}

\lref\JHKT{J. Van der Jeugt, J.W.B. Hughes, R.C. King and
J. Thierry-Mieg, Comm. Algebra, {\bf 18}, 3453 (1990);
J. Math. Phys. {\bf 31}, 2278 (1990) .} 

\lref\Jeu{J. Van der Jeugt,
Comm. Algebra, {\bf 19}, 199 (1991).}

\lref\Serga{V. Serganova, 
 Selecta Math. {\bf 2}, 607 (1996).} 

\lref\VZ{J. van der Jeugt and R.B. Zhang, 
 Lett. Math. Phys. {\bf 47}, 49 (1999).}

\lref\Bru{J. Brundan, 
 J. Amer. Math. Soc. {\bf 16} (2002) 185; 
Adv. Math. {\bf 182}, 28 (2004).} 

\lref\SuZh{Yucai Su and R.B. Zhang, Character and dimension formulae for
general linear superalgebra, math.QA/0403315.}

\lref\DoZh{V.K. Dobrev and R.B. Zhang, 
Phys. Atom. Nuclei, {\bf 68} (2005) 1660-1669;\nl
V.K. Dobrev, A.M. Miteva, R.B. Zhang and B.S. Zlatev,
Czech. J. Phys. {\bf 54} (2004) 1249-1256.}

\lref\Dobpds{V.K. Dobrev,
J. Phys. A  {\bf A41} (2008) 425206; arXiv:0712.4375 [hep-th].}


\hfill CERN-PH-TH/2012-232

\vskip 2truecm

\centerline{{\tfont Explicit Character Formulae for Positive Energy UIRs}}
\vskip 2truemm
\centerline{{\tfont of D=4 Conformal Supersymmetry}}

\vskip 1.5cm

\centerline{{\bf V.K. Dobrev}}
\vskip 0.5cm

\centerline{Theory Division, Department of Physics, CERN}
\centerline{CH-1211 Geneva 23, Switzerland}
\centerline{and}
\centerline{Institute of Nuclear Research and Nuclear Energy}
\centerline{Bulgarian Academy of Sciences,}
\centerline{72 Tsarigradsko Chaussee, 1784 Sofia, Bulgaria}
\centerline{(permanent address)}
\vskip 1.5cm

\centerline{{\bf Abstract}}

\midinsert\narrower\narrower{\male
This paper continues the project of constructing
the character formulae for the
positive energy unitary irreducible
representations of the $N$-extended $D=4$ conformal
superalgebras $su(2,2/N)$. In the first paper we
gave the bare characters which represent the
defining odd entries of the characters. Now we give
the full explicit character formulae for $N=1$ and
for several important examples for $N=2$ and $N=4$.
}\endinsert

\vskip 1.5cm

\newsec{Introduction}

\nt Recently, superconformal field theories in various dimensions
are attracting more interest, in particular, due to their duality to
AdS supergravities.  Mostly studied are those for ~$D\leq
6$~ since in these cases the relevant superconformal algebras
satisfy \Nahm{} the Haag-Lopuszanski-Sohnius theorem \HLS. This
makes the classification of the UIRs of these superalgebras very
important. In the 1980s  such classification was done for
the ~$D=4$~ superconformal algebra ~$su(2,2/1)$ \FF{} and then
for the extended superconformal algebras
~$su(2,2/N)$ \DPu\ (for arbitrary $N$). Then in the 1990s
the classification for ~$D=3$ (for even $N$), $D=5$, and $D=6$
(for $N=1,2$)~ was given in \Min{} (some results being
conjectural). Later,  the $D=6$ case (for arbitrary $N$) was
finalized in \Dosix. Then  in \DoZh\ the classification was done
for  $osp(1|2n,R)$, which is relevant for ~$D=9,10,11$.

Once we know the UIRs of a (super-)algebra the next question is to find
their characters, since these give the spectrum which is
important for the applications, especially, in super-Yang-Mills theories, super conformal theories,
higher spin symmetries, spin chains, superconformal QCD, etc.
cf. e.g., \refs{\BBMS,\BGHR,\HJS,\KMMR,\Dolb,\BBMR,\GPR}.
Some results on the spectrum
of $su(2,2/N)$ were given in the early papers \refs{\Sie,\GuMa,\DPu} but it is
necessary to have systematic results for which the character formulae
are needed. From the mathematical
point of view this question is clear only for representations
with conformal dimension above the unitarity threshold
viewed as irreps of the corresponding complex superalgebra
$sl(4/N)$. But for $su(2,2/N)$ even the UIRs above the unitarity
threshold are truncated for small values of spin and isospin.
This question was addressed in \Dobch{}   for the UIRs of
$D=4$ conformal superalgebras $su(2,2/N)$. There the
general theory for the characters of $su(2,2/N)$ was developed  in great detail.
For the general theory  we used the odd reflections introduced in \DPf\
(see also \Serg).   In \Dobch\ we also pin-pointed
the difference between character formulae for $sl(4,N)$ and
$su(2,2/N)$ since for the latter we needed to introduce and use the
notion of counter-terms.\foot{For an alternative approach to character formulae
see \refs{\Dola,\BDHO}.}

The general formulae given in \Dobch\ are for the so-called bare characters
(or superfield decompositions) and are valid for arbitrary $N$.
However, the even entries are not given explicitly.

Thus, the present paper is  a follow-up of \Dobch. The idea is to give
   the complete explicit character formulae including the even parts of the
   characters.

We need detailed knowledge about the structure of the UIRs
from the representation-theoretical point of view
which is contained in \refs{\DPu,\DPf,\DPm,\DPp}. Following
these   papers in Section 2 we recall the basic ingredients of the
representation theory of the D=4 superconformal algebras.
In Section 3 we recall the necessary ingredients of character theory,
including the character formulae of ~$su(2,2)$~ and ~$su(N)$, for which we
give  explicitly all formulae that we need.
In section 4 we give the explicit complete character formulae for $N=1$ and
for a number of important examples for $N=2,4$.

\newsec{Representations of D=4 conformal supersymmetry}

\subsec{The setting}

\nt The superconformal algebras in $D=4$ are ~$\cg ~=~ su(2,2/N)$.
The even subalgebra of ~$\cg$~ is the algebra ~$\cg_0 ~=~ su(2,2)
\oplus u(1) \oplus su(N)$. We label their physically relevant
representations of ~$\cg$~ by the signature: \eqn\sgn{\chi ~=~
[\,d\,;\, j_1\,,\,j_2\,;\,z\,;\,r_1\,,\ldots,r_{N-1}\,]} where ~$d$~
is the conformal weight, ~$j_1,j_2$~ are non-negative
(half-)integers which are Dynkin labels of the finite-dimensional
irreps of the $D=4$ Lorentz subalgebra ~$so(3,1)$~ of dimension
~$(2j_1+1)(2j_2+1)$, ~$z$~ represents the ~$u(1)$~ subalgebra which
is central for ~$\cg_0$~ (and for $N=4$ is central for $\cg$
itself), and ~$r_1,\ldots,r_{N-1}$~ are non-negative integers which
are Dynkin labels of the finite-dimensional irreps of the internal
(or $R$) symmetry algebra ~$su(N)$.

We recall that the   approach to $D=4$ conformal
supersymmetry developed in \refs{\DPu,\DPf,\DPm,\DPp} involves two related
constructions - on function spaces and as Verma modules. The first
realization employs the explicit construction of
representations of ~$\cg$~ (and of the corresponding supergroup ~$G
~=~ SU(2,2/N)$) induced from parabolic subalgebras (subgroups)
in spaces of functions (superfields) over superspace
which are called elementary representations (ER), cf. \DPf. The UIRs of $\cg$
are realized as irreducible components of ERs, and then they
coincide with the usually used superfields in indexless notation.
This construction is canonical, yet we should mention that some
of the resulting superspaces were obtained  in the papers \GIKOS,\GIOS,\AFSZ,\FS,\FSa, using the notions of
'harmonic superspace analyticity' and 'Grassmann analyticity'. The relation between the latter approach
and ours were commented on in \DPu,\DPf,\Dobch.\nl
The Verma module realization is also very useful as it provides
simpler and more intuitive picture for the relation between
reducible ERs, for the construction of the irreps, in particular, of
the UIRs.   Here we shall actually use the second - Verma module - construction, though
we shall mention occasionally the superfield approach.

\vskip 5mm

\subsec{Verma modules}

\nt To introduce Verma modules one needs the standard triangular
decomposition: \eqn\trig{ \cg^\bac ~=~ \cg^+ \oplus \ch \oplus
\cg^-} where ~$\cg^\bac ~=~ sl(4/N)$~ is the complexification of
$\cg$, ~~$\cg^+$, $\cg^-$, resp., are the subalgebras corresponding
to the positive, negative, roots of ~$\cg^\bac$, resp., and $\ch$
denotes the Cartan subalgebra of ~$\cg^\bac$.

We consider lowest weight Verma modules, so that ~$V^\L ~ \cong
U(\cg^+) \otimes v_0\,$,
 where ~$U(\cg^+)$~ is the universal enveloping algebra of $\cg^+$,
~$\L\in\ch^*$~ is the lowest weight, and ~$v_0$~ is the lowest
weight vector $v_0$ such that: \eqn\low{\eqalign{
 X \ v_0\ =&\ 0 \ , \quad X\in \cg^- \ , \cr
 H \ v_0 \ =&\ \L(H)\ v_0 \ , \quad H\in \ch \ .}}
Further, for simplicity we omit the sign ~$\otimes \,$, i.e., we
write $P\,v_0\in V^\L$ with $P\in U(\cg^+)$.

The lowest weight $\L$ is characterized by its values on the Cartan
subalgebra ~$\ch$, or, equivalently, by its products with the simple
roots (given explicitly below). In general, these would be ~$N+3$~
complex numbers, however, in order to be useful for the
representations of the real form ~$\cg$~ these values would be
restricted to be real and furthermore to correspond to the
signatures ~$\chi\,$, and we shall write ~$\L = \L(\chi)$, or ~$\chi
= \chi(\L)$. Note, however, that there are Verma modules to which
correspond no ERs, cf. \DPf{} and below.

If a Verma module ~$V^\L$~ is irreducible then it gives the lowest
weight irrep ~$L_\L$~ with the same weight. If a Verma module
~$V^\L$~ is reducible then it contains a maximal invariant submodule
~$I^\L$~ and the lowest weight irrep ~$L_\L$~ with the same weight
is given by factorization: ~$L_\L ~=~ V^\L\,/\,I^\L$ \Di,\Kac,\Kacl.

Thus, we need first to know which Verma modules are reducible. The
reducibility conditions for highest weight Verma modules over basic
classical Lie superalgebra were given by Kac \Kacl. Translating his
conditions to lowest weight Verma modules we have \DPf: ~~~A lowest
weight Verma module ~$V^\L$~ is reducible only if at least one of
the following conditions is true:\foot{Many statements below are
true for any basic classical Lie superalgebra, and would require
changes only for the superalgebras $osp(1/2N)$.} \eqna\redg
$$\eqalignno{ (\r - \L, \b) ~=& ~m (\b,\b)/2 \ , \qquad \b \in \D^+ \ ,
~~(\b,\b) \neq 0\ ,\quad m\in\bbn \ , &\redg a\cr (\r - \L, \b) ~=&
~0 \ , \qquad \b \in \D^+ \ , ~~(\b,\b) = 0\
 , &\redg b\cr} $$
where ~$\Delta^+$~ is the positive root system of ~$\cg^\bac$,
~$\r\in\ch^*$~ is the very important in representation theory
element given by ~$\r = \r_\o - \r_\I\,$, where ~$\r_\o\,,\r_\I$~
are the half-sums of the even, odd, resp., positive roots,
~$(\cdot,\cdot)$~ is the standard bilinear product in ~$\ch^*$.

If a condition from \redg{a} is fulfilled then ~$V^\L$~ contains a
submodule which is a Verma module ~$V^{\L'}$~ with shifted weight
given by the pair ~$m,\b$~: ~$\L' ~=~ \L + m\b$. The embedding of
~$V^{\L'}$~ in ~$V^\L$~ is provided by mapping the lowest weight
vector ~$v'_0$~ of ~$V^{\L'}$~ to the singular vector ~$v_s^{m,\b}$~
in ~$V^\L$~ which is completely determined by the conditions:
\eqn\lowp{\eqalign{
 X \ v_s^{m,\b}\ =&\ 0 \ , \quad X\in \cg^- \ , \cr
 H \ v_s^{m,\b} \ =&\ \L'(H)\ v_0 \ , \quad H\in \ch \ ,
~~~\L' ~=~ \L + m\b\ .}} Explicitly, ~$v_s^{m,\b}$~ is given by an
even polynomial in the positive root generators: \eqn\sing{
v_s^{m,\b} ~=~ P^{m,\b} \,v_0 \ , \quad P^{m,\b}\in U(\cg^+)\ .}
Thus, the submodule of ~$V^\L$~ which is isomorphic to ~$V^{\L'}$~
is given by ~$U(\cg^+)\, P^{m,\b} \,v_0\,$. [More on the even case
following the same approach may be seen in, e.g., \Doo,\Dor.]

If a condition from \redg{b} is fulfilled then ~$V^\L$~ contains a
submodule ~$I^{\b}$~ obtained from the Verma module ~$V^{\L'}$~ with
shifted weight ~$\L' ~=~ \L + \b$ as follows. In this situation
~$V^\L$~ contains a singular vector \eqn\lows{\eqalign{
 X \ v_s^{\b}\ =&\ 0 \ , \quad X\in \cg^- \ , \cr
 H \ v_s^{\b} \ =&\ \L'(H)\ v_0 \ , \quad H\in \ch \ ,
~~ \L' ~=~ \L + \b }} Explicitly, ~$v_s^{\b}$~ is given by an odd
polynomial in the positive root generators: \eqn\sing{ v_s^{\b} ~=~
P^{\b} \,v_0 \ , \quad P^{\b}\in U(\cg^+) \ .} Then we have:
\eqn\subm{ I^{\b} ~=~ U(\cg^+)\, P^{\b} \,v_0 } which is smaller
than ~$V^{\L'} ~=~ U(\cg^+)\, v'_0$~ since this polynomial is
Grassmannian: \eqn\gras{\( P^{\b} \)^2 ~=~ 0 \ .} To describe this
situation we say that ~$V^{\L'}$~ is ~{\bf oddly embedded}~ in
~$V^{\L}$.

Note, however, that the above formulae describe also more general
situations when the difference ~$\L'-\L = \b $~ is not a root, as
used in \DPf, and below.

The weight shifts ~$\L' = \L + \b$, when ~$\b$~ is an odd root are
called ~{\bf odd~reflections}~ in \DPf, (see also \Serg).
 We recall from \DPf\ the
definition of the odd reflection ~$s_\alpha\,$,
~$\alpha\in\Delta_{\bar 1}\,$, ~acting on ~$\g\in\Delta_{\bar 1}$~:
\eqna\oddr
$$\eqalignno{
&s_\alpha\cdot \g = \g - (\g,\alpha^\vee)\,\alpha\ ,
\qquad (\alpha,\alpha)\neq 0 \ ,&\oddr{}\cr &s_\alpha\cdot \g = \g -
(-1)^{p(\g)}(\g,\alpha)\, \alpha\ , \qquad (\alpha,\alpha)= 0\ ,\quad
\alpha\neq \g \ , \cr &s_\alpha\cdot \alpha = -\alpha\ , \qquad
(\alpha,\alpha)= 0\ , }
$$ where $(\cdot,\cdot)$ is the standard bilinear product in
$\ch^*$, ~$p(\g)$~ is the parity of $\g$~:
\eqn\parr{ p(\g) = \cases{0 & if ~$\g\in\Delta_{\bar 0}\,$ \cr
1 & if ~$\g\in\Delta_{\bar 1}\,$ }}

Further, to be more explicit we need to recall the root
system of ~$\cg^\bac$~ - for definiteness - as used in \DPf{}. The
positive root system ~$\D^+$~ is comprised from ~$\a_{ij}\,$,
~$1\leq i <j \leq 4+N$, ~
the even positive root system ~$\D^+_\o$~
is comprised from ~$\a_{ij}\,$, with\ $i,j\leq 4$~ and ~$i,j\geq
5$; ~the odd positive root system ~$\D^+_\I$~
is comprised from ~$\a_{ij}\,$, with ~$i\leq 4, j \geq 5$.
The even system is actually the root system of ~$sl(4) \otimes sl(N)$~
with simple roots ~$\{\a_1,\a_2,\a_3\}$, ~$\{\a_5,\ldots,\a_{3+N}\}$,
resp., where ~$\a_j \equiv \a_{j,j+1}\,$.
The simple roots of the superalgebra are chosen as in (2.4) of \DPf{}:
\eqn\smplr{ \g_1 = \a_{12}\,, ~\g_2 = \a_{34}\,, ~\g_3 = \a_{25}\,,
~\g_4 = \a_{4,4+N}\,, ~\g_k = \a_{k,k+1}\,, ~ 5\leq k\leq 3+N}
Thus, the Dynkin diagram is:
\eqn\dynk{
\vbox{\offinterlineskip\baselineskip=10pt
\halign{\strut#
\hfil
& #\hfil
\cr &\cr
& $\white{{1}} \riga \gray{{3}} \riga \white{{5}} \riga
\cdots \riga \white{{3+N}} \riga
\gray{{4}} \riga \white{{2}}$
\cr }}}
This is a non-distinguished simple root system with two odd
simple roots (for the various root systems of the basic classical
superalgebras we refer to \Kacs).

We choose this diagram since it has a mirror symmetry (conjugation):
\eqna\mirrs
$$\eqalignno{
& \g_1 \lra \g_2\ , &\mirrs{a}\cr
& \g_3 \lra \g_4\ , &\mirrs{b}\cr
& \g_j \lra \g_{N+8-j}\ , \quad j\geq 5 \ ,&\mirrs{c}\cr
}$$
and furthermore it is consistent with the mirror symmetry
of the $sl(4)$ and $sl(N)$ root systems by identifying: $\g_1\mt \a_1\,,\g_2 \mt \a_3\,$,
and $\g_j \mt \a_j\,$, $j\geq 5$, resp.

The reducibility conditions w.r.t. to the positive roots coming
from ~$sl(4)(su(2,2))$~ coming from \redg{} (denoting $m\to
n_{j}$ for $\b\to \a_{j}$) are:
\eqna\redd
$$\eqalignno{
n_{1} ~~=&~~ 1 + 2j_1
&\redd a\cr
n_{2} ~~=&~~ 1 - d - j_1 - j_2
&\redd b\cr
n_{3} ~~=&~~ 1 + 2j_2 &\redd c\cr
n_{12} ~~=&~~ 2 - d + j_1 - j_2
~~=~~ n_1 + n_2 &\redd d\cr
n_{23} ~~=&~~ 2 - d - j_1 + j_2
~~=~~ n_2 + n_3 &\redd e\cr
n_{13} ~~=&~~ 3 - d + j_1 + j_2
~~=~~ n_1 + n_2 + n_3 &\redd f\cr}$$

Thus, reducibility conditions \redd{a,c} are fulfilled automatically
for ~$\L(\chi)$~ with $\chi$ from \sgn{} since we always have:
~$n_1,n_3\in\bbn$.
 There are no such conditions for the ERs since they are induced
from the finite-dimensional irreps of the Lorentz subalgebra
(parametrized by $j_1,j_2\,$.) However, to these two conditions
correspond differential operators of order ~$1+2j_1$~ and ~$1+2j_2$~
(as we mentioned above) and these annihilate all functions of the
ERs with signature $\chi$.

The reducibility conditions w.r.t. to the positive roots coming
from ~$sl(N)(su(N))$~ are all fulfilled for ~$\L(\chi)$~
with $\chi$ from \sgn.
In particular, for the simple roots from those
condition \redg{} is fulfilled with ~$\b\to\g_j\,$, $m=1+r_{N+4-j}\,$.
for every $j=5,6,\ldots,N+3$.
There are no such conditions for the ERs since they are induced
from the finite-dimensional UIRs of $su(N)$.
However, to these $N-1$ conditions correspond $N-1$ differential
operators of orders $1+r_k$ (as we mentioned) and
the functions of our ERs are annihilated by all
these operators \DPf.\foot{Note that there are actually as many
operators as positive roots of $sl(N)$ but all are expressed in
terms of the $N-1$ above corresponding to the simple roots \DPf.}

For future use we note also the following decompositions:
\eqna\lamde
$$\eqalignno{ \L ~=&~ \sum_{j=1}^{N+3}\ \l_j\ \a_{j,j+1} ~=~ \L^s + \L^z + \L^u
&\lamde a\cr
\L^s \equiv& \sum_{j=1}^{3}\ \l_j\ \a_{j,j+1}\ , \quad \L^z \equiv \l_4\ \a_{45} \ , \quad
\L^u \equiv \sum_{j=5}^{N+3}\ \l_j\ \a_{j,j+1}\ &\lamde b\cr}$$
which actually employ the distinguished root system with one odd root $\a_{45}\,$.

The reducibility conditions for the
~$4N$~ odd positive roots of $\cg$ are \DPu,\DPf:
\eqna\redu
$$\eqalignno{
& d ~=~ d^1_{Nk} - z \d_{N4} &\redu {a.k}\cr
& d^1_{Nk} ~\equiv~ 4-2k +2j_2 +z+2m_k -2m/N \cr &&\cr
& d ~=~ d^2_{Nk} - z \d_{N4} &\redu {b.k}\cr
& d^2_{Nk} ~\equiv~ 2-2k -2j_2 +z+2m_k -2m/N \cr &&\cr
& d ~=~ d^3_{Nk} + z \d_{N4} &\redu {c.k}\cr
& d^3_{Nk} ~\equiv~ 2+2k-2N +2j_1 -z-2m_k +2m/N \cr &&\cr
& d ~=~ d^4_{Nk} + z \d_{N4} &\redu {d.k}\cr
& d^4_{Nk} ~\equiv~ 2k-2N -2j_1 -z-2m_k +2m/N \cr }$$
where in all four cases of \redu{} ~$k=1,\ldots,N$, ~$m_N\equiv 0$, and
\eqn\mkm{
m_k \equiv \sum_{i=k}^{N-1} r_i \ , \quad
m \equiv \sum_{k=1}^{N-1} m_k = \sum_{k=1}^{N-1} k r_k}
$m_k$ is the number of cells of the $k$-th row of the standard Young tableau,
$m$ is the total number of cells.
Condition \redu{a.k} corresponds to the root ~$\a_{3,N+5-k}\,$,
\redu{b.k} corresponds to the root ~$\a_{4,N+5-k}\,$,
\redu{c.k} corresponds to the root ~$\a_{1,N+5-k}\,$,
\redu{d.k} corresponds to the root ~$\a_{2,N+5-k}\,$.

Note that for a fixed module and fixed ~$i=1,2,3,4$~ only one of
the odd ~$N$~ conditions involving ~$d^i_{Nk}$~ may be satisfied. Thus, no
more than four (two, for $N=1$) of the conditions \redu{} may hold for a given Verma module.

\nt
{\it Remark:}~~ {\iti Note that for ~$n_2\in\bbn$~ (cf. \redd{})
 the corresponding irreps of ~su(2,2)~ are finite-dimensional
 (the necessary and sufficient condition for this is:
~$n_1,n_2,n_3\in\bbn$).
Then the irreducible LWM ~L$_\L$~ of ~su(2,2/N)~ are also
finite-dimensional (and non-unitary) independently on whether
the corresponding Verma module ~V$^\L$~
is reducible w.r.t. any odd root.
If ~V$^\L$~ is not reducible w.r.t. any odd root, then these
finite-dimensional irreps are called 'typical' \Kacl, otherwise,
the irreps are called 'atypical' \Kacl. In our considerations
~$n_2\notin\bbn$~ in all cases, except the trivial 1-dimensional
UIR (for which $n_2=1$, cf. below).}\dia

We shall consider quotients of Verma modules
factoring out the ~{\it even}~ submodules for which the reducibility
conditions are always fulfilled. Before this we recall
the root vectors following \DPf. The positive (negative) root vectors
corresponding to ~$\a_{ij}$, ($-\a_{ij}$), are denoted by
~$X^+_{ij}$, ($X^-_{ij}$). The simple root vectors ~$X^+_i$~
follow the notation of the simple roots: ~$X^+_j \equiv X^+_{j,j+1}\,$.

The mentioned submodules are generated by the singular vectors
related to the even simple roots
~$\a_1,\a_3,\a_5,\ldots,\a_{N+3}$~\DPf{}:
\eqna\sings
$$\eqalignno{ v^1_s ~=&~ (X^+_1)^{1+2j_1}\, v_0 \ ,&\sings a\cr
 v^3_s ~=&~ (X^+_3)^{1+2j_2}\, v_0 \ , &\sings b\cr
v^j_s ~=&~ (X^+_j)^{1+r_{N+4-j}}\, v_0 \ ,
\quad j=5,\ldots,N+3 &\sings c\cr}$$
(for ~$N=1$~ \sings{c} being empty).
The corresponding submodules are ~$I^\L_k ~=~ U(\cg^+)\,v^k_s\,$,
and the invariant submodule to be factored out is:
\eqn\subb{ I^\L_c ~=~ \bigcup_k\, I^\L_k }
Thus, instead of ~$V^\L$~ we shall consider the factor-modules:
\eqn\fcc{ \tV^\L ~=~ V^\L\, /\, I^\L_c }
which are closer to the structure of the ERs.\foot{For
explicit expressions for the $sl(n)$ singular vectors we refer to \Dobsin.}
In the factorized modules the singular vectors \sings{} become
null conditions, i.e., denoting by ~$\widetilde{|\L\rg}$~ the
lowest weight vector of ~$\tV^\L$, we have:
\eqna\nulm
$$\eqalignno{ &(X^+_1)^{1+2j_1}\, \widetilde{|\L\rg} ~=~ 0 \ , &\nulm a\cr
&(X^+_3)^{1+2j_2}\, \widetilde{|\L\rg} ~=~ 0 \ ,
&\nulm b\cr
&(X^+_j)^{1+r_{N+4-j}}\, \widetilde{|\L\rg} ~=~ 0 \ ,
\quad j=5,\ldots,N+3 &\nulm c\cr}$$

\vskip 5mm

\subsec{Singular vectors and invariant submodules
at the unitary reduction points}

\nt
We first recall the result of \DPu{} (cf. part (i) of the Theorem
there) that the following is the complete list of lowest weight
(positive energy) UIRs of $su(2,2/N)$~:
\eqna\unitt
$$\eqalignno{
&d ~\geq~ d_{\rm max} ~=~ \max (d^1_{N1}, d^3_{NN})\ , &\unitt a\cr
&d ~=~ d^4_{NN} \geq d^1_{N1}\ , ~~ j_1=0\ , &\unitt b\cr
&d ~=~ d^2_{N1} \geq d^3_{NN}\ , ~~ j_2=0\ ,&\unitt c\cr
&d ~=~ d^2_{N1} = d^4_{NN}\ , ~~ j_1 = j_2=0\ ,&\unitt d\cr }$$
where ~$d_{\rm max}$~ is the threshold of the continuous unitary
spectrum.

 Note that in case (d) we have $d=m_1$, $z=2m/N -m_1\,$,
and that it is trivial for $N=1$ since then the internal symmetry
algebra $su(N)$ is trivial and by definition $m_1=m=0$
(the resulting irrep is 1-dimensional with
$d=z=j_1=j_2=0$). The UIRs for N=1 were first given in \FF.

Next we note that if ~$d ~>~ d_{\rm max}$~ the factorized Verma
modules are irreducible and coincide with the UIRs ~$L_\L\,$.
These UIRs are called ~${\bf long}$~ in the modern literature,
cf., e.g., \FGPW,\CDDF,\FS,\FSa,\AES,\BKRSa,\EdSo,\HeHoa.
Analogously, we shall use for the cases when ~$d= d_{\rm
max}\,$, i.e., \unitt{a}, the terminology of ~{\bf semi-short}~ UIRs,
introduced in \FGPW,\CDDF,\FSa,
while the cases \unitt{b,c,d} are also called ~{\bf
short}~ UIRs, cf., e.g., \CDDF,\FS,\FSa,\AES,\BKRSa,\EdSo,\HeHoa.

Next consider in more detail the UIRs at the four distinguished
reduction points determining the list above:
\eqn\diss{\eqalign{
& d^1_{N1} ~=~ 2 +2j_2 +z+2m_1 -2m/N\ , \cr
& d^2_{N1} ~=~ z+2m_1 -2m/N \ , \quad (j_2=0)\ ,\cr
& d^3_{NN} ~=~ 2 +2j_1 -z +2m/N \ ,\cr
& d^4_{NN} ~=~ -z +2m/N \ , \quad (j_1=0)\ . }}

The singular vectors corresponding
to these points were given in \DPf, (8.9a), (8.7b), (8.8a), (8.7a).
{}From the expressions of the singular vectors follow, using
\subm{}, the explicit formulae for the corresponding
invariant submodules ~$I^\b$~ of the modules ~$\tV^\L$~:
\eqna\isub
$$\eqalignno{
I^1 ~=&~
 U(\cg^+)  \( 2j_2
X^+_{3,4+N} - X^+_4 X^+_2 \)   \widetilde{|\L\rg},\qquad
d=d^1_{N1}\,,\ j_2 > 0  &\isub {a}\cr I^2 ~=&~ U(\cg^+)\,
X^+_4\, \widetilde{|\L\rg} \ ,\qquad d=d^2_{N1} \ , &\isub b\cr I^3
~=&~
 U(\cg^+)\, \( 2j_1 X^+_{15} - X^+_3 X^+_1 \)\,
\widetilde{|\L\rg} \ ,\qquad d=d^3_{NN}\ , \ j_1> 0  &\isub
{c}\cr I^4 ~=&~ U(\cg^+)\, X^+_3\, \widetilde{|\L\rg}\ ,\qquad
d=d^4_{NN} \ , &\isub d\cr I^{12} ~=&~ U(\cg^+)\,
\tv^{12} ~=~
X^+_4\,X^+_2\,X^+_4\, \widetilde{|\L\rg} \ ,\qquad d=d^1_{N1} \ ,\
j_2 = 0 \ , &\isub e\cr I^{34} ~=&~ U(\cg^+)\, \tv^{34} ~=~
X^+_3\,X^+_1\,X^+_3 \, \widetilde{|\L\rg} \ ,\qquad d=d^3_{NN}\ , \
j_1 = 0\ . &\isub f\cr }$$

In the cases \isub{a-d} to the singular vectors in the ER picture there
correspond first-order super-differential operators given explicitly in
formulae (7) of \DPu. The invariant submodules are the kernels of these
super-differential operators.\nl
 Note that there is a subtlety for $d^1_{N1}$, $d^3_{NN}$, when
$j_2=0$, $j_1=0$, resp., since in these cases the invariant
submodules in \isub{a}, \isub{c}, resp., trivialize (as $X^+_2  \,
\widetilde{|\L\rg}=0$, $X^+_1  \, \widetilde{|\L\rg}=0$, resp.). The
embeddings which replace them are given in \isub{e,f} and they arise
from singular vectors ~$\tv^{12}$, $\tv^{34}$, which correspond to
~$\b=\a_{3,4+N}+\a_{4,4+N}$, ~$\b =\a_{15}+\a_{25}$ (both sums are
not roots!), see \DPf\ the formula before (D.4), formula (D.1),
resp.\foot{ Note that w.r.t. $V^\L$ the analogues of the vectors
$\tv^{34}$ and $\tv^{12}$ are not singular, but subsingular vectors,
cf. \Dobch.}
 To the last two singular vectors in the ER picture correspond
second-order super-differential operators given explicitly in
formulae (11a,b) of \DPu, and in formulae (D3),(D5) of \DPf.

The invariant submodules \isub{} were used in \DPu{} in the construction
of the UIRs.

\vskip 5mm


 We note a partial ordering of the four
distinguished points \diss{} of reducibility of Verma modules:
 \eqn\parto{ d^1_{N1} ~>~ d^2_{N1} \ , \qquad d^3_{NN}
~>~ d^4_{NN} \ .}  Due to this ordering at most two of these four points
may coincide. Thus, we have two possible situations: of Verma
modules (or ERs) reducible at one and at two reduction points from
\diss{}.

 First  we deal with the situations in which ~{\it no
two}~ of the points in \diss{} coincide. According to \DPu{}
(Theorem) there are four such situations involving UIRs:
\eqna\dist
$$\eqalignno{
&{\bf a}\qquad d ~=~ d_{\rm max} ~=~ d^1_{N1} > d^3_{NN}\ , &\dist a\cr
&{\bf b}\qquad d ~=~ d^2_{N1} > d^3_{NN}\ , ~~ j_2=0\ ,&\dist b\cr
&{\bf c}\qquad d ~=~ d_{\rm max} ~=~ d^3_{NN} > d^1_{N1}\ , &\dist c\cr
&{\bf d}\qquad d ~=~ d^4_{NN} > d^1_{N1}\ , ~~ j_1=0\ . &\dist d\cr}$$

We shall call these cases ~{\bf single-reducibility-condition
(SRC)}~ Verma modules or UIRs, depending on the context.
In addition, as already stated, we use for the cases when ~$d= d_{\rm
max}\,$, i.e., \dist{a,c}, the terminology of semi-short UIRs,
  while the cases \dist{b,d} are also called
short UIRs.

The factorized Verma modules ~$\tV^\L$~ with the unitary signatures
from \dist{} have only one invariant (odd) submodule which has to be
factorized in order to obtain the UIRs.
These odd embeddings are given explicitly as:
\eqna\embs
$$\eqalignno{&\tV^\L ~\rightarrow~ \tV^{\L+\b} \ ,&\embs{}\cr
&\L+\b ~=~ s_\b \cdot \L }$$
where we use the convention \DPm{} that arrows point to the
oddly embedded module, and there are the following cases for $\b$~:
\eqna\paorz
$$\eqalignno{
\b ~=&~ \a_{3,4+N}\ , \quad {\rm for}~ \dist{a}, \quad j_2> 0 ,
&\paorz{a} \cr
=&~ \a_{4,4+N}\ , \quad {\rm for}~ \dist{b},
&\paorz{b} \cr
=&~ \a_{15}\ ,\quad {\rm for}~ \dist{c}, \quad
j_1> 0, &\paorz{c}\cr
=&~ \a_{25}\ ,
\quad {\rm for}~ \dist{d}, &\paorz{d}\cr
=&~ \a_{3,4+N}+\a_{4,4+N}\ , \quad {\rm for}~ \dist{a}, \quad
j_2=0, &\paorz{e} \cr
=&~ \a_{15}+\a_{25}\ ,
\quad {\rm for}~ \dist{c},\quad j_1=0 \quad &\paorz{f}
 }$$
Note that in \embs{} in the cases \paorz{e,f}
we have extended the action of the odd reflections, defined in \oddr{} for
root elements, to sums of odd roots which are not roots.

The diagram \embs{} gives the UIR ~$L_\L$~ contained in ~$\tV^\L$~ as follows:
\eqn\genun{
L_\L ~=~ \tV^\L/I^\b \ , }
where ~$I^\b$~ is given by ~$I^1$, $I^2$, $I^3$, $I^4$,
$I^{12}$, $I^{34}$, resp., (cf. \isub{}),
in the cases \paorz{a,b,c,d,e,f}, resp.

It is useful to record the signatures of the shifted lowest weights,
i.e., ~$\chi' ~=~ \chi(\L+\b)$. In fact, for
future use we give the signature changes for arbitrary roots.
The explicit formulae are \DPm,\DPf:
\eqna\sgnn \eqna\sgnnn
$$\eqalignno{
\b=\a_{3,N+5-k}~: & ~\chi' ~=~ [d+\half;\, j_1,j_2-\half;\,
z + \eps_N\,;\,r_1,\ldots,r_{k-1}-1,r_k+1,\ldots,r_{N-1}], \qquad &\sgnn a\cr
& j_2> 0 \ , \quad r_{k-1}> 0 &\sgnn {a'}\cr
&&\cr
\b=\a_{4,N+5-k}~: & ~\chi' ~=~ [d+\half;\, j_1,j_2+\half;\,
z + \eps_N\,;\,r_1,\ldots,r_{k-1}-1,r_k+1,\ldots,r_{N-1}], \qquad &\sgnn b\cr
& r_{k-1}> 0 &\sgnn {b'}\cr
&&\cr
\b=\a_{1,N+5-k}~: & ~\chi' ~=~ [d+\half;\, j_1-\half,j_2\,;\,
z - \eps_N\,;\,r_1,\ldots,r_{k-1}+1,r_k-1,\ldots,r_{N-1}], \qquad &\sgnn c\cr
& j_1> 0 \ , \quad r_{k}> 0 &\sgnn {c'}\cr
&&\cr
\b=\a_{2,N+5-k}~: & ~\chi' ~=~ [d+\half;\, j_1+\half,j_2\,;\,
z - \eps_N\,;\,r_1,\ldots,r_{k-1}+1,r_k-1,\ldots,r_{N-1}],
\qquad\qquad &\sgnn d\cr
& r_{k}> 0 &\sgnn {d'}\cr
&&\cr
&k = 1,\ldots,N\ , \qquad
\eps_N ~\equiv ~ \halfn - \half &\sgnnn {}\ .
 }$$
For each fixed ~$\chi$~ the lowest weight ~$\L(\chi')$~ fulfills
the same odd reducibility condition as ~$\L(\chi')$.
We need also the special cases used in \paorz{e,f}:
$$\eqalignno{
\b_{12}=\a_{3,4+N} + \a_{4,4+N}~: & ~\chi'_{12} ~=~ [d+1;\, j_1,0;\,
z +2 \eps_N\,;\,r_1+2,r_2,\ldots,r_{N-1}], &\sgnn {e}\cr
& j_2= 0,\ d=d^1_{N1} \cr
&&\cr
\b_{34}=\a_{15}+\a_{25}~: & ~\chi'_{34} ~=~ [d+1;\, 0,j_2\,;\,
z -2 \eps_N\,;\,r_1,...,r_{N-2},r_{N-1}+2],\qquad &\sgnn {f}\cr
& j_1= 0,\ d=d^3_{NN} \cr }
$$
The lowest weight $\L(\chi'_{12})$ fulfils \dist{b},
while the lowest weight $\L(\chi'_{34})$ fulfils \dist{d}.

\vskip 5mm


We consider now the situations in which ~{\it two}~ of the points in
\diss{} coincide. According to \DPu{} (Theorem) there are four
such situations involving UIRs:
\eqna\disd
$$\eqalignno{
&{\bf ac}\qquad d ~=~ d_{\rm max} ~=~ d^{ac} ~\equiv~ d^1_{N1} = d^3_{NN}\ , &\disd a\cr
&{\bf ad}\qquad d ~=~ d^1_{N1} = d^4_{NN}\ , ~~ j_1=0\ , &\disd b\cr
&{\bf bc}\qquad d ~=~ d^2_{N1} = d^3_{NN}\ , ~~ j_2=0\ ,&\disd c\cr
&{\bf bd}\qquad d ~=~ d^2_{N1} = d^4_{NN}\ , ~~ j_1=j_2=0\ .&\disd d\cr
}$$

We shall call these ~{\bf double-reducibility-condition (DRC)}
Verma modules or UIRs. As in the previous subsection we shall
use for the cases when ~$d= d_{\rm max}\,$, i.e., \disd{a}, also
the terminology of semi-short UIRs, \FGPW,\FSa,
while the cases \disd{b,c,d} shall also be called
short UIRs \FS,\FSa,\AES,\BKRSa,\EdSo,\HeHoa.

To finalize the structure we should check the even reducibility
conditions \redd{b,d,e,f}. This analysis was done in \Dobch, and the
results are as follows.

The embedding diagrams for the corresponding modules ~$\tV^\L$~
 ~{\it without even embeddings}~ are:
\eqn\embd{
\matrix{
\tV^{\L+\b'}&& \cr
&&\cr
\uparrow &&\cr
&&\cr
\tV^\L &\rightarrow & \tV^{\L+\b} \cr
}}
where ~$\L+\b = s_\b \cdot \L$, ~$\L+\b' = s_{\b'} \cdot \L$,
\eqna\paor
$$\eqalignno{
(\b,\b') ~=&~\cr ~=&~
(\a_{15},\a_{3,4+N}), \quad {\rm for}~ \disd{a}, \quad m_1j_1j_2> 0
&\paor{a} \cr
~=&~ (\a_{15},\a_{3,4+N}+\a_{3,4+N}),
\quad {\rm for}~ \disd{a},\quad j_1> 0,\ j_2=0 &\paor{b} \cr
~=&~ (\a_{15}+\a_{25},\a_{3,4+N}),\quad {\rm for}~ \disd{a}, \quad
j_1=0,\ j_2> 0 &\paor{c} \cr
~=&~
(\a_{15}+\a_{25},\a_{3,4+N}+\a_{3,4+N}),
\quad {\rm for}~ \disd{a},\quad j_1=j_2=0\quad &\paor{d}
\cr ~=&~
(\a_{25},\a_{3,4+N}), \quad {\rm for}~ \disd{b}, \quad j_2> 0 ,
\ 2j_2 +m_1 \geq 2
&\paor{e} \cr ~=&~
(\a_{25},\a_{3,4+N}+\a_{4,4+N}), \quad {\rm for}~ \disd{b}, \quad
j_2=0 , \ m_1 >0
&\paor{f} \cr ~=&~
(\a_{15},\a_{4,4+N}), \quad {\rm for}~ \disd{c}, \quad j_1> 0 ,
\ 2j_1 +m_1 \geq 2
&\paor{g} \cr ~=&~
(\a_{15}+\a_{25},\a_{4,4+N}), \quad {\rm for}~ \disd{c}, \quad
j_1= 0 , \ m_1 >0 &\paor{h} \cr ~=&~
(\a_{25},\a_{4,4+N}), \quad {\rm for}~ \disd{d},
\quad m_1 \neq 1
&\paor{i} }$$
This diagram gives the UIR ~$L_\L$~ contained in ~$\tV^\L$~ as follows:
\eqna\genuna
$$\eqalignno{
L_\L ~=&~ \tV^\L /I^{\b,\b'} \ , \quad
I^{\b,\b'} ~=~ I^\b \cup I^{\b'} &\genuna {}}$$
where ~$I^\b$, $I^{\b'}$ are given in \isub{},
accordingly to the cases in \paor{}.

The embedding diagrams for the corresponding modules ~$\tV^\L$~
~{\it with  even embeddings}~   are:
\eqn\embde{
\matrix{
&&\tV^{\L+\b'}&& \cr
&&&&\cr
&&\uparrow &&\cr
&&&&\cr
\tV^{\L+\b_e} & \leftarrow & \tV^\L & \rightarrow & \tV^{\L+\b} \cr
}}
where ~$\L+\b_e = s_{\b_e} \cdot \L$, (note $n_{\b_e}=1$, \Dobch),
\eqna\paory
$$\eqalignno{
(\b,\b',\b_e) ~=&~\cr ~=&~
(\a_{15},\a_{3,4+N},\a_{14}), \quad {\rm for}~ \disd{a}, \quad j_1j_2> 0,
~~m_1=0
&\paory{a} \cr
~=&~
(\a_{25},\a_{3,4+N},\a_{24}),
\quad {\rm for}~ \disd{b}, \quad
j_2=\ha\,,\ m_1=0\qquad
&\paory{b} \cr
~=&~
(\a_{25},\a_{3,4+N}+\a_{4,4+N},\a_{23}+\a_{14}),
\quad {\rm for}~ \disd{b}, \quad j_2=m_1 = 0
&\paory{c} \cr
~=&~
(\a_{15},\a_{4,4+N},\a_{13}), \quad {\rm for}~ \disd{c}, \quad
j_1=\ha\,,\ m_1=0
&\paory{d} \cr ~=&~
(\a_{15}+\a_{25},\a_{4,4+N},\a_{23}+\a_{14}),
 \quad {\rm for}~ \disd{c}, \quad j_1=m_1 = 0 \qquad
&\paory{e} \cr ~=&~
(\a_{25},\a_{4,4+N},\a_{23}+\a_{14}), \quad {\rm for}~ \disd{d},
\quad m_1 = 1
&\paory{f} }$$
This diagram gives the UIR ~$L_\L$~ contained in ~$\tV^\L$~ as follows:
\eqna\genunaz
$$\eqalignno{
L_\L ~=&~ \tV^\L /I^{\b,\b',\b_e} \ , \quad
I^{\b,\b'} ~=~ I^\b \cup I^{\b'}\cup \tV^{\L+\b_e} &\genunaz {}}$$

Naturally, the two odd embeddings in
\embd{} or \embde{} are the combination of the different cases of
\embs{}.

\bigskip

\newsec{Character formulae of positive energy UIRs}

\subsec{Character formulae: generalities}

\nt
In the beginning of this subsection we follow \Di.
Let ~$\hcg$~ be a ~{\it simple Lie algebra}~ of rank ~$\ell$~
with Cartan subalgebra
~$\hch$, root system ~$\hD$, simple root system ~$\hpi$.
Let ~$\G$, (resp. $\G_+$), be the set of all integral, (resp.
integral dominant), elements of $\hch^*$, i.e., $\l
\in \hch^*$ such that $(\l , \a_i^\vee) \in \bbz$, (resp. $\bbz_+$),
for all simple roots $\a_i\,$, ($\a_i^\vee \equiv 2\a_i/(\a_i,\a_i)$).
Let $V$ be a lowest weight module with lowest weight $\L$ and
lowest weight vector $v_0\,$. It has the following decomposition:
\eqn\wei{ V ~=~ \mathop{\oplus}\limits_{\mu\in\G_+} V_\mu ~~, \ \
~~~V_\mu ~=~ \{ u \in V ~\vert ~Hu = (\l + \mu)(H)u, \
\forall ~H\in\ch \} }
(Note that $V_0 = \bbc v_0\,$.) Let $E(\ch^*)$ be
the associative abelian algebra consisting of the series
$\sum_{\mu \in \ch^*} c_{\mu} e(\mu)$ , where $c_{\mu} \in \bbc ,
 ~c_{\mu} = 0$ for $\mu$ outside the union of a finite number of
sets of the form $D(\l) = \{ \mu \in \ch^* \vert \mu \geq
\l \}$~, ~using some ordering of $\ch^*$,
e.g., the lexicographic one; the formal
exponents $e(\mu)$ have the properties:~
$e(0) = 1, \ e(\mu) e(\nu) = e(\mu + \nu)$.

Then the (formal) character of $V$ is defined by:
\eqn\cha{ch_0~V ~~=~~ \sum_{\mu \in\G_+} (\dim \ V_\mu) ~e(\L+\mu)
 ~~=~~ e(\L) \sum_{\mu\in\G_+} (\dim \ V_\mu) ~e(\mu) }
(We shall use subscript '0' for the even case.)

For a Verma module, i.e.,
~$V = V^\L$~ one has $\dim \ V_\mu = P(\mu)$,
where ~$P(\mu)$ is a generalized
partition function, $P(\mu) = \#$ of ways $\mu$ can be presented
as a sum of positive roots $\beta$, each root taken with its
multiplicity $\dim \CG_{\beta}$ ($=1$ here),
$P(0)\equiv 1$. Thus, the character formula for Verma modules is:
\eqn\chv{ch_0~V^\L ~~=~~ e(\L)\sum_{\mu\in\G_+} P(\mu) e(\mu) ~~=~~ e(\L)
\prod_{\a \in\D^+}(1 - e(\a))^{-1} }

Further we recall the
standard reflections in $\hch^*$~:
\eqn\rfl{s_\a(\l) ~=~ \l - (\l, \a^\vee)\a \,,
\quad \l \in \hch^* \,, \quad \a\in\hD }
The Weyl group ~$W$~ is generated by the simple reflections
$s_i \equiv s_{\a_i}$, $\a_i\in\hpi\,$.
Thus every element $w\in W$ can
be written as the product of simple reflections. It is said that
$w$ is written in a reduced form if it is written with the
minimal possible number of simple reflections; the number of
reflections of a reduced form of $w$ is called the length of $w$,
denoted by $\ell (w)$.

The Weyl character
formula for the finite-dimensional irreducible LWM $L_\L$
over ~$\hcg$, i.e., when ~$\L\in -\G_+\,$,
has the form:\foot{A more general character formula involves
the Kazhdan--Lusztig polynomials  \KL.}
\eqn\chm{
ch_0~L_\L ~~=~~ \sum_{w\in W}(-1)^{\ell(w)} ~ch_0~V^{w\cdot\L} \,, \quad
\L\in -\G_+}
where the dot ~$\cdot$~ action is defined by $w\cdot \l = w(\l -
\r) +\r$.
For future reference we note:
\eqn\rft{ s_\a ~\cdot ~\L ~~=~~ \L ~+~ n_\a \a }
where
\eqn\nchs{ n_\a ~~=~~ n_\a(\L) ~~\doteq~~ (\r-\L,\a^\vee)
~~=~~ (\rho-\L)(H_\a)
\,, \quad \a\in\D^+ }

\vskip 5mm

In the case of ~{\it basic classical Lie superalgebras}~ the first
character formulae were given by Kac \Kacl,\Kacc.\foot{Kac considers
highest weight modules but his results are immediately
transferable to lowest weight modules.} For all such
superalgebras (except $osp(1/2N)$)
the character formula for Verma modules is \Kacl,\Kacc:
\eqn\chv{ch~V^\L
~=~ e(\L)
\( \prod_{\a \in\D^+_\o}(1 - e(\a))^{-1} \)
\( \prod_{\a \in\D^+_\I}(1 + e(\a)) \)
}
Note that the factor ~$\prod_{\a \in\D^+_\o}(1 - e(\a))^{-1}$~
represents the states of the even sector: ~$V_0^\L ~\equiv~
U((\cg^\bac_+)_{(0)})\,v_0\,$ (as above in the even case),
while ~$\prod_{\a \in\D^+_\I}(1 + e(\a))$
represents the states of the odd sector: ~$\hV^\L ~\equiv~
\(U(\cg^\bac_+)/U((\cg^\bac_+)_{(0)})\)\,v_0\,$. Thus, we
may introduce a character for ~$\hV^\L$~ as follows:
\eqn\chv{ch~\hV^\L ~\equiv~ \prod_{\a \in\D^+_\I}(1 + e(\a)) . }

In our case,
~$\hV^\L$~ may be viewed as the result of all possible applications
of the $4N$ odd generators $X^+_{a,4+k}$ on $v_0\,$, i.e., ~$\hV^\L$~
has ~$2^{4N}$~ states (including the vacuum). Explicitly, the
basis of ~$\hV^\L$~ may be chosen as in \DPp: \eqn\bas{ \eqalign{ \Psi_{\bar \ve}
~=&~ \( \ \prod_{k=N}^1\ (X^+_{1,4+k})^{\ve_{1,4+k}} \) \ \( \
\prod_{k=N}^1\ (X^+_{2,4+k})^{\ve_{2,4+k}} \) \ \times \cr
&\times\ \( \ \prod_{k=1}^N\ (X^+_{3,4+k})^{\ve_{3,4+k}} \) \ \( \
\prod_{k=1}^N\ (X^+_{4,4+k})^{\ve_{4,4+k}} \) \, v_0 \ , \cr &
\ve_{aj} = 0,1}} where ~${\bar \ve}$~ denotes the set of all
~$\ve_{ij}\,$.\foot{The order chosen in \bas{} was important in
the proof of unitarity in \DPu,\DPp{} and for that purpose one
may choose also an order in which the vectors on the first row are
exchanged with the vectors on the second row.}
 Thus, the
character of ~$\hV^\L$~ may be written as: \eqna\chvv
$$\eqalignno{
ch~\hV^\L ~=&~ \sum_{\bar \ve}~ e(\Psi_{\bar \ve}) ~= &\chvv a\cr
 =&~ \sum_{\bar \ve}~
\( \ \prod_{k=1}^N\ e(\a_{1,4+k})^{\ve_{1,4+k}} \)
\ \( \ \prod_{k=1}^N\ e(\a_{2,4+k})^{\ve_{2,4+k}} \) \ \times \cr
&\times\ \( \ \prod_{k=1}^N\ e(\a_{3,4+k})^{\ve_{3,4+k}} \)
\ \( \ \prod_{k=1}^N\ e(\a_{4,4+k})^{\ve_{4,4+k}} \) ~= &\chvv
b\cr ~=&~ \sum_{\bar \ve}~ e \(\ \sum_{k=1}^N \sum_{a=1}^4\
\ve_{a,4+k}\,\a_{a,4+k} \)
&\chvv c\cr
}$$
(Note that in the above formula there is no actual dependence from
$\L$.)

We shall use the above to write for the character of $V^\L$~:
\eqna\chv
$$\eqalignno{
ch~V^\L ~=&~ ch\ \hV^\L\ \cdot\ ch_0\ V_0^{\L} ~= &\chv{}\cr
=&~ \sum_{\bar \ve}~ e \(\ \sum_{k=1}^N \sum_{a=1}^4\
\ve_{a,4+k}\,\a_{a,4+k} \)\ \cdot\ e(\L)\
\( \prod_{\a \in\D^+_\o}(1 - e(\a))^{-1} \) ~= \cr
=&~ \sum_{\bar \ve}~ e \(\ \L \ +\ \sum_{k=1}^N \sum_{a=1}^4\
\ve_{a,4+k}\,\a_{a,4+k} \)\
\( \prod_{\a \in\D^+_\o}(1 - e(\a))^{-1} \) ~=
\cr =&~ \sum_{\bar \ve}~ ch_0\ V_0^{\L \ +\ \sum_{k=1}^N \sum_{a=1}^4\
\ve_{a,4+k}\,\a_{a,4+k} }
}$$
where $ch_0\ V_0^{\L}$ is the character obtained by
restriction of $V^\L$ to $V_0^{\L}$:
\eqn\chevv{ch_0\ V_0^{\L} ~=~ e(\L^z)\ \cdot\ ch_0\ V^{\L^s}\ \cdot\ ch_0\ V^{\L^u} }
where we use the decomposition ~$\L = \L^s + \L^z + \L^u$ from \lamde{a},
and $V^{\L^s}$, $V^{\L^u}$, resp., are Verma modules over the complexifications of
$su(2,2)$, $su(N)$, resp., cf. next Subsection.

 Analogously, for the factorized
Verma modules ~$\tV^\L$~ the character formula is:
\eqn\chlr{\eqalign{
ch~\tV^\L ~=&~ ch\ \hV^\L\ \cdot\ ch_0\ \tV_0^\L ~= \cr
=&~ \sum_{\bar \ve}~ ch_0\ \tV_0^{\L \ +\ \sum_{k=1}^N \sum_{a=1}^4\
\ve_{a,4+k}\,\a_{a,4+k} }}}
where ~$ch_0\ \tV_0^\L$~ is the character obtained by
restriction of ~$\tV^\L$~ to ~ $\tV_0^\L ~\equiv~ U((\cg^\bac_+)_{(0)})\,\wt\,$,
or more explicitly:
\eqn\cheve{ch_0\ \tV_0^\L ~=~ e(\L^z)\ \cdot\ ch_0\ L_{\L^s}\ \cdot\ ch_0\ L_{\L^u} }
where we use the decomposition ~$\L = \L^s + \L^z + \L^u$ from \lamde{a},
and character formulae  for the irreps of the even subalgebra from
next subsection.

Formula \chlr{} represents the expansion of
the corresponding superfield in components, and each component
has its own even character. We see that this expansion is given exactly
by the expansion of the odd character \chvv{}.

We have already displayed how the UIRs ~$L_\L$~ are obtained as
factor-modules of the (even-submodules-factorized) Verma modules
$\tV^\L$. Of course, this factorization means that the odd singular
vectors of $\tV^\L$  are becoming null conditions in
$L_\L\,$. However, this is not enough to determine the character
formulae even when considering our UIRs as irreps of the
complexification $sl(4/N)$. The latter is a well known feature
even in the bosonic case. Here the situation is much more
complicated and much more refined analysis is necessary.

The most important aspect of this analysis is the determination
of the superfield content. This analysis was used in \DPu,\DPp{} but
was not explicated enough. It was made more explicit in \Dobch, which
we use in the present paper.

\vskip 5mm

\subsec{Characters of the even subalgebra}

\nt For the characters of the even subalgebra:
~$\cg^\bac_0 ~=~ sl(4) \oplus gl(1) \oplus sl(N)$~ of
~$\cg^\bac$, we use formulae \chevv,\cheve.  In fact,
since the subalgebra ~$\cg^\bac_0$~ is reductive the
corresponding character formulae will be given by the
products of the character formulae of the two simple factors
~$sl(4)$~ and ~$sl(N)$.

We start with the ~$sl(4)$~ case. We denoted the six positive
roots of $sl(4)$ by $\a_{ij}\,$, $1\leq i<j\leq 4$. (Of course, they are
part of the root system of $sl(4/N)$.) For the simplification of
the character formulae we use notation for the formal
exponents corresponding to the $sl(4)$ simple roots:
~$t_j ~\equiv~ e(\a_{j})$, $j=1,2,3$; then for the three
non-simple roots we have: $e(\a_{13}) = t_1t_2$, $e(\a_{24}) = t_2t_3$,
$e(\a_{14}) = t_1t_2t_3\,$. In terms of these the
character formula for a Verma module over $sl(4)$ is:
\eqn\chas{ch_0~V^{\L^s} ~~=~~ {e(\L^s)\over (1-t_1) (1-t_2) (1-t_3)
(1-t_1t_2) (1-t_2t_3) (1-t_1t_2t_3)} }
where by ~$\L^s$~ we denote the $sl(4)$ lowest weight.

The representations of $sl(4)$ which we consider are
infinite-dimensional. When ~$d>d_{\rm max}$~ then all the numbers:
~$n_2,n_{12},n_{23},n_{13}\,$~ from \redd{} cannot be positive
integers. Then
the only reducibilities of the $sl(4)$ Verma module are related
to the complexification of the Lorentz subalgebra of $su(2,2)$,
i.e., with $sl(2) \oplus sl(2)$,
and the character formula is given by the product of the two
character formulae for finite-dimensional $sl(2)$ irreps.
In short, the $sl(4)$ character formula is:
\eqna\slgen
$$\eqalignno{
ch\ L^2_{d;j_1,j_2} ~=&~ ch_0~V^{\L^s} ~-~
ch_0~V^{\L^s + n_1\a_{12}} ~-~ ch_0~V^{\L^s + n_3\a_{34}}
~+~ ch_0~V^{\L^s + n_1\a_{12} + n_3\a_{34}} ~=\cr
=&~ {e(\L^s)\ (1-t_1^{n_1})\ (1-t_3^{n_3})
 \over (1-t_1) (1-t_2) (1-t_3)
(1-t_1t_2) (1-t_2t_3) (1-t_1t_2t_3)} ~=\cr
=&~ {e(\L^s)
 \over (1-t_2) (1-t_1t_2) (1-t_2t_3) (1-t_1t_2t_3)}\ \left( \sum_{j=0}^{n_1-1}\ t_1^j \right) \
\left( \sum_{k=0}^{n_3-1}\ t_3^k \right) ~=\cr
=&~ {e(\L^s)\over (1-t_2) (1-t_1t_2) (1-t_2t_3) (1-t_1t_2t_3)} \ \cq_{n_1,n_3} (t_1,t_3)\ , \cr
&n_1 = 2j_1+1,\ n_3 = 2j_2+1,\ ~d > d_{\rm max} \ , &\slgen{}
}$$
and we have introduced for later use notation $\cq_{n_1,n_2}$
for the character factorized by $e(\L^s)/((1-t_2) (1-t_1t_2) (1-t_2t_3) (1-t_1t_2t_3))$.
The above formula obviously has the form $\chm$ replacing ~$W
~\mt~ W_2 \times W_2\,$, where ~$W_2$~ is the two-element Weyl
group of $sl(2)$.

When ~$d\leq d_{\rm max}$~ there are additional even reducibilities,
cf.  \embde{}, \paory{}.

First we consider the case when ~$d=2+j_1+j_2\,$, i.e., the
unitarity threshold when $j_1j_2\neq 0$. Using the definitions in
\redd{} we have: \eqn\thres{ \eqalign{ & n_{1} ~=~ 1 + 2j_1 \ ,
\quad n_2 ~=~ -1 -2 j_1 - 2j_2 \ , \quad
 n_3 ~=~ 1 + 2j_2\ , \cr
&n_{12} ~=~ - 2j_2 \ , \quad n_{23} ~= ~ - 2 j_1 \ , \quad
n_{13} ~=~ 1 \ .}} The corresponding character formula is given in
\Dobsy\ (formula (4.32c) (one has to make the changes: $m_{23}\mt
n_{1}$, $m_{12}\mt n_{3}$, since that formula is parametrized w.r.t.
some referent dominant weight, and set $m_2=1$):
\eqna\charthr
$$\eqalignno{ ch~L_{d^{ac};j_1,j_2} ~=&~ ch~V^{\L_{ac}} ~\Big( 1 -
t_1^{n_{1}} - t_3^{n_{3}} - t_{13} + t_1^{n_{1}} t_3^{n_{3}} +
t_1^{n_{1}} t_{23} + t_{12} t_{3}^{n_{3}} - t_1^{n_{1}} t_{2}
t_{3}^{n_{3}}
 \Big) \ = \cr ~=&~
ch~V^{\L_{ac}} ~\Big( (1-t_1^{n_1}) (1-t_3^{n_3}) ~-~ t_{13}
(1-t_1^{n_1-1}) (1-t_3^{n_3-1}) ~\Big) \ = \cr = &~
 {e(\L_{ac})\over (1-t_2) (1-t_1t_2) (1-t_2t_3) (1-t_1t_2t_3)} \ \cp_{n_1,n_3} (t_1,t_2,t_3)
\ , &\charthr{}\cr
& \qquad d^{ac} = 2 +j_1 +j_2 \
, ~~~n_{1} ~=~ 1 + 2j_1 \ , ~~~n_3 ~=~ 1 + 2j_2 \ , \cr
\cp_{n_1,n_3} ~=&~ \cq_{n_1,n_3} ~+~ t_{13}\, \cq_{n_1-1,n_3-1} \ .
 \cr}$$
Note that this formula is valid also for ~$j_1=0$, ($n_1=1$) and/or ~$j_2=0$, ($n_3=1$)
when the second term disappears, ($\cq_{0,n} = \cq_{n,0} =0$),
and then the formula coincides with \slgen{}. This may be explained with
the fact that when ~$j_1j_2=0$, then the value ~$d=2+j_1+j_2$~ is not a threshold, instead
~$d=1+j_1+j_2$~ is the threshold.

Next we consider the case when ~$d=1+j_1+j_2\,$, i.e., the
massless unitarity threshold when $j_1j_2= 0$. First we take ~$j_1=0$. Using the definitions in
\redd{} we have: \eqn\thres{ \eqalign{ & n_{1} ~=~ 1 \ ,
\quad n_2 ~=~ - 2j_2 \ , \quad
 n_3 ~=~ 1 + 2j_2\ , \cr
&n_{12} ~=~ 1 - 2j_2 \ , \quad n_{23} ~= ~ 1 \ , \quad
n_{13} ~=~ 2 \ .}} The corresponding character formula is given in
\Dobsy\ (formula (4.32b) (one has to make the changes: $m_{13}\mt
n_{3}$, $m_{2}\mt n_{1}=1$, $m_{3}\mt n_{23}=1$, $m_{23}\mt
n_{13}=2$, since that formula is parametrized w.r.t.
some more general referent dominant weight):
\eqna\charthrz
$$\eqalignno{ ch~L_{d^{ad};0,j_2} ~=&~ ch~V^{\L_{ad}}
 ~\Big( ~1
~-~ t_1
~-~ t_3^{n_{3}} ~+~ t_1 t_3^{n_{3}} ~-\cr
& -~ t_{23}
~+~ t_2 t_3^{n_{3}}
~+~ t_1^{2} t_{23}
~-~ t_1^{2} t_2 t_3^{n_{3}} ~+\cr
& +~ t_1 t_{23}^{2}
~-~ t_1 t_2^{2} t_3^{n_{3}}
~-~ t_{13}^{2}
~+~ t_{12}^{2} t_3^{n_{3}}
 ~\Big) \ =\cr
=&~ ch\,V^{\L_{ad}} (1-t_1)
 \,\Big( 1-t_3^{n_3} \,-\, t_{23} (1+t_1) (1-t_3^{n_3-1})
\,+\, t_2 t_{13} (t_3- t_3^{n_3-1}) \Big) \
= \cr = &~
 {e(\L_{ad})\over (1-t_2) (1-t_1t_2) (1-t_2t_3) (1-t_1t_2t_3)} \
 \cp_{n_3} (t_1,t_2,t_3) \ , &\charthrz {}\cr &\cr
\cp_{n_3} ~=&~ \cases{ \cq_{n_3}\ -\ t_{23} (1+t_1) \cq_{n_3-1}
\ +\ t_{23} t_{13} \cq_{n_3-2}\ , ~&~ $n_3\geq 2$ \cr &\cr
1- t_2 t_{13}\ , ~&~ $n_3=1$ } }$$
This formula simplifies considerably for $j_2=0$, ($n_3=1$)
and $j_2=\han$, ($n_3=2$):
\eqn\chxxx{
 ch~L_{d^{ad}=1;\,0,0} ~= ~ {e(\L_{ad}) \ ( 1 - t_2 t_{13} )\over ( 1 - t_{2} )
( 1 - t_{12}) ( 1 - t_{23}) ( 1 - t_{13} )} \ .}
\eqn\chyyy{
 ch~L_{d^{ad}=\trh;\,0,\han} ~= ~ {e(\L_{ad}) \ ( 1+t_3 - t_{23} - t_{13} )\over ( 1 - t_{2} )
( 1 - t_{12}) ( 1 - t_{23}) ( 1 - t_{13} )} \ . }

The case ~$j_2=0$~ is obtained from the above by the changes $n_3\mt n_1$ and $t_1 \lra t_3$, and the
character formula is:
\eqna\charthrzy{
$$\eqalignno{ ch~L_{d^{bc};j_1,0} ~=&~ ch~V^{\L_{ad}}
 ~\Big( ~1
~-~ t_3
~-~ t_1^{n_{1}} ~+~ t_1^{n_{1}}t_3 ~-\cr
& -~ t_{12}
~+~ t_1^{n_{1}} t_2
~+~ t_{12} t_3^{2}
~-~ t_1^{n_{1}} t_2 t_3^{2} ~+\cr
& +~ t_{12}^{2} t_3
~-~ t_1^{n_{1}} t_2^{2} t_3
~-~ t_{13}^{2}
~+~ t_1^{n_{1}} t_{23}^{2}
 ~\Big) \ =\cr
=&~ ch\,V^{\L_{bc}} (1-t_3)
 \,\Big( 1-t_1^{n_1} \,-\, t_{12} (1+t_3) (1-t_1^{n_1-1})
\,+\, t_1 t_{2}^{2} t_3 (t_1- t_1^{n_1-1}) \Big) \ = \cr
 = &~
 {e(\L_{bc})\over (1-t_2) (1-t_1t_2) (1-t_2t_3) (1-t_1t_2t_3)} \
 \cp'_{n_1} (t_1,t_2,t_3) \ , &\charthrzy {}\cr &\cr
\cp'_{n_1} ~=&~ \cases{ \cq'_{n_1}\ -\ t_{12} (1+t_3) \cq'_{n_1-1}
\ +\ t_{12} t_{13} \cq'_{n_1-2}\ , ~&~ $n_1\geq 2$ \cr &\cr
1- t_2 t_{13}\ , ~&~ $n_1=1$ } }$$
This formula simplifies considerably for $j_1=0$, ($n_1=1$) when it coincides with
\chxxx\ since ~$\cp'_1 = \cp_1\,$, while in the case $j_1=\han$, ($n_1=2$) we have:
\eqn\chzzz{
 ch~L_{d^{bc}=\trh;\,\han,0} ~= ~ {e(\L_{ad}) \ ( 1+t_1 - t_{12} - t_{13} )\over ( 1 - t_{2} )
( 1 - t_{12}) ( 1 - t_{23}) ( 1 - t_{13} )} \ .}

\nt
{\it Remark:}~~ {\iti It is not surprising that the three cases \chxxx,\chyyy,\chzzz
(when ~$j_1+j_2\leq \han$) are special since (unlike the other massless irreps)
they are not related to finite-dimensional irreps. The easiest way to see this is by the value of the Casimir,
which in terms of the parameters $n_{ik}$ is given as follows:
\eqn\casi{
C_2 ~~=~~ \ha \left( n_{13}^2 + n_2^2 + \ha (n_1-n_3)^2\right) -5
}
and is normalized so that for each finite-dimensional irrep (when $n_k\in\bbn$)
it is non-negative and zero
only for the one-dimensional irrep ($n_k=1$).
It is easy to see that for the massless cases given in \thres\ one has:
\eqn\casim{ C_2 = 3 (j_2^2 -1) \ , \quad j_2 \in \han\bbz_+ \ ,}
which is indeed negative for $j_2=0,\han$ and non-negative for $j_2\geq 1$.
In fact, the finite-dimensional irrep, related to a massless case with $j_2\geq 1$, has dimension \Dobsu:
\eqn\fds{ \dim \L_{{\rm fdir},j_2} ~=~ \third\, j_2 (4 j^2_2 -1)
, \quad j_2 \in 1 + \han\bbz_+ \ .}
For the conjugate
 massless case we have similarly
\eqn\casimp{ C_2 = 3 (j_1^2 -1) \ , \quad j_1 \in \han\bbz_+ \ ,}
\eqn\fdsp{ \dim \L_{{\rm fdir'},j_1} ~=~ \third\, j_1 (4 j^2_1 -1)
, \quad j_1 \in 1 + \han\bbz_+ \ ,}
with the same conclusions.\nl
These three representations are the so-called minimal UIRs of ~$su(2,2)$, cf., e.g., \Koba,
and for other of their properties we refer to \Dobcond,\Dobpds.} \dia

\vskip 5mm

In the case of ~$sl(N)$~ the representations are
finite-dimensional since we induce from UIRs of $su(N)$.
The character formula is \chm, which we repeat in order to
introduce the corresponding notation:
\eqn\chu{
\cs_{r_1,\ldots,r_{N-1}}
~~=~~ \sum_{w\in W_u}(-1)^{\ell(w)} ~ch_0~V^{w\cdot\L^u} \,, \quad
\L^u\in -\G^u_+}
The index $u$ is to distinguish the quantities pertinent to the case.

We shall write down explicitly the cases that we shall need, namely, $sl(2)$ and $sl(4)$.
In the $sl(2)$ case the Weyl group has only two elements and the character formula
is very simple:
\eqn\chrf{ \eqalign{ \cs_{r} ~=&~ ch\, V^\L \, ( 1 - t^{r+1} ) ~=~
e(\L) { 1 - t^{r+1} \over 1-t} ~=\cr
~=&~ e(\L) (1 + t + t^2 + \cdots + t^r) \ , \cr
&~ e(\L) = t^{-r/2} \ , \qquad r\in\bbz_+ \ . }}

In the $sl(4)$ case the Weyl group has 24 elements and the character formula
is:
\eqna\chrf
$$\eqalignno{
\cs_{r_1,r_2,r_3} ~=&~
 ch\ V^{\L^u} ~\Big( 1 - t_1^{n_1} - t_2^{n_2} -
t_3^{n_3} + t_1^{n_1} t_3^{n_3} + t_1^{n_1} t_2^{n_{12}}
+ t_2^{n_{23}} t_3^{n_3}\ + &\chrf{}\cr
&+~ t_1^{n_{12}} t_2^{n_2} + t_2^{n_2} t_3^{n_{23}}
- t_1^{n_1} t_2^{n_{13}} t_3^{n_3}
- t_1^{n_{12}} t_2^{n_2} t_3^{n_{23}}
- t_1^{n_{13}} t_2^{n_{23}} t_3^{n_3}\ - \cr
&-~ t_1^{n_1} t_2^{n_{12}} t_3^{n_{13}}
- (t_1 t_2)^{n_{12}} - (t_2t_3)^{n_{23}}
+ t_1^{n_{12}} t_2^{n_2 + n_{13}} t_3^{n_{23}}\ + \cr
&+~ t_1^{n_1} (t_2 t_3)^{n_{13}}
+ (t_1 t_2)^{n_{12}} t_3^{n_{13}}
+ t_1^{n_{13}} (t_2 t_3)^{n_{23}}
+ (t_1 t_2)^{n_{13}} t_3^{n_3}\ - \cr
&-~ t_1^{n_{12}} t_2^{n_2 + n_{13}} t_3^{n_{13}} - t_1^{n_{13}}
t_2^{n_2 + n_{13}} t_3^{n_{23}}
- (t_1 t_2 t_3)^{n_{13}}
+ (t_1 t_2 t_3)^{n_{13}} t_2^{n_2} \Big) \ , \cr
& \qquad n_k = r_k+1, ~n_{12} = r_1+r_2 + 2, ~n_{23} = r_2+r_3 + 2,\cr
& \qquad n_{13} = r_1+r_2 + r_3 +3 }$$
The expression for ~$\L^u$ in terms of the $sl(4)$ simple roots is:
\eqna\exprl
$$\eqalignno{ \L^u ~=&~ -\han(r_2+\han(3r_1+r_3))\,\b_1 - (r_2+\han(r_1+r_3))\,\b_2 -
\han(r_2+\han(r_1+3r_3))\,\b_3 \ = \cr
=&~ \sum_{k=1}^3\ \l_k\, \b_k \ , &\exprl{}}$$
where the minus signs are due to the fact that $\L^u$ is assumed to be lowest weight,
for highest weight the minuses become pluses.
Furthermore, the notation for the simple roots depends on the application, e.g., when
applied to the compact part of the even subalgebra for $N=4$: ~$\cg_0 = su(2,2) \oplus u(1)
\oplus su(4)$ then the roots are mapped: ~$\b_1 \mt \a_7\,$, ~$\b_2 \mt \a_6\,$,
~$\b_3 \mt \a_5\,$.

We give some explicit examples that would be actually used:
\eqna\charslf
$$\eqalignno{ & \cs_{0,0,0} ~=~ 1 \ ,&\charslf{}\cr
& \cs_{1,0,0} ~=~ e(\L^u)\ (1 + t_1 + t_{12} + t_{13}) \ , \cr
& \cs_{0,0,1} ~=~ e(\L^u)\ (1 + t_3 + t_{23} + t_{13}) \ , \cr
& \cs_{0,1,0} ~=~ e(\L^u)\ (1 + t_2 + t_{12} + t_{23} + t_{13} + t_{13} t_2) \ , \cr
& \cs_{2,0,0} ~=~ e(\L^u)\ (1 + t_1 + t_1^2 + t_{12} + t_{12}^2 + t_{13} + t_{13}^2
+ t_{12} t_{1} + t_{13} t_{1} + t_{13} t_{12}
) \ , \cr
& \cs_{0,0,2} ~=~ e(\L^u)\ (1 + t_3 + t_3^2 + t_{23} + t_{23}^2 + t_{13} + t_{13}^2
+ t_{23} t_{3} + t_{13} t_{3} + t_{13} t_{23}
) \ , \cr
&e(\L^u) ~=~ \sum_{k=1}^3\ t_k^{\l_k} \ .
}$$
Naturally, the number of terms in the character formula is equal to the dimension
of the corresponding irrep:
\eqn\dimu{ \dim_{r_1,r_2,r_3} ~=~ \twelvth\ (r_1+1) (r_2+1)(r_3+1) (r_1+r_2+2)
(r_2+r_3+2) (r_1+r_2+r_3+3) \ .}

\vskip 5mm

\newsec{Explicit character formulae for ~N=1,2,4}

\subsec{{\bf N=1}}

\nt\bu{\bf ~~ Long superfields}

\nt If ~$d ~>~ d_{\rm max}\,$, ~$j_1j_2> 0$~ then ~$\hL_\L$~ has the
maximum possible number of states: ~$16$.\nl
The bare character formula is (3.19) from \Dobch{}:
\eqn\chlla{\eqalign{
ch\ \hL_\L ~=&~ \prod_{\a
\in\D^+_\I}(1 + e(\a)) ~= \cr ~=&~ (1+e(\a_{15})) (1 +
e(\a_{25})) (1 + e(\a_{35})) (1 + e(\a_{45}))   \ , }}
where we use \smplr~:
\eqn\chllb{\eqalign{
& \a_{15} ~=~ \g_1+\g_3 = \a_1+\g_3 \ , \cr
& \a_{35} ~=~ \g_2+\g_4 = \a_3 +\g_4 \ , \cr
& \a_{25} ~=~ \g_3    \ , \quad \a_{45}~=~   \g_{4} }}
The mirror symmetry of \chlla\ follows from \mirrs{}:
$$\g_{1} \lra \g_{2}\ , ~~~\g_3 \lra \g_4 $$
For the characters we shall need also the following notation related
to the odd roots:
\eqn\notcone{ \eqalign{
&\vth ~\equiv~ e(\a_{15})\ , \qquad
\tth ~\equiv~ e(\a_{25})\ ,  \cr
&\bvh ~\equiv~ e(\a_{35})\ , \qquad
\bth ~\equiv~ e(\a_{45})\ , }}
(the bar respecting the mirror symmetry),
and for the $sl(4)$--related variables : ~$t_k ~\equiv~ e(\a_k)\,$, $k=1,2,3$, cf.
Subsection 2.2. Note the relation: \eqn\relthe{\vth~=~ t_1\,\tth \ ,
\quad \bvh ~=~ t_3\,\bth \ ,} which is also mirror symmetric.

The full character formula takes into account the characters of the
conformal algebra entries, i.e., we have: \eqna\chllb
$$\eqalignno{
ch\,L_\L ~=&~ e(\L) \ \Big\{\ ch' \ L^2_{d;j_1,j_2} ~+~ e(\a_{15})\,
ch'\,L^2_{d+\han;j_1-\han,j_2} ~+~ e(\a_{25})\,
ch'\,L^2_{d+\han;j_1+\han,j_2} ~+ \cr &+~ e(\a_{35})\,
ch'\,L^2_{d+\han;j_1,j_2-\han} ~+~ e(\a_{45})\,
ch'\,L^2_{d+\han;j_1,j_2+\han} ~+\cr &+~ e(\a_{15})\,e(\a_{25})\,
ch'\,L^2_{d+1;j_1,j_2} ~+~ e(\a_{15})\,e(\a_{35})\,
ch'\,L^2_{d+1;j_1-\han,j_2-\han} ~+\cr &+~ e(\a_{15})\,e(\a_{45})\,
ch'\,L^2_{d+1;j_1-\han,j_2+\han} ~+~ e(\a_{25})\,e(\a_{35})\,
ch'\,L^2_{d+1;j_1+\han,j_2-\han} ~+\cr &+~ e(\a_{25})\,e(\a_{45})\,
ch'\,L^2_{d+1;j_1+\han,j_2+\han} ~+~ e(\a_{35})\,e(\a_{45})\,
ch'\,L^2_{d+1;j_1,j_2} ~+\cr &+~
e(\a_{15})\,e(\a_{25})\,e(\a_{35})\, ch'\,L^2_{d+\trh;j_1,j_2-\han}
~+~ e(\a_{15})\,e(\a_{25})\,e(\a_{45})\,
ch'\,L^2_{d+\trh;j_1,j_2+\han} ~+ \cr &+~
e(\a_{15})\,e(\a_{35})\,e(\a_{45})\, ch'\,L^2_{d+\trh;j_1-\han,j_2}
~+~ e(\a_{25})\,e(\a_{35})\,e(\a_{45})\,
ch'\,L^2_{d+\trh;j_1+\han,j_2} ~+\cr &+~
e(\a_{15})\,e(\a_{25})\,e(\a_{35})\,e(\a_{45})\,
ch'\,L^2_{d+2;j_1,j_2} \ \Big\} &\chllb {}}$$
 where ~$ch'\ L^2_{d;j_1,j_2}$~ is the conformal (actually $sl(4)$) character
formula  \slgen{}, the prime denoting that we have omitted the
prefactor $e(\L^s)$ since above it is distributed among the other
prefactors. Substituting the explicit expressions for ~$ch'\
L^2_{d;j_1,j_2}$~ we obtain: \eqna\chllba
$$\eqalignno{
ch\,L_\L ~=&~ {e(\L)\over (1-t_2) (1-t_1t_2) (1-t_2t_3)
(1-t_1t_2t_3)}
 \ \times \cr &\times \ \Big\{\
(1 + \vth \,\tth)\, (1 + \bvh\,\bth)\, \cq_{n_1,n_3} ~+\cr &+ ~
\vth\, \,\cq_{n_1-1,n_3} ~+~ \tth\, \,\cq_{n_1+1,n_3} ~+\cr &+~
\bvh\, \,\cq_{n_1,n_3-1} ~+~ \bth\, \,\cq_{n_1,n_3+1} ~+\cr &+~
\vth\,\bvh\, \,\cq_{n_1-1,n_3-1} ~+ ~ \vth\,\bth\,
\,\cq_{n_1-1,n_3+1} ~+\cr &+~ \tth\,\bvh\, \,\cq_{n_1+1,n_3-1} ~+ ~
\tth\,\bth\, \,\cq_{n_1+1,n_3+1}
 ~+\cr
&+~ \vth\,\tth\,\bvh\, \,\cq_{n_1,n_3-1} ~+~ \vth\,\tth\,\bth\,
\,\cq_{n_1,n_3+1} ~+ \cr &+~ \vth\,\bvh\,\bth\, \,\cq_{n_1-1,n_3}
~+~ \tth\,\bvh\,\bth\, \,\cq_{n_1+1,n_3} \ \Big\} &\chllba {}}$$
where   the polynomials ~$\cq$~ are in terms of the
$sl(4)$--related variables ~$t_1,t_2,t_3$.

When ~$j_1j_2=0$~ there are less terms in the character formula,
since ~$\cq_{0,\cdot}=0= \cq_{\cdot,0}\,$, and further the entries
simplify.

If ~$j_1=0,j_2> 0$~ then the generator ~$X^+_{15}$~ can appear only
together with the generator ~$X^+_{25}\,$, and ~$\hL_\L$~ has $12$
states = 3(chiral)$\times${}4(anti-chiral) states.\foot{In
statements like this each sector includes the vacuum.} The bare
character formula is (3.26) from \Dobch{} \eqna\chlll
$$\eqalignno{
ch~\hL_\L ~=&~ ch~\hV^\L ~-~ \car \ , \qquad d ~>~ d_{\rm
max} \ ,&\chlll{}\cr \car ~\equiv&~ e(\hV^\L_{\rm excl}) ~=~ \sum_{\rm
excluded\atop states}\ e(\Psi_{\bar \ve})\ ,
  }$$
where the counter-term  ~$\car$~ in our case is:
\eqn\chnoa{ \car ~=~ e(\a_{15}) (1 + e(\a_{35})) (1 + e(\a_{45}))}
and the full character formula is:
\eqna\chllbb
$$\eqalignno{ ch\,L_\L
~=&~ {e(\L)\over (1-t_2) (1-t_1t_2) (1-t_2t_3) (1-t_1t_2t_3)}
 \ \times &\chllbb{}\cr &\times \ \Big\{\
(1 + \vth\,\tth)\, (1 + \bvh\,\bth)\, \cq_{n_3} ~+ \cr &+~
 \tth\, (1 + \bvh\,\bth) \, (1+t_1)\,\cq_{n_3} ~+\cr
&+~ \bvh\, (1 + \vth\,\tth) \, \cq_{n_3-1} ~+\cr &+~ \bth\, (1 +
\vth\,\tth) \, \cq_{n_3+1} ~+\cr &+~ \tth\,\bvh\, (1+t_1)\,
\cq_{n_3-1} ~+~
 \tth\,\bth\, (1+t_1)\,\cq_{n_3+1}
\ \Big\} }$$ where we use:
$$\cq_{1,n_3} ~=~ \sum_{k=0}^{n_3-1}\ t_3^k ~\equiv~ \cq_{n_3} \ , \qquad
\cq_{2,n_3} (t_1,t_3) ~=~ (1+t_1)\ \cq_{n_3} (t_3)\ .$$

The next case is conjugate. If ~$j_1> 0,j_2= 0$~ then the generator
~$X^+_{35}$~ can appear only together with the generator
~$X^+_{45}\,$, and ~$\hL_\L$~ has $12$ states. The bare character
formula is again \chlll{} with counter-term: \eqn\chnob{ \car ~=~
e(\a_{35}) (1 + e(\a_{15})) (1 + e(\a_{25}))} and the full character
formula is:
\eqna\chllcc
$$\eqalignno{ ch\,L_\L ~=&~ {e(\L)\over (1-t_2)
(1-t_1t_2) (1-t_2t_3) (1-t_1t_2t_3)}
 \ \times &\chllcc{}\cr &\times \ \Big\{\
(1 + \vth\,\tth)\, (1 + \bvh\,\bth)\, \cq'_{n_1} ~+ \cr &+~
 \bth\, (1 + \vth\,\tth) \, (1+t_3)\,\cq'_{n_1} ~+\cr
&+~ \vth\, (1 + \bvh\,\bth) \, \cq'_{n_1-1} ~+\cr &+~ \tth\, (1 +
\bvh\,\bth) \, \cq'_{n_1+1} ~+\cr &+~ \vth\,\bth\, (1+t_3)\,
\cq'_{n_1-1} ~+~
 \tth\,\bth\, (1+t_3)\,\cq'_{n_1+1}
\ \Big\} }$$ where we use:
$$\cq_{n_1,1} ~=~ \sum_{j=0}^{n_1-1}\ t_1^j ~\equiv~ \cq'_{n_1} \ , \qquad
\cq_{n_1,2} (t_1,t_3) ~=~ (1+t_3)\ \cq'_{n_1} (t_1)\ .$$

The next case combines the previous two. If ~$j_1=j_2= 0$~ then the
generator ~$X^+_{15}$~ can appear only together with the generator
~$X^+_{25}\,$, the generator ~$X^+_{35}$~ can appear only together
with the generator ~$X^+_{45}\,$, and ~$\hL_\L$~ has $9$ states =
3(chiral)$\times${}3(anti-chiral) states. The character formula is
\chlll{} with: \eqn\chnoc{ \car ~=~ \vth (1 + e(\a_{35})) (1 +
e(\a_{45})) ~+~ e(\a_{35}) (1 + e(\a_{15}) ) (1 + e(\a_{25})) ~-~
e(\a_{15}) e(\a_{35}) \ ,} i.e., we combine the counter-terms of the
previous two cases, but need to subtract a counter-term that is
counted twice. The bare character formula is: \eqn\chlld{ ch\ \hL_\L
~= ~ (1+ e(\a_{25}) + \vth\, e(\a_{25})) (1+ e(\a_{45}) +
e(\a_{35})\, e(\a_{45}))} and the full character formula may be
obtained from \chllbb{} setting $n_3=1$ (or from \chllcc{} setting
$n_1=1$):
\eqna\chlldd
$$\eqalignno{ ch\,L_\L ~=&~ {e(\L)\over (1-t_2)
(1-t_1t_2) (1-t_2t_3) (1-t_1t_2t_3)}
 \ \times &\chlldd{}\cr &\times \ \Big\{\
(1 + \vth\,\tth)\, (1 + \bvh\,\bth) ~+\cr &+~
 \tth\, (1 + \bvh\,\bth) \, (1+t_1) ~+\cr
&+~ \bth\, (1 + \vth\,\tth) \, (1+t_3) ~+\cr &+~
 \tth\,\bth\, (1+t_1)\, (1+t_3)
\ \Big\} }$$ using $\cq_{1} = 1$, $\cq_{2} = 1+t_3\,$, (or $
\cq'_{1}=1$, $\cq'_{2} = 1+t_1\,$).

\vskip 10mm

\nt\bu ~~~{\bf SRC cases}

\nt\bu{\bf a} ~~~$d ~=~ d_{\rm max} ~=~ d^1_{11} ~=~ 2 +2j_2 +z ~>~
d^3_{11} $ \ .\nl The generator $X^+_{35}$ is eliminated \Dobch\ and
for $z=0$ (then $j_2>j_1$) these are $\quarter$-BPS cases \Dobbps.\nl
These are called semi-conserved superfields in \CDDF. For ~$j_2> 0$~ they obey the
first-order super-differential operator given explicitly in
formulae (7a) of \DPu. When ~$j_2=0$~  that first-order super-differential operator
has trivial kernel and is replaced by second-order super-differential operator
given in (11b) of \DPu.

\nt\bu ~~$j_1> 0$. Here there are only $8$ states.\foot{For brevity,
here and often below we shall say "there are M states" meaning
"there are M states in ~$\hL_\L$~".} The bare character formula is
(3.36)  (or equivalently (3.39)) from \Dobch{} without
counter-terms: \eqn\chsnoa{ ch~\hL_\L ~= ~
 \prod_{\a \in\D^+_\I \atop \a\neq \a_{35}}
\ (1 +e(\a)) \ .} The full character formula follows from \chllba{}:
\eqna\chsrca
$$\eqalignno{ ch\,L_\L ~=&~ {e(\L)\over (1-t_2) (1-t_1t_2)
(1-t_2t_3) (1-t_1t_2t_3)}
 \ \times &\chsrca{}\cr &\times \ \Big\{\
(1 + \vth\,\tth)\, \cq_{n_1,n_3} ~+ ~ \vth\, \,\cq_{n_1-1,n_3} ~+\cr
&+~ \tth\, \,\cq_{n_1+1,n_3} ~+ ~ \bth\, \,\cq_{n_1,n_3+1} ~+\cr &+~
\vth\,\bth\, \,\cq_{n_1-1,n_3+1} ~+ ~ \tth\,\bth\,
\,\cq_{n_1+1,n_3+1}
 ~+\cr
&+~ \vth\,\tth\,\bth\, \,\cq_{n_1,n_3+1} \ \Big\} }$$

\nt{\it Remark:}~~{\iti
For the finite-dimensional irreps of ~sl(4/N)~   the SRC situations are called
'singly atypical' and the character formulae are written as \chsnoa{},
 cf. \BL,\JHKT,\Jeu.\foot{For
character formulae of finite-dimensional irreps beyond the singly
atypical case cf. \Serga, \VZ, \Bru, \SuZh, and references therein.}}\dia

\nt\bu ~~$j_1=0$. The generator ~$X^+_{15}$~ can appear only
together with the generator ~$X^+_{25}$~ and there are only $6$
states. Then the bare character formula is (3.36)  (or equivalently
(3.39)) from \Dobch{} with counter-term: \eqn\chsrcaa{\eqalign{
ch~\hL_\L ~=&~
 \prod_{\a \in\D^+_\I
\atop { \atop \a\neq \a_{35} }} \ (1 +e(\a)) ~-~ \car ~=\cr =&~ (1+
e(\a_{25}) + \vth\, e(\a_{25})) (1 + e(\a_{45}))
 , \cr
&\car ~=~ \vth (1 + e(\a_{45}))
 \ .}}
The full character formula follows from \chsrca{}:
\eqn\chsrcaa{\eqalign{ ch\,L_\L ~=&~ {e(\L)\over (1-t_2) (1-t_1t_2)
(1-t_2t_3) (1-t_1t_2t_3)}
 \ \times \cr &\times \ \Big\{\
(1 + \vth\,\tth)\, \cq_{n_3} ~+\cr &+~ \tth\, (1+t_1) \,( \cq_{n_3}
+ \bth\,\cq_{n_3+1}) ~+\cr
 &+~ (1 + \vth\,\tth)\,\bth\, \,\cq_{n_3+1}
\ \Big\} }}

\vskip 5mm

\nt{\bu{\bf b}}~~~~~ $d ~=~ d^2_{11} ~=~ z ~>~ d^3_{11}\ , ~~
j_2=0$\ .\nl
These UIRs are called chiral since all anti-chiral
generators are eliminated. They obey the
first-order super-differential operator given explicitly in
formulae (7b) of \DPu.

\nt\bu ~~$j_1> 0$. The generators $X^+_{35}$ and $X^+_{45}$ are
eliminated and there are only $4$ states. The bare character formula
is (3.65) from \Dobch{} (for $i_0=0$) without counter-terms:
\eqn\chszz{ ch~\hL_\L ~= ~
 (1 +e(\a_{15})) \ (1 +e(\a_{25})) }
The full character formula follows from \chsrca{}:
\eqn\chsrcb{\eqalign{ ch\,L_\L ~=&~ {e(\L)\over (1-t_2) (1-t_1t_2)
(1-t_2t_3) (1-t_1t_2t_3)}
 \ \times \cr &\times \ \Big\{\
(1 + \vth\,\tth)\, \cq'_{n_1} ~+ ~ \vth\, \cq'_{n_1-1} ~+~ \tth\,
\,\cq'_{n_1+1} \ \Big\} }}

\nt\bu ~~$j_1= 0$. The generators $X^+_{35}$ and $X^+_{45}$ are
eliminated, the generator ~$X^+_{15}$~ can appear only together with
the generator ~$X^+_{25}\,$, and there are only $3$ states. The bare
character formula is (3.65) from \Dobch{} (for $i_0=0$) with
counter-term ~$\car = e(\a_{15})$~: \eqn\chsnobz{ ch~\hL_\L ~=~
 1 +e(\a_{25}) + e(\a_{15}) e(\a_{25})}
The full character formula follows from \chsrcb{}:
\eqn\chsrcbb{\eqalign{ ch\,L_\L ~=&~ {e(\L)\over (1-t_2) (1-t_1t_2)
(1-t_2t_3) (1-t_1t_2t_3)}
 \ \times \cr &\times \ \Big\{\
1 + \vth\,\tth ~+ ~
 \tth\, (1+t_1)
\ \Big\} }}

\vskip 5mm

The next two cases {\bf c,d} are conjugate to the above {\bf a,b}.
The character formulae are obtainable by the changes ~$j_1\lra j_2$,
$n_1\lra n_3$, $t_1\lra t_3$, $\vth \lra \bvh$, $\tth \lra \bth$,
($\a_{15} \lra \a_{35}$, $\a_{25} \lra \a_{45}$). Thus, we list the
character formulae without explanations.

\vskip 5mm

\nt\bu{\bf c} ~~~$d ~=~ d_{\rm max} ~=~ d^3_{11} ~=~ 2 +2j_1 -z ~>~
d^1_{11}$ \ .\nl The generator $X^+_{15}$ is eliminated \Dobch\ and
for $z=0$ (then $j_1>j_2$) these are $\quarter$-BPS cases \Dobbps.\nl
These are called semi-conserved superfields in \CDDF. For ~$j_1> 0$~ they obey the
first-order super-differential operator given explicitly in
formulae (7c) of \DPu. When ~$j_1=0$~  that first-order super-differential operator
has trivial kernel and is replaced by second-order super-differential operator
given in (11a) of \DPu.

\nt\bu ~~$j_2> 0$. The character formulae are: \eqn\chsnoc{
ch~\hL_\L ~= ~
 \prod_{\a \in\D^+_\I \atop \a\neq \a_{15}}
\ (1 +e(\a)) \ .}
\eqna\chsrcc
$$\eqalignno{ ch\,L_\L ~=&~ {e(\L)\over
(1-t_2) (1-t_1t_2) (1-t_2t_3) (1-t_1t_2t_3)}
 \ \times \chsrcc{}\cr &\times \ \Big\{\
(1 + \bvh\,\bth)\, \cq_{n_1,n_3} ~+ ~ \bvh\, \,\cq_{n_1,n_3-1} ~+\cr
&+~ \bth\, \,\cq_{n_1,n_3+1} ~+ ~ \tth\, \,\cq_{n_1+1,n_3} ~+\cr &+~
\tth\,\bvh\, \,\cq_{n_1+1,n_3-1} ~+ ~ \tth\,\bth\,
\,\cq_{n_1+1,n_3+1}
 ~+\cr &+~
\tth\,\bvh\,\bth\, \,\cq_{n_1+1,n_3} \ \Big\} }$$

\nt\bu ~~$j_2=0$.~~ The character formulae are: \eqn\chsnocc{
ch~\hL_\L ~=~ (1 + e(\a_{25})) (1 + e(\a_{45}) + e(\a_{35})\,
e(\a_{45}))\ .} \eqn\chsrccc{\eqalign{ ch\,L_\L ~=&~ {e(\L)\over
(1-t_2) (1-t_1t_2) (1-t_2t_3) (1-t_1t_2t_3)}
 \ \times \cr &\times \ \Big\{\
(1 + \bvh\,\bth)\, \cq'_{n_1} ~+\cr &+~ \bth\, (1+t_3) \,(
\cq'_{n_1} + \tth\,\cq'_{n_1+1}) ~+\cr
 &+~ (1 + \tth\,\bvh)\,\bth\, \,\cq'_{n_1+1}
\ \Big\} }}

\vskip 5mm

\nt{\bu{\bf d}}~~~~~ $d ~=~ d^4_{11} ~=~ -z ~>~ d^1_{11}\ , ~~
j_1=0$\ .\nl
These UIRs are called anti-chiral since all chiral
generators are eliminated. They obey the
first-order super-differential operator given explicitly in
formulae (7d) of \DPu.

\nt\bu ~~$j_2> 0$. The character formulae are: \eqn\chsnd{ ch~\hL_\L
~= ~
 (1 +e(\a_{35})) \ (1 +e(\a_{45})) }
\eqn\chsrcd{\eqalign{ ch\,L_\L ~=&~ {e(\L)\over (1-t_2) (1-t_1t_2)
(1-t_2t_3) (1-t_1t_2t_3)}
 \ \times \cr &\times \ \Big\{\
(1 + \bvh\,\bth)\, \cq_{n_3} ~+ ~ \bvh\, \cq_{n_3-1} ~+~ \bth\,
\,\cq_{n_3+1} \ \Big\} }}

\nt\bu ~~$j_2= 0$. The character formulae are: \eqn\chsnodz{
ch~\hL_\L ~=~
 1 +e(\a_{45}) + e(\a_{35}) e(\a_{45})}
\eqn\chsrcdd{\eqalign{ ch\,L_\L ~=&~ {e(\L)\over (1-t_2) (1-t_1t_2)
(1-t_2t_3) (1-t_1t_2t_3)}
 \ \times \cr &\times \ \Big\{\
1 + \bvh\,\bth ~+ ~
 \bth\, (1+t_3)
\ \Big\} }}

\vskip 10mm

\nt\bu ~~~{\bf DRC cases}

\nt{\bu{\bf ac}}~~~ $d ~=~ d^{ac} ~=~ d_{\rm max} ~=~ d^1_{11} =
d^3_{11} ~=~ d^{ac} ~=~ 2 + j_1 + j_2$\ , ~~$z=z_{ac}=j_1-j_2$ \ .

\nt These are the conserved superfields. For ~$j_1j_2\neq 0$~ they obey the two
first-order super-differential operators given explicitly in
formulae (7a,c) of \DPu. These semi-short UIRs may be called Grassmann-analytic
following \FSa, since odd generators from different chiralities
are eliminated.\nl
When ~$j_1=0$~ ($j_2=0$) the first-order super-differential operator
from (7c) of \DPu\ ((7a) of \DPu, resp.)
has trivial kernel and is replaced by second -order super-differential operator
given in (11a) of \DPu\ ((11b) of \DPu, resp.).

The generators $X^+_{15}$ and $X^+_{35}$ are eliminated (though
for different reasons for $j_1> 0$ and $j_1 =0$, resp., for $j_2> 0$
and $j_2 =0$). For $j_2=j_2$ (then $z=0$) these are $\han$-BPS cases
\Dobbps.\foot{There are only three BPS cases for $N=1$, the other
two were mentioned above in cases {\bf a,c}.} There are only $4$
states and the bare character formula is (3.84) from \Dobch{} (for
$i_0=i'_0=0$) without counter-terms: \eqn\chdnoac{\eqalign{
ch~\hL_\L ~=&~ (1 +e(\a_{25})) \ (1 +e(\a_{45})) ~=\cr =&~ ch~\hV^\L
~-~ {1\over 1 +e(\a_{15})}\ ch~\hV^{\L+\a_{15}} ~-~ {1\over 1
+e(\a_{35})}\ ch~\hV^{\L+\a_{35}} ~+\cr &+~ {1\over (1 +e(\a_{15}))
(1 +e(\a_{35}))}\ ch~\hV^{\L+\a_{15}+\a_{35}} \ ,}} where the terms
with minus may be interpreted as taking out states, while the last
term indicates adding back what was taken two times. The
corresponding decomposition of ~$L_\L$~ is given by:
\eqn\chdnoacz{\eqalign{ L_\L ~=&~ L_{d^{ac};j_1,j_2} \otimes
L^z_{z_{ac}} ~+~ L_{d^{ac}+\han;j_1+\han,j_2} \otimes
L^z_{z_{ac}-\trh} \ +\cr &+ ~ L_{d^{ac}+\han;j_1,j_2+\han} \otimes
L^z_{z_{ac}+\trh} ~+~ L_{d^{ac}+1;j_1+\han,j_2+\han} \otimes
L^z_{z_{ac}} }} Note that for all four conformal entries is
fulfilled the relation: ~$d= 2+j_1+j_2$, which for $j_1j_2\neq 0$ is
the conformal unitarity threshold. Thus, for the conformal
characters we have to use formula \charthr{}, and then
 the full character formula is: \eqn\chdnoacy{\eqalign{
\chh\ L_\L ~=&~ e(\L) \Big\{ \chh'\ L_{d^{ac};j_1,j_2} ~+~
e(\a_{25})\ \chh'\ L_{d^{ac}+\han;j_1+\han,j_2} ~+\cr &+~
e(\a_{45})\ \chh'\ L_{d^{ac}+\han;j_1,j_2+\han} ~+~ e(\a_{25})
e(\a_{45}) \ \chh'\ L_{d^{ac}+1;j_1+\han,j_2+\han} \Big\} \ = \cr
~=&~ {e(\L) \over (1-t_1) (1-t_2) (1-t_3) (1-t_{12}) (1-t_{23})
(1-t_{13}) } \Big\{ \cp_{n_1,n_3} ~+\cr &+~ \tth \, \cp_{n_1+1,n_3}
+ \bth \, \cp_{n_1,n_3+1} ~+~ \tth\,\bth\, \cp_{n_1+1,n_3+1} \Big\}
}} where $\chh'$ is the character formula \charthr{} without the
prefactor $e(\L_{ac})$ - this prefactor is subsumed in the overall
prefactor $e(\L)$ since the relative difference weights between the
four terms of \chdnoacy\ are taken into account by the prefactors
$e(\a_{k5})$.

\vskip 5mm

\nt{\bu{\bf ad}}~~~ $d ~=~ d^{ad} = d^1_{11} = d^4_{11} ~=~ 1 + j_2
= -z\ , ~~ j_1=0$.

\nt The generators $X^+_{15}$, $X^+_{25}$ and $X^+_{35}$ are
eliminated (for the latter for different reasons for $j_2> 0$ and
$j_2 =0$). These are the first series of massless UIRs, and
everything is already explicit in the general formulae. There are
only $2$ states and the bare character formula is (3.97) from
\Dobch{} for $N=1$~: \eqn\chdnoac{ch~\hL_\L ~=~ 1 +e(\a_{45}) \ . }
The corresponding decomposition of ~$L_\L$~ is given by:
\eqn\chdnoadz{\eqalign{ L_\L ~=&~ L_{d^{ad};0,j_2} \otimes
L^z_{z_{ad}} ~+~
 L_{d^{ad}+\han;0,j_2+\han} \otimes L^z_{z_{ad}+\trh} }} Note that
for both conformal entries is fulfilled the relation: ~$d=
1+j_1+j_2$, which for $j_1j_2= 0$ is the conformal unitarity
threshold. For the characters we have to use formula \charthrz{},
(or its reductions for $j_2=0,\han$), and then
 the full character formula is: \eqn\chdnoady{\eqalign{
\chh\ L_\L ~=&~ e(\L) \Big\{ \chh'\ L_{d^{ad};0,j_2} ~ +~
e(\a_{45})\ \chh'\ L_{d^{ad}+\han;0,j_2+\han} \Big\}\ =\cr =&~
e(\L)\ \Big\{ \cp_{n_3} ~ +~ \bth \, \cp_{n_3+1} \Big\}\ . }} where
$\chh'$ is the character formula \charthrz{} without the prefactor
$e(\L_{ad})$ - cf. the explanation above. This formula simplifies
for $j_2=0$, ($n_3=1$) (using both \chxxx{} and \chyyy{}):
\eqn\chdnobcyz{\eqalign{ \chh\ L_\L ~=&~ {e(\L)\over ( 1 - t_{2} ) (
1 - t_{12}) ( 1 - t_{23}) ( 1 - t_{13} )}\ \Big( 1 - t_1 t_{2}^{2}
t_3 + \bth\,
 ( 1+t_3 - t_{23} - t_{13} ) \Big)}}

The next case is conjugate.

\vskip 5mm

\nt{\bu{\bf bc}}~~~ $d ~=~ d^{bc} = d^2_{11} = d^3_{11} ~=~ 1 + j_1
= z\ , ~~ j_2=0$.

\nt The generators $X^+_{15}$, $X^+_{35}$ and $X^+_{45}$ are
eliminated (for the first for different reasons for $j_1> 0$ and
$j_1 =0$). These are the second series of massless UIRs. There are
only $2$ states and the bare character formula is (3.100) from
\Dobch{} for $N=1$~: \eqn\chdnoac{ ch~\hL_\L ~=~ 1 +e(\a_{25}) \ .}
For the full characters we have to use formula \charthrzy{}, (or its
reductions for $j_1=0,\han$), and then
 we have: \eqn\chdnobcy{\eqalign{
\chh\ L_\L ~=&~ e(\L) \Big\{ \chh'\ L_{d^{bc};j_1,0} ~ +~
e(\a_{25})\ \chh'\ L_{d^{bc}+\han;\,j_1+\han,0} \Big\}\ =\cr =&~
e(\L)\ \Big\{ \cp'_{n_1} ~ +~ \tth \, \cp'_{n_1+1} \Big\}\ . }} This
formula simplifies for $j_1=0$, ($n_1=1$) (using \chxxx{} and
\chzzz{}): \eqn\chdnobcyz{\eqalign{ \chh\ L_\L ~=&~ {e(\L)\over ( 1
- t_{2} ) ( 1 - t_{12}) ( 1 - t_{23}) ( 1 - t_{13} )}\ \Big( 1 - t_1
t_{2}^{2} t_3 + \tth \, ( 1+t_1 - t_{12} - t_{13} ) \Big)}}

\vskip 5mm

\vskip 3mm\nt{\bu{\bf bd}}~~~ $d ~=~ d^2_{11} = d^4_{11} ~=~
j_1=j_2=z ~=0$

\nt As we explained  this is the trivial 1-dimensional
irrep consisting of the vacuum.

\vskip 10mm

\subsec{N=2}

\nt
For $N=2$ we consider only the {\bf DRC cases}.

First we introduce notation for the odd roots when $N=2$ using \smplr~:
\eqn\relntwo{ \eqalign{
&\a_{15} ~=~ \a_1 +\g_3 \ , \quad \a_{16} ~=~ \a_1 +\g_3 +\a_5 \ , \cr
& \a_{25} ~=~ \g_3 \ , \quad \a_{26} ~=~ \g_3 +\a_5 \ , \cr
& \a_{35} ~=~ \a_2 +\g_4 +\a_5 \ , \quad \a_{36} ~=~ \a_2 +\g_4 \ , \cr
& \a_{45} ~=~ \g_4 +\a_5\ , \quad \a_{46} ~=~ \g_4 \ . \cr
}}
Note that as a consequence of our mirror symmetry \mirrs{} we have:
\eqn\mirtwo{\eqalign{ & \a_{15} ~\lra ~ \a_{36}\ , \quad \a_{16}~ \lra ~\a_{35}\cr
&\a_{25}~ \lra~ \a_{46} \ , \quad \a_{26} ~\lra~ \a_{45} }}
For the characters we shall need also the following notation:
\eqn\notcth{ \eqalign{
&\vth_k ~\equiv~ e(\a_{1,7-k})\ , \quad
\tth_k ~\equiv~ e(\a_{2,7-k})\ , \quad k=1,2 \cr
&\bvh_k ~\equiv~ e(\a_{3,4+k})\ , \quad
\bth_k ~\equiv~ e(\a_{4,4+k})\ , \quad k=1,2 \ ,}}
the bar respecting the mirror symmetry of \mirtwo.

\nt{\bu{\bf ac}}~~~
$d ~=~ d^{ac} ~=~ d_{\rm max}~=~ d^1_{21} = d^3_{22} ~=~ 2 + j_1 + j_2 +
r$\ , ~~~$z ~=~ j_1-j_2$\ .\nl
~The maximal number of states is
$64 = 8$(chiral)$\times 8$(anti-chiral), achieved for
$r\geq 4$. The 8 anti-chiral, chiral,
states are as described in \bu{\bf a},\bu{\bf c}, resp.,
(differing for $j_2> 0$ and $j_2=0$, $j_1> 0$ and $j_1=0$,
resp.).

\nt\bu ~~$j_1j_2> 0\,$.\nl
Here hold the bare character formulae (3.84) from \Dobch{} (without counter-terms for
$r\geq 4$). The states ~$X^+_{15}\, |\L\rg$,
 ~$X^+_{36}\,|\L\rg$~ and their descendants are
eliminated. (These semi-short UIRs may be called Grassmann-analytic
following \FSa, since odd generators from different chiralities
are eliminated.)  For ~$r=0$~ also the generators ~$X^+_{35}$~ and ~$X^+_{16}$~ are
eliminated. Thus, when ~$j_1=j_2$ (then $z=0$) for $r>0$ we have $\quarter$-BPS cases,
and for $r=0$ we have $\han$-BPS cases.
We recall from \Dobch\ that correspondingly to the values of $r$: ~$r\geq 4$, $r=3$, $r=2$, $r=1$,
$r=0$, there are, respectively, ~$64,63,57,42,11$~ terms in the superfield.

We shall present the character only for the last case which
is the shortest semi-short $N=2$ superfield.
The 11 corresponding states are:
\eqna\dtacd
$$\eqalignno{
&|\L\rg \ , \quad X^+_{25} \, |\L\rg \ , \quad X^+_{46} \, |\L\rg \ ,\cr
&X^+_{26} \, X^+_{25} \, |\L\rg \ , \quad X^+_{45} \, X^+_{46} \, |\L\rg \ , \cr
&X^+_{25} \,X^+_{46} \, |\L\rg \ , \quad
 X^+_{26} \,X^+_{46} \, |\L\rg \ , \quad X^+_{45} \, X^+_{25} \, |\L\rg \ , \cr
&X^+_{26} \, X^+_{25} \,X^+_{46} \, |\L\rg \ , \quad
X^+_{45} \, X^+_{25} \, X^+_{46} \, |\L\rg \ , \cr
& X^+_{26} \, X^+_{45} \, X^+_{25} \, X^+_{46} \, |\L\rg \
. &\dtacd{}}$$
The corresponding signatures - conformal and $su(2)$ - in format
~$[d,j_1,j_2\,;\,r]$~ are:
\eqna\signac
$$\eqalignno{ &[d\equiv d^{ac},j_1,j_2\,;\,0] \ , \quad
[d+\han,j_1+\han,j_2\,;\,1] \ , \quad [d+\han,j_1,j_2+\han\,;\,1] \ , \cr
&[d+1,j_1+1,j_2\,;\,0] \ , \quad [d+1,j_1,j_2+1\,;\,0] \ , \cr
&[d+1,j_1+\han,j_2+\han\,;\,2] \ ,\quad [d+1,j_1+\han,j_2+\han\,;\,0] \ , \quad
[d+1,j_1+\han,j_2+\han\,;\,0] \ , \cr
& [d+\trh,j_1+1,j_2+\han\,;\,1] \ , \quad [d+\trh,j_1+\han,j_2+1\,;\,1]\ , \cr
&[d+2,j_1+1,j_2+1\,;\,0] \
 .&\signac{}}$$
Note that all conformal entries are on the conformal unitarity threshold ~$d=2+j_1+j_2$,
($j_1j_2>0$), i.e., shall use as input formula \charthr{}. Thus, the character formula is:
\eqna\chdrcac
$$\eqalignno{
ch\,L_\L ~=&~ {e(\L)\over (1-t_2) (1-t_1t_2) (1-t_2t_3) (1-t_1t_2t_3)}
 \ \times \cr \times \ &\Big\{\
 \cp_{n_1,n_2} ~+~ e(\a_{25})\, \cp_{n_1+1,n_2}\,\cs_1
 ~+~ e(\a_{46})\, \cp_{n_1,n_2+1}\,\cs_1 ~+\cr
 +~ &e(\a_{26})\,e(\a_{25})\, \cp_{n_1+2,n_2} ~+~
e(\a_{45})\,e(\a_{46})\, \cp_{n_1,n_2+2}\, ~+\cr
 +~&e(\a_{25})\,e(\a_{46})\,\cp_{n_1+1,n_2+1} \,\cs_2 ~+\cr
+~& \Big(e(\a_{26})\,e(\a_{46}) + e(\a_{45})\,e(\a_{25})\Big) \,\cp_{n_1+1,n_2+1} ~+\cr
+~ &\Big(e(\a_{26})\,\cp_{n_1+2,n_2+1} + e(\a_{45})\,\cp_{n_1+1,n_2+2}\Big)
\, e(\a_{25})\,e(\a_{46})\, \cs_1
 ~+\cr
+~ &e(\a_{26})\,e(\a_{45})\,e(\a_{25})\,e(\a_{46})\, \cp_{n_1+2,n_2+2} \
 \Big\} \ ~=
 &\chdrcac{a}\cr
~=&~ {e(\L)\over (1-t_2) (1-t_1t_2) (1-t_2t_3) (1-t_1t_2t_3)}
 \ \times \cr \times \ &\Big\{\
 \cp_{n_1,n_2} ~+~ \tth_2\, \cp_{n_1+1,n_2}\,(1+t_5)
 ~+~ \bth_2\, \cp_{n_1,n_2+1}\,(1+t_5) ~+\cr
 +~ &\tth_1\,\tth_2\, \cp_{n_1+2,n_2} ~+~
\bth_1\,\bth_2\, \cp_{n_1,n_2+2}\, ~+\cr
 +~&\tth_2\,\bth_2\,\cp_{n_1+1,n_2+1} \,(1+t_5+t_5^2) ~+\cr
+~& \Big(\tth_1\,\bth_2 + \bth_1\,\tth_2\Big) \,\cp_{n_1+1,n_2+1} ~+\cr
+~ &\Big(\tth_1\,\cp_{n_1+2,n_2+1} + \bth_1\,\cp_{n_1+1,n_2+2}\Big)
\, \tth_2\,\bth_2\,\,(1+t_5)
 ~+\cr
+~ &\tth_1\,\bth_1\,\tth_2\,\bth_2\, \cp_{n_1+2,n_2+2} \
 \Big\}
&\chdrcac{b} }
$$
where we have used the characters of ~$su(2)$~:~ $\cs_1 = 1+t_5\,$,
~$\cs_2 = 1+t_5+t_5^2\,$, ($t_5 = e(\a_5))$.

\nt\bu ~~$j_1> 0\,,j_2=0\,$. ~~
Here hold bare character formulae (3.86) from \Dobch{} (without counter-terms for
$r\geq 4$). The states ~$X^+_{36}\,X^+_{46}\, |\L\rg$,
 ~$X^+_{15}\,|\L\rg$~ and their descendants are
eliminated. We recall from \Dobch\
that correspondingly to the values of $r$: ~$r\geq 4$, $r=3$, $r=2$, $r=1$,
$r=0$, there are, respectively, ~$64,63,58,45,16$~ states.
In the last case, where $r=0$, we eliminate the generator ~$X^+_{16}$~
and exclude the generators ~$X^+_{3,4+k}$~
from the anti-chiral sector. Then for $z=j_1$ this is a ~$\quarter$-BPS case.

We shall present the character only for the last case.
The 16 states of the superfield are:
\eqna\dtaca
$$\eqalignno{
&|\L\rg \ , \quad X^+_{25} \, |\L\rg \ , \quad X^+_{46} \, |\L\rg \ ,\cr
&X^+_{26} \, X^+_{25} \, |\L\rg \ , \quad X^+_{45} \, X^+_{46} \, |\L\rg \ , \cr
&X^+_{25} \,X^+_{46} \, |\L\rg \ , \quad
 X^+_{26} \,X^+_{46} \, |\L\rg \ , \quad X^+_{45} \, X^+_{25} \, |\L\rg \ , \cr
&X^+_{26} \, X^+_{25} \,X^+_{46} \, |\L\rg \ , \quad
X^+_{45} \, X^+_{25} \, X^+_{46} \, |\L\rg \ , \cr
& X^+_{26} \, X^+_{45} \, X^+_{25} \, X^+_{46} \, |\L\rg \ , \cr
& X^+_{35} \, X^+_{25} \,X^+_{46} \, |\L\rg \ , \quad
 X^+_{36} \, X^+_{45} \, X^+_{25} \,|\L\rg \ , \cr
&X^+_{35} \, X^+_{45} \, X^+_{25} \,X^+_{46}\, |\L\rg \ , \quad
X^+_{26} \, X^+_{35} \, X^+_{25} \, X^+_{46} \, |\L\rg \ , \cr
&X^+_{26} \, X^+_{36} \, X^+_{45} \, X^+_{25} \,|\L\rg
\ . &\dtaca{}}$$
The states of \dtacd{} appear as the first 11 of \dtaca{},
though the content is different:
\eqna\signacc
$$\eqalignno{ &[d\equiv 2+j_1,j_1,0\,;\,0] \ , \quad
[d+\han,j_1+\han,0\,;\,1] \ , \quad [d+\han,j_1,\han\,;\,1] \ , \cr
&[d+1,j_1+1,0\,;\,0] \ , \quad [d+1,j_1,1\,;\,0] \ , \cr
&[d+1,j_1+\han,\han\,;\,2] \ ,\quad [d+1,j_1+\han,\han\,;\,0] \ , \quad
[d+1,j_1+\han,\han\,;\,0] \ , \cr
& [d+\trh,j_1+1,\han\,;\,1] \ , \quad [d+\trh,j_1+\han,1\,;\,1]\ , \cr
&[d+2,j_1+1,1\,;\,0] \ , \cr
&[d+\trh,j_1+\han,0\,;\,1] \ , \quad [d+\trh,j_1+\han,0\,;\,1] \ ,\cr
&[d+2,j_1+\han,\han\,;\,0] \ , \quad [d+2,j_1+1,0;\,0] \ , \cr
&[d+2,j_1+1,0\,;\,0] \ .
&\signacc{}}$$
Since in all entries the value of ~$d$~ is above the conformal unitarity threshold, then we
use formula \slgen{} for the conformal part of the character formula:
\eqna\chdrcacc
$$\eqalignno{
ch\,L_\L ~=&~ {e(\L)\over (1-t_2) (1-t_1t_2) (1-t_2t_3) (1-t_1t_2t_3)}
 \ \times \cr \times \ &\Big\{\
 \cq'_{n_1} ~+~ \tth_2\, \cq'_{n_1+1}\,(1+t_5)
 ~+~ \bth_2\, \cq'_{n_1}\,(1+t_3)\,(1+t_5) ~+\cr
 +~ &\tth_1\,\tth_2\, \cq'_{n_1+2} ~+~
\bth_1\,\bth_2\, \cq'_{n_1}\, (1+t_3+t_3^2) ~+\cr
 +~&\tth_2\,\bth_2\,\cq'_{n_1+1}\,(1+t_3) \,(1+t_5+t_5^2) ~+\cr
+~& \Big(\tth_1\,\bth_2 + \bth_1\,\tth_2\Big) \,\cq'_{n_1+1}\,(1+t_3) ~+\cr
+~ &\Big(\tth_1\,\cq'_{n_1+2}\,(1+t_3) + \bth_1\,\cq'_{n_1+1}\, (1+t_3+t_3^2)\Big)
\, \tth_2\,\bth_2\,\,(1+t_5)
 ~+\cr
+~ &\tth_1\,\bth_1\,\tth_2\,\bth_2\, \cq'_{n_1+2}\,(1+t_3+t_3^2) ~+\cr
+~ &\Big(\bvh_1\,\bth_2 + \bvh_2\,\bth_1\Big)\, \tth_2\,
\cq'_{n_1+1}\,(1+t_5) ~+\cr
+~ &\Big(\bvh_1\,\bth_2 + \bvh_2\,\bth_1\Big)\, \tth_1\, \tth_2\,
 \cq'_{n_1+2} ~+\cr
+~ &\bvh_1 \,\bth_1\,\tth_2\,\bth_2\, \cq'_{n_1+1}\,(1+t_3)
 \Big\} \ .
&\chdrcacc {}}$$
where we have used ~$\cq_{n,1} =\cq'_{n}\,$, ~$\cq_{n,2} =(1+t_3)\,\cq'_{n}\,$,
~$\cq_{n,3} = (1+t_3+t_3^2)\,\cq'_{n}\,$.

\medskip

The next case is conjugate to the preceding.

\medskip

\nt\bu ~~$j_1=0,j_2>0\,$. ~~
Here hold bare character formulae (3.89) from \Dobch{} (without counter-terms for
$r\geq 4$).
The states ~$X^+_{15}\,X^+_{25}\, |\L\rg$,
 ~$X^+_{36}\,|\L\rg$~ and their descendants are
eliminated. When $r=0$, we eliminate the generator ~$X^+_{35}$~
and exclude the generators ~$X^+_{1,4+k}$~
from the chiral sector. Then for $z=-j_2$ this is a ~$\quarter$-BPS case.
We consider only the latter case.
The 16 states of the superfield are:
\eqna\dtacb
$$\eqalignno{
&|\L\rg \ , \quad X^+_{25} \, |\L\rg \ , \quad X^+_{46} \, |\L\rg \ ,\cr
&X^+_{26} \, X^+_{25} \, |\L\rg \ , \quad X^+_{45} \, X^+_{46} \, |\L\rg \ , \cr
&X^+_{25} \,X^+_{46} \, |\L\rg \ , \quad
 X^+_{26} \,X^+_{46} \, |\L\rg \ , \quad X^+_{45} \, X^+_{25} \, |\L\rg \ , \cr
&X^+_{26} \, X^+_{25} \,X^+_{46} \, |\L\rg \ , \quad
X^+_{45} \, X^+_{25} \, X^+_{46} \, |\L\rg \ , \cr
& X^+_{26} \, X^+_{45} \, X^+_{25} \, X^+_{46} \, |\L\rg \ , \cr
& X^+_{16} \, X^+_{25} \,X^+_{46} \, |\L\rg \ , \quad
 X^+_{15} \, X^+_{26} \, X^+_{46} \,|\L\rg \ , \cr
&X^+_{16} \, X^+_{45} \, X^+_{25} \,X^+_{46}\, |\L\rg \ , \quad
X^+_{26} \, X^+_{15} \, X^+_{45} \, X^+_{46} \,|\L\rg \ , \cr
&X^+_{26} \, X^+_{16} \, X^+_{25} \, X^+_{46} \, |\L\rg
\ . &\dtacb{}}$$
The states of \dtacd{} appear as the first 11 of \dtacb{}, the same states as in
\dtaca{}, though the content is different:
\eqna\signaccc
$$\eqalignno{ &[d\equiv 2+j_2,0,j_2\,;\,0] \ , \quad
[d+\han,\han,j_2\,;\,1] \ , \quad [d+\han,0,j_2+\han\,;\,1] \ , \cr
&[d+1,1,j_2\,;\,0] \ , \quad [d+1,0,j_2+1\,;\,0] \ , \cr
&[d+1,\han,j_2+\han\,;\,2] \ ,\quad [d+1,\han,j_2+\han\,;\,0] \ , \quad
[d+1,\han,j_2+\han\,;\,0] \ , \cr
& [d+\trh,1,j_2+\han\,;\,1] \ , \quad [d+\trh,\han,j_2+1\,;\,1]\ , \cr
&[d+2,1,j_2+1\,;\,0] \ , \cr
&[d+\trh,0,j_2+\han\,;\,1] \ , \quad [d+\trh,0,j_2+\han\,;\,1] \ ,\cr
&[d+2,0,j_2+1\,;\,0] \ , \quad [d+2,0,j_2+1\,;\,0] \ , \cr
& [d+2,\han,j_2+\han;\,0] \ .
&\signaccc{}}$$
The character formula is:
\eqna\chdrcaccc
$$\eqalignno{
ch\,L_\L ~=&~ {e(\L)\over (1-t_2) (1-t_1t_2) (1-t_2t_3) (1-t_1t_2t_3)}
 \ \times \cr \times \ &\Big\{\
 \cq_{n_3} ~+~ \tth_2\, \cq_{n_3}\,(1+t_1)\,(1+t_5)
 ~+~ \bth_2\, \cq_{n_3+1}\,(1+t_5) ~+\cr
 +~ &\tth_1\,\tth_2\, \cq_{n_3}\, (1+t_1+t_1^2) ~+~
\bth_1\,\bth_2\, \cq_{n_3+2} ~+\cr
 +~&\tth_2\,\bth_2\,\cq_{n_3+1}\,(1+t_1) \,(1+t_5+t_5^2) ~+\cr
+~& \Big(\tth_1\,\bth_2 + \bth_1\,\tth_2\Big) \,\cq_{n_3+1}\,(1+t_1) ~+\cr
+~ &\Big(\tth_1\,\cq_{n_3+1}\, (1+t_1+t_1^2) + \bth_1\,\cq_{n_3+2}\,(1+t_1)\Big)
\, \tth_2\,\bth_2\,\,(1+t_5)
 ~+\cr
+~ &\tth_1\,\bth_1\,\tth_2\,\bth_2\, \cq_{n_3+2}\,(1+t_1+t_1^2) ~+\cr
+~ &\Big(\vth_1\,\tth_2 + \vth_2\,\tth_1\Big)\, \bth_2\,
\cq_{n_3+1}\,(1+t_5) ~+\cr
+~ &\Big(\vth_1\,\tth_2 + \vth_2\,\tth_1\Big)\, \bth_1\, \bth_2\,
 \cq_{n_3+2} ~+\cr
+~ &\tth_1\,\tth_2\,\bth_1\,\bth_2\, \cq_{n_3+1}\,(1+t_1)
 \Big\}
&\chdrcaccc {}}$$

 \bigskip

\nt\bu ~~$j_1=j_2=0$. ~~
Here hold bare character formulae (3.92) from \Dobch{} (without counter-terms for
$r\geq 4$). The states ~$X^+_{15}\,X^+_{25}\, |\L\rg$,
 ~$X^+_{36}\,X^+_{46}\, |\L\rg$~ and their descendants are
eliminated. We recall from \Dobch\
that correspondingly to the values of $r$: ~$r\geq 4$, $r=3$, $r=2$, $r=1$,
$r=0$, there are, respectively, ~$64,63,59,47,24$~ states.
In the last case, when $r=0$, we exclude the generators ~$X^+_{3,4+k}$~
from the anti-chiral sector and the generators ~$X^+_{1,4+k}$~
from the chiral sector and also the combination of impossible
states:
\eqn\impt{X^+_{15}\, X^+_{26} \,X^+_{36}\, X^+_{45} \, |\L\rg\ .}

We shall consider only the 24 states of the UIR at ~$r=0$~:
\eqna\dtac
$$\eqalignno{
&|\L\rg \ , \quad X^+_{25} \, |\L\rg \ , \quad X^+_{46} \, |\L\rg \ ,\cr
&X^+_{26} \, X^+_{25} \, |\L\rg \ , \quad X^+_{45} \, X^+_{46} \, |\L\rg \ , \cr
&X^+_{25} \,X^+_{46} \, |\L\rg \ , \quad
 X^+_{26} \,X^+_{46} \, |\L\rg \ , \quad X^+_{45} \, X^+_{25} \, |\L\rg \ , \cr
&X^+_{26} \, X^+_{25} \,X^+_{46} \, |\L\rg \ , \quad
X^+_{45} \, X^+_{25} \, X^+_{46} \, |\L\rg \ , \cr
& X^+_{26} \, X^+_{45} \, X^+_{25} \, X^+_{46} \, |\L\rg \ , \cr
& X^+_{35} \, X^+_{25} \,X^+_{46} \, |\L\rg \ , \quad
 X^+_{36} \, X^+_{45} \, X^+_{25} \,|\L\rg \ , \cr
&X^+_{35} \, X^+_{45} \, X^+_{25} \,X^+_{46}\, |\L\rg \ , \quad
X^+_{26} \, X^+_{35} \, X^+_{25} \, X^+_{46} \, |\L\rg \ , \cr
&X^+_{26} \, X^+_{36} \, X^+_{45} \, X^+_{25} \,|\L\rg \ , \cr
& X^+_{16} \, X^+_{25} \,X^+_{46} \, |\L\rg \ , \quad
 X^+_{15} \, X^+_{26} \, X^+_{46} \,|\L\rg \ , \cr
&X^+_{16} \, X^+_{45} \, X^+_{25} \,X^+_{46}\, |\L\rg \ , \quad
X^+_{26} \, X^+_{15} \, X^+_{45} \, X^+_{46} \,|\L\rg \ , \cr
&X^+_{26} \, X^+_{16} \, X^+_{25} \, X^+_{46} \, |\L\rg \ , \cr
& X^+_{16} \, X^+_{25} \,X^+_{35} \, X^+_{46}\,|\L\rg
\ , ~~~X^+_{16} \, X^+_{25} \,X^+_{36} \, X^+_{45}\,|\L\rg\cr
&X^+_{15} \, X^+_{26}\,X^+_{35} \, X^+_{46}\,|\L\rg \ . &\dtac{}}$$

The states of \dtacd{} appear as the first 11 of \dtac{},
the states of \dtaca{} as the first 16 of \dtac{},
the states of \dtacb{} as as the first 11 plus states 17-21 of \dtac{},
though of course the contents is different:
\eqna\signacz
$$\eqalignno{ &[2,0,0\,;\,0] \ , \quad
[\frh,\han,0\,;\,1] \ , \quad [\frh,0,\han\,;\,1] \ , \cr
&[\trh,1,0\,;\,0] \ , \quad [\trh,0,1\,;\,0] \ , \cr
&[\trh,\han,\han\,;\,2] \ ,\quad [\trh,\han,\han\,;\,0] \ , \quad
[\trh,\han,\han\,;\,0] \ , \cr
& [\srh,1,\han\,;\,1] \ , \quad [\srh,\han,1\,;\,1]\ , \cr
&[4,1,1\,;\,0] \ , \cr
&[\srh,\han,0\,;\,1] \ , \quad [\srh,\han,0\,;\,1] \ ,\cr
&[4,\han,\han\,;\,0] \ , \quad [4,1,0;\,0] \ , \cr
&[4,1,0\,;\,0] \ , \cr
&[\srh,0,\han\,;\,1] \ , \quad [\srh,0,\han\,;\,1] \ ,\cr
&[4,0,1\,;\,0] \ , \quad [4,0,1\,;\,0] \ , \cr
& [4,\han,\han;\,0] \ , \cr
& [4,0,0;\,0] \ , \quad [4,0,0;\,0] \ , \quad [4,0,0;\,0] &\signacz {}}$$
The character formula is:
\eqna\chdrcaccz
$$\eqalignno{
ch\,L_\L ~=&~ {e(\L)\over (1-t_2) (1-t_1t_2) (1-t_2t_3) (1-t_1t_2t_3)}
 \ \times \cr \times \ &\Big\{\
 1 ~+~ \tth_2\, (1+t_1)\,(1+t_5)
 ~+~ \bth_2\, (1+t_3)\,(1+t_5) ~+\cr
 +~ &\tth_1\,\tth_2\, (1+t_1+t_1^2) ~+~
\bth_1\,\bth_2\, (1+t_3+t_3^2) ~+\cr
 +~&\tth_2\,\bth_2\,(1+t_1)\,(1+t_3) \,(1+t_5+t_5^2) ~+\cr
+~& \Big(\tth_1\,\bth_2 + \bth_1\,\tth_2\Big) \,(1+t_1)\,(1+t_3) ~+\cr
+~ &\Big(\tth_1\,(1+t_1+t_1^2)\,(1+t_3) + \bth_1\,(1+t_1)\, (1+t_3+t_3^2)\Big)
\, \tth_2\,\bth_2\,\,(1+t_5)
 ~+\cr
+~ &\tth_1\,\bth_1\,\tth_2\,\bth_2\, (1+t_1+t_1^2)\,(1+t_3+t_3^2) ~+\cr
+~ &\Big(\bvh_1\,\bth_2 + \bvh_2\,\bth_1\Big)\, \tth_2\,
(1+t_1)\,(1+t_5) ~+\cr
+~ &\Big(\bvh_1\,\bth_2 + \bvh_2\,\bth_1\Big)\, \tth_1\, \tth_2\,
 (1+t_1+t_1^2) ~+\cr
+~ &\bvh_1\,\bth_1\,\tth_2\,\bth_2\, (1+t_1)\,(1+t_3)\cr
+~ &\Big(\vth_1\,\tth_2 + \vth_2\,\tth_1\Big)\, \bth_2\,
(1+t_3)\,(1+t_5) ~+\cr
+~ &\Big(\vth_1\,\tth_2 + \vth_2\,\tth_1\Big)\, \bth_1\, \bth_2\,
 (1+t_3+t_3^2) ~+\cr
+~ &\vth_1\,\tth_1\,\tth_2\,\bth_2\, (1+t_3)\,(1+t_1) ~+\cr
+~ & \vth_1 \, \tth_2 \,\bvh_1 \, \bth_2 ~+~  \vth_1 \, \tth_2 \,\bvh_2 \, \bth_1
~+~  \vth_2 \, \tth_1\,\bvh_1 \, \bth_2
 \Big\} \ .
&\chdrcaccz{}}$$

\vskip 3mm

\nt{\bu{\bf ad}}~~~
$d ~=~ d^1_{21} = d^4_{22} ~=~ 1 + j_2 + r\ , ~~ j_1=0$,
~~~$z ~=~ -1-j_2$\ .\nl
Here hold bare character formulae (3.95) from \Dobch{} when $j_2r> 0$,
(3.96)  when $j_2=0,r> 0$, (both these cases without
counter-terms for $r\geq4$), and finally when $r=0$ holds (3.97)
independently of the value of $j_2$ - these are the anti-chiral
massless UIRs.

The generators $X^+_{15}$,
$X^+_{25}$, and in addition $X^+_{36}$ for $j_2> 0$
(resp. the state ~$X^+_{36}\,X^+_{46}\, |\L\rg$,
 and its descendants for $j_2=0$) are eliminated.
The maximal number of states is $24 = 3$(chiral)$\times
8$(anti-chiral), achieved for $r\geq 4$.
The chiral sector for $r>0$
 consists of the states:
\eqn\chirrz{\eqalign{
&X^+_{26} \, |\L\rg \ , \quad r\geq 1 \ , \cr &X^+_{16} \, X^+_{26}
\, |\L\rg \ , \quad r\geq 2 \ , }}
and the vacuum,
while the anti-chiral sector is given by
\eqna\incz
$$\eqalignno{
&X^+_{46} \ , \qquad \ve^a_r =1\ , ~\ve^a_j =1\ , \cr &X^+_{45} \,
X^+_{46} \ , \qquad \ve^a_r =0\ , ~\ve^a_j = 2\ , \cr &{\bf 1}\
,\quad X^+_{35}\, X^+_{46} \ , \qquad \ve^a_r =0\ , ~\ve^a_j = 0\ ,
&\incz{a}\cr &X^+_{45} \ , \quad X^+_{45} \, X^+_{35} \, X^+_{46} \
, \qquad \ve^a_r =-1\ , ~\ve^a_j = 1\ , \cr &X^+_{35} \ , \qquad
\ve^a_r =-1\ , ~\ve^a_j = -1\ , \cr &X^+_{35} \, X^+_{45} \ , \qquad
\ve^a_r =-2\ , ~\ve^a_j = 0\ \cr }$$
for $j_2>0$ and by
\eqn\inczz{\eqalign{ &X^+_{46} \
, \qquad \ve^a_r =1\ , ~\ve^a_j =1\ , \cr &X^+_{45} \, X^+_{46} \ ,
\qquad \ve^a_r =0\ , ~\ve^a_j = 2\ , \cr &{\bf 1}\ ,\quad X^+_{35}\,
X^+_{46} \ ,\quad ~X^+_{45} \, X^+_{36} \ , \qquad \ve^a_r =0\ ,
~\ve^a_j = 0\ , \cr &X^+_{45} \ , \quad X^+_{45} \, X^+_{35} \,
X^+_{46} \ , \qquad \ve^a_r =-1\ , ~\ve^a_j = 1\ , \cr &X^+_{35} \,
X^+_{45} \ , \qquad \ve^a_r =-2\ , ~\ve^a_j = 0\ \cr }}
 for $j_2=0$.
Finally, for $r=0$ also the generators $X^+_{16}$,
$X^+_{26}$, $X^+_{35}$ are eliminated.

The possible 24 states for $j_2>0$ are given explicitly as:
\eqna\tada
$$\eqalignno{
&|\L\rg\ ,~
X^+_{46} \,|\L\rg\ , ~ X^+_{45} \, X^+_{46} \,|\L\rg\ ,
\qquad r\geq0\ ,&\tada{}\cr
&X^+_{26}\,|\L\rg\ , ~~X^+_{45} \,|\L\rg\ , \qquad r\geq1 \cr
&X^+_{35}\, X^+_{46} \,|\L\rg\ ,~
~~ X^+_{26}\, X^+_{46} \,|\L\rg\ ,
\qquad r\geq1\ ,\cr
&X^+_{26}\, X^+_{45} \, X^+_{46} \,|\L\rg\ ,
~~X^+_{26}\, X^+_{35}\, X^+_{46} \,|\L\rg\ ,\qquad r\geq1\ ,\cr
&X^+_{16} \, X^+_{26}\, X^+_{46} \,|\L\rg\ ,
~~X^+_{45} \, X^+_{35} \, X^+_{46} \,|\L\rg\ ,\qquad r\geq1\ ,\cr
&X^+_{35} \,|\L\rg\ , \qquad r\geq1\ ,& (*)\cr
&X^+_{26}\, X^+_{45} \,|\L\rg\ , ~~X^+_{16} \, X^+_{26}\,|\L\rg\ ,
~~X^+_{35} \, X^+_{45} \,|\L\rg\ , \qquad r\geq2\ ,\cr
&X^+_{26}\, X^+_{45} \, X^+_{35} \, X^+_{46} \,|\L\rg\ ,
 ~~X^+_{16} \, X^+_{26}\, X^+_{45} \, X^+_{46} \,|\L\rg\ ,
~~X^+_{16} \, X^+_{26}\, X^+_{35}\, X^+_{46} \,|\L\rg\ ,
\qquad r\geq2\ ,\cr
&X^+_{26}\, X^+_{35} \,|\L\rg\ ,
\qquad r\geq2\ ,& (*)\cr
&X^+_{26}\, X^+_{35} \, X^+_{45} \,|\L\rg\ ,
~~X^+_{16} \, X^+_{26}\, X^+_{45} \,|\L\rg\ ,
~~X^+_{16} \, X^+_{26}\, X^+_{45} \, X^+_{35} \, X^+_{46} \,|\L\rg\ ,
\quad r\geq3\ ,\cr
&X^+_{16} \, X^+_{26}\, X^+_{35} \,|\L\rg\ ,
\quad r\geq3\ ,& (*)\cr
&X^+_{16} \, X^+_{26}\, X^+_{35} \, X^+_{45} \,|\L\rg\ ,
\qquad r\geq4\ . }$$
Correspondingly to the values of $r$: ~$r\geq 4$, $r=3$, $r=2$, $r=1$,
$r=0$, there are, respectively, ~$24,23,19,12,3$~ states.
Three states are marked with $(*)$ - these states are not present when $j_2=0$.
Thus, the 24 states for $j_2=0$ are given as the 21 from \tada{} without $(*)$
and the following 3 states:
\eqna\tad
$$\eqalignno{
& X^+_{36} \, X^+_{45} \, |\L\rg\ ,
~~~ X^+_{26}\, X^+_{36} \, X^+_{45} \, |\L\rg\ ,
\qquad r\geq1\ ,\cr
& X^+_{16} \, X^+_{26}\, X^+_{36} \,X^+_{45} \,|\L\rg\ ,
\qquad r\geq2\ .
 &\tad{}}$$
 Thus, for $j_2=0$ correspondingly to the values of $r$ there are,
respectively, ~$24,23,20,13,3$~ states.
The content of the states in \tada{} is as follows:
\eqna\signad
$$\eqalignno{ &[d\equiv 1+j_2+r,0,j_2\,;\,r] \ , \quad
[d+\han,0,j_2+\han\,;\,r+1] \ , \quad [d+1,0,j_2+1\,;\,r] \ ,\qquad r\geq0 \cr
&[d+\han,\han,j_2\,;\,r-1] \ , \quad [d+\han,0,j_2+\han\,;\,r-1] \ , \qquad r\geq1\cr
& [d+1,0,j_2\,;\,r] \ , \quad [d+1,\han,j_2+\han\,;\,r]\ , \qquad r\geq1\cr
&[d+\trh,\han,j_2+1\,;\,r-1] \ , \quad [d+\trh,\han,j_2\,;\,r-1] \ , \qquad r\geq1\cr
&[d+\trh,0,j_2+\han\,;\,r-1]\ , \quad
[d+\trh,0,j_2+\han\,;\,r-1] \ , \qquad r\geq1\cr
&[d+\han,0,j_2-\han\,;\,r-1] \ , \qquad r\geq1& (*)\cr
&[d+1,\han,j_2+\han\,;\,r-2] \ , \quad [d+1,0,j_2\,;\,r-2] \ , \quad
[d+1,0,j_2\,;\,r-2] \ ,\qquad r\geq2\cr
&[d+2,\han,j_2+\han\,;\,r-2] \ ,\quad
 [d+2,0,j_2+1\,;\,r-2] \ , \quad [d+2,0,j_2\,;\,r-2] \ ,\qquad r\geq2\cr
&[d+1,\han,j_2-\han\,;\,r-2] \ , \qquad r\geq2\ , & (*)\cr
&[d+\trh,\han,j_2\,;\,r-3] \ , \quad
[d+\trh,0,j_2+\han\,;\,r-3] \ , \quad
[d+\frh,0,j_2+\han\,;\,r-3] \ , \qquad r\geq3\ ,\cr
&[d+\trh,0,j_2-\han\,;\,r-3] \ , \qquad r\geq3\ ,& (*)\cr
&[d+2,0,j_2\,;\,r-4] \ , \qquad r\geq4 \ .
&\signad{}}$$
and for the states in \tad{}:
\eqna\signadd
$$\eqalignno{
& [d+1,0,j_2\,;\,r] \ , \quad [d+1,\han,j_2\,;\,r-1]\ , \qquad r\geq1\cr
&[d+2,0,j_2\,;\,r-2] \ , \qquad r\geq2 \ . &\signadd{}}$$
The character formula for ~$r>0$~ is:
\eqna\chdrcad
$$\eqalignno{
&ch\,L_\L ~=~ {e(\L)\over (1-t_2) (1-t_1t_2) (1-t_2t_3) (1-t_1t_2t_3)}
 \ \times \cr &\times \ \Big\{\
 \cq_{n_3}\,\cs_{r} ~+~
\bth_2 \,\cq_{n_3+1}\,\cs_{r+1} ~+~ \bth_1 \, \bth_2 \, \cq_{n_3+2}\,\cs_{r} ~+
 \cr
&+~\tth_1\, (1+t_1)\,\cq_{n_3}\,\cs_{r-1} ~+~ \bth_1 \, \cq_{n_3+1}\,\cs_{r-1} ~+
 \cr
&+~\bvh_1\, \bth_2 \,\cq_{n_3}\,\cs_{r} ~+~
 \tth_1\, \bth_2 \, (1+t_1)\,\cq_{n_3+1}\,\cs_{r} ~+\cr
&+~\tth_1\, \bth_1 \, \bth_2 \,(1+t_1)\,\cq_{n_3+2}\,\cs_{r-1} ~+~
\tth_1\, \bvh_1\, \bth_2 \,(1+t_1)\,\cq_{n_3}\,\cs_{r-1} ~+ \cr
&+~\vth_1 \, \tth_1\, \bth_2 \,\cq_{n_3+1}\,\cs_{r-1} ~+~
\bth_1 \, \bvh_1 \, \bth_2 \,\cq_{n_3+1}\,\cs_{r-1} ~+ \cr
&+~\bvh_1 \,\cq_{n_3-1}\,\cs_{r-1} ~+ \cr
&+~\tth_1\, \bth_1 \, (1+t_1)\,\cq_{n_3+1}\,\cs_{r-2} ~+~
\vth_1 \, \tth_1\, \cq_{n_3}\,\cs_{r-2} ~+\cr
&+\bvh_1 \, \bth_1 \, \cq_{n_3}\,\cs_{r-2} ~+ \cr
&+~\tth_1\, \bth_1 \, \bvh_1 \, \bth_2 \,(1+t_1)\,\cq_{n_3+1}\,\cs_{r-2} ~+\cr
&+~\vth_1 \, \tth_1\, \bth_1 \, \bth_2 \,\cq_{n_3+2}\,\cs_{r-2} ~+\cr
&+~ \vth_1 \, \tth_1\, \bvh_1\, \bth_2 \,\cq_{n_3}\,\cs_{r-2} ~+ \cr
&+~\tth_1\, \bvh_1 \, (1+t_1)\,\cq_{n_3-1}\,\cs_{r-2} ~+\cr
&+~\tth_1\, \bvh_1 \, \bth_1 \,(1+t_1)\,\cq_{n_3}\,\cs_{r-3} ~+~
\vth_1 \, \tth_1\, \bth_1 \,\cq_{n_3+1}\,\cs_{r-3} ~+\cr
&+~\vth_1 \, \tth_1\, \bth_1 \, \bvh_1 \, \bth_2 \,
\cq_{n_3+1}\,\cs_{r-3} ~+ \cr
&+~\vth_1 \, \tth_1\, \bvh_1 \,\cq_{n_3-1}\,\cs_{r-3} ~+ \cr
&+~\vth_1 \, \tth_1\, \bvh_1 \, \bth_1\, \cq_{n_3}\,\cs_{r-4} ~+ \cr
&+~\d_{j_2,0}\,\bvh_2\, \bth_1 \,\cq_{n_3}\,\cs_{r} ~+~
\d_{j_2,0}\, \tth_1\,\bvh_2\, \bth_1 \,
(1+t_1)\,\cq_{n_3}\,\cs_{r-1} ~+\cr
&+~\d_{j_2,0}\, \vth_1\, \tth_1\,\bvh_2\, \bth_1 \,
\cq_{n_3}\,\cs_{r-2}\
\Big\} &\chdrcad {}}$$
where we use the ~$su(2)$ character factors
~$\cs_{p} = \sum_{s=0}^{p}\, t_5^s$~ for ~$p\in\bbz_+\,$, and for continuity
we use: ~$\cs_{p} =0$~ for ~$p\in -\bbn$,
we also use the convention ~$\cq_{p}=0$~ for $p\leq 0$.
Thus, the formula is valid for all $r>0$ and for all ~$j_2\,$.

For ~$r=0$~ we have the character formula for the anti-chiral
massless UIRs and we have to use the massless conformal characters \charthrz{},\chxxx{},\chyyy{}:
\eqna\chdrcadd
$$\eqalignno{
&ch\,L_\L ~=~ {e(\L)\over (1-t_2) (1-t_1t_2) (1-t_2t_3) (1-t_1t_2t_3)}
 \ \times \cr &\times \ \Big\{\
 \cp_{n_3} ~+~
\bth_2 \,\cp_{n_3+1}\, (1+t_5) ~+~ \bth_1 \, \bth_2 \, \cp_{n_3+2}
\ \Big\} \ . &\chdrcadd {}}$$

\vskip 3mm

\nt{\bu{\bf bc}}~~~
$d ~=~ d^2_{21} = d^3_{22} ~=~ 1 + j_1 + r\ , ~~ j_2=0$,
~~~$z ~=~ 1+j_1$\ .\nl
Here hold bare character formulae (3.98) from \Dobch{} when $j_1r> 0$,
(3.99)  when $j_1=0,r> 0$, (both these cases without
counter-terms for $r\geq4$), and finally when $r=0$ holds (3.100)
independently of the value of $j_1$ - these are the chiral
massless UIRs.

This case is conjugate to the previous one ~{\bf ad}~ and everything may be obtained
from it by the mirror symmetry. We give only the character of the chiral massless case:
\eqna\chdrchir
$$\eqalignno{
&ch\,L_\L ~=~ {e(\L)\over (1-t_2) (1-t_1t_2) (1-t_2t_3) (1-t_1t_2t_3)}
 \ \times \cr &\times \ \Big\{\
 \cp'_{n_1} ~+~
\tth_2 \,\cp'_{n_1+1}\, (1+t_5) ~+~ \tth_1 \, \tth_2 \, \cp'_{n_1+2}
\ \Big\}\ . &\chdrchir {}}$$

\vskip 3mm

\nt{\bu{\bf bd}}~~~
$d ~=~ d^2_{21} = d^4_{22} ~=~ r \ , ~~ j_1=j_2=0=z$\ .\nl
~~The generators
~$X^+_{15}$, ~$X^+_{25}$, ~$X^+_{36}$,
~$X^+_{46}$ are eliminated. Thus, for $r>1$ these are ~$\ha$-BPS cases.
For ~$r=1$~ also the generators
~$X^+_{16}$, ~$X^+_{35}$ are eliminated. Thus, the latter is a ~$\trq$-BPS case, and
it is also the $N=2$ mixed massless irrep.
For ~$r=0$~ the remaining two generators
~$X^+_{26}$, ~$X^+_{45}$ are eliminated and we have the trivial
irrep as explained in general.

For ~$r> 0$~ the bare character formula is (3.101) from \Dobch{} with
~$i_0=i'_0=0$. The maximal number of states is
nine and the list of states together with the conditions when
they exist are:
\eqn\tbd{\eqalign{
&|\L\rg \ , \qquad r\geq 0 \ , \cr
&X^+_{26} \, |\L\rg \ , \quad
X^+_{45} \, |\L\rg\ , \qquad r\geq 1 \ , \cr
&X^+_{16} \, X^+_{26} \, |\L\rg\ , \quad
X^+_{35} \, X^+_{45} \, |\L\rg\ , \quad
X^+_{26} \, X^+_{45} \, |\L\rg\ , \qquad r\geq 2 \ , \cr
&X^+_{16} \, X^+_{26} \, X^+_{45} \, |\L\rg\ , \quad
X^+_{26} \, X^+_{35} \, X^+_{45} \, |\L\rg\ ,
\qquad r\geq 3 \ , \cr
&X^+_{16} \, X^+_{26} \, X^+_{35} \,
X^+_{45} \, |\L\rg\ , \qquad r\geq 4 \
. }}
Thus, correspondingly to the values of $r$ we have ~$9,8,6,3,1$~ states in the
superfield decomposition.
The mixed massless $\trq$-BPS irrep is obtained for ~$d=r=1$~ and consists
of the first three states above (as was shown in general).

The explicit character formula for $r>1$ is:
\eqn\chdrcbd{\eqalign{
ch\,L_\L ~=&~ {e(\L)\over (1-t_2) (1-t_1t_2) (1-t_2t_3) (1-t_1t_2t_3)}
 \ \times \cr \times \ &\Big\{\
 \cs_r ~+~ \tth_1\, (1+t_1)\, \cs_{r-1} ~+~ \bth_1\, (1+t_3)\, \cs_{r-1} ~+\cr
 +~ &\vth_1\,\tth_1\, \cs_{r-2} ~+~
\bvh_1\,\bth_1\, \cs_{r-2} ~+~ \tth_1\,\bth_1\,(1+t_1)\,(1+t_3)\, \cs_{r-2} ~+\cr
+~ &\vth_1\,\tth_1\,\bth_1\,(1+t_3)\,\cs_{r-3} ~+~
\tth_1\,\bvh_1\,\bth_1\,(1+t_1)\,\cs_{r-3} ~+\cr
+~
&\vth_1\,\tth_1\,\bvh_1\,\bth_1\, \cs_{r-4}
 \Big\}
 }}
where we have used the conformal factor ~$\cq_{n_1,n_3}$ from \slgen{} (only for $n_1,n_3\leq 2$),

For the $\trq$-BPS mixed massless irrep, $r=1$, we have to use the massless conformal characters
from \chxxx{},\chyyy{},\chzzz{}, and we have:
\eqn\chdrcbd{\eqalign{
ch\,L_\L ~=&~ {e(\L)\over (1-t_2) (1-t_1t_2) (1-t_2t_3) (1-t_1t_2t_3)}
 \ \times \cr \times \ &\Big\{\
 (1-t_2t_{13})\, (1+t_5) ~+~ \tth_1\, (1+t_1-t_{12} -t_{13})
 ~+~ \bth_1\, (1+t_3 -t_{23} -t_{13}) \Big\} }}

\vskip 5mm

\subsec{N=4}

\nt
For $N=4$ we consider only some important examples.

First we introduce notation for the odd roots when $N=4$ using \smplr~:
\eqn\relntwo{ \eqalign{
& \a_{25} = \g_3 \ , ~~ \a_{26} = \g_3 +\a_5 \ ,
~~ \a_{27} = \g_3 +\a_5 +\a_6\ , ~~ \a_{28} = \g_3 +\a_5 +\a_6 +\a_7\ ,\cr
&\a_{1k} ~=~ \a_1 + \a_{2k}\ ,  \quad k =5,6,7,8; \cr
& \a_{45} = \g_4 +\a_5+\a_6 +\a_7\ , ~~\a_{46} = \g_4 +\a_5+\a_6\ ,
~~\a_{47} = \g_4 +\a_5\ , ~~ \a_{48} = \g_4 \ , \cr
&\a_{3k} ~=~ \a_3 + \a_{4k}\ ,  \quad k =5,6,7,8 \ .
}}
Note that as a consequence of our mirror symmetry \mirrs{} we have:
\eqn\mirtwo{\eqalign{ & \a_{15} ~\lra ~ \a_{38}\ ,  ~~ \a_{16}~ \lra ~\a_{37}
\ ,  ~~ \a_{17}~ \lra ~\a_{36} \ ,  ~~ \a_{18}~ \lra ~\a_{35}\ ,\cr
&\a_{25}~ \lra~ \a_{48} \ , ~~ \a_{26} ~\lra~ \a_{47}\ ,
~~ \a_{27} ~\lra~ \a_{46}\ , ~~ \a_{28} ~\lra~ \a_{45} \ .
}}
For the characters we shall need also the following notation:
\eqn\notcth{ \eqalign{
&\vth_k ~\equiv~ e(\a_{1,9-k})\ , \quad
\tth_k ~\equiv~ e(\a_{2,9-k})\ , \quad k=1,2,3,4 \cr
&\bvh_k ~\equiv~ e(\a_{3,4+k})\ , \quad
\bth_k ~\equiv~ e(\a_{4,4+k})\ , \quad k=1,2,3,4 \ .}}

\md

\bu First we consider the ~{\it massless}~ multiplets. As in all cases when $N>1$ there are
~{\it three} cases of massless multiplets. In our classification
they are  DRC cases ~{\bf ad}, {\bf bc}, {\bf bd}.

\md

\nt{\bu{\bf ad}}~~~
$d ~=~ d^1_{41} = d^4_{44} ~=~ d^{ad} ~ =
 ~ 1 + j_2 = -z \ ,
~~ j_1=0 \ , \quad r_i=0, \ \forall\, i \ ,$\nl
and all generators
~$X^+_{1,4+k}\,$, ~$X^+_{2,4+k}\,$,
~$X^+_{3,4+k}\,$ are eliminated.
These anti-chiral irreps were
denoted ~$\chi^+_s\,$, $s=j_2=0,\half,1,\ldots$, in Section 3 of \DPu.
Besides the vacuum they contain only ~$N$~ states in ~$\hL_\L$~ and
 should be called ultrashort UIRs.
The bare character formula can be written
in the most explicit way \Dobch:
\eqna\chdadz
$$\eqalignno{
ch~\hL_\L ~=&~
  1 ~+~ e(\a_{48}) ~+~ e(\a_{47})\,e(\a_{48}) ~+~ e(\a_{46})\, e(\a_{47})\,e(\a_{48})
  \ +\cr &+~ e(\a_{45})\,e(\a_{46})\, e(\a_{47})\,e(\a_{48})\ .
 &\chdadz {}
}$$
Their signatures are:
\eqna\chdadzz
$$\eqalignno{
&[d\equiv 1 + j_2;0,j_2\,;\,0,0,0] \ , \qquad [d+\ha;0,j_2+\ha\,;\,1,0,0] \ ,
\qquad [d+1;0,j_2+1\,;\,0,1,0] \ , \cr
&[d+\trh;0,j_2+\trh\,;\,0,0,1] \ , \qquad [d+2;0,j_2+2\,;\,0,0,0] \ , &\chdadzz {}
}$$
All conformal entries are on the massless unitarity threshold, thus for the
conformal entries we must use formula \charthrz{}. For the $su(4)$ entries we
use formulae from \charslf{}. Then we have the explicit character formula for the
anti-chiral massless case:
\eqna\chlessfour
$$\eqalignno{
ch\,L_\L ~=&~ {e(\L)\over (1-t_2) (1-t_1t_2) (1-t_2t_3) (1-t_1t_2t_3)}
 \ \times &\chlessfour  {}\cr \times \ &\Big\{\
\cp_{n_3} ~+~ \bth_4\,\cp_{n_3+1}\,\cs_{100} ~+~
\bth_3\bth_4\,\cp_{n_3+2}\,\cs_{010} ~+ \cr
&~+~ \bth_2\bth_3\bth_4\,\cp_{n_3+3}\,\cs_{001}
~+~ \bth_1\bth_2\bth_3\bth_4\,\cp_{n_3+4}\ \Big\}
}$$

\md

\nt{\bu{\bf bc}}~~~
$d ~=~ d^2_{41} = d^3_{44} ~=~ d^{bc} ~ =
 ~ 1 + j_1 = z \ ,
~~ j_2=0 \ , \quad r_i=0, \ \forall\, i \ ,$\nl
and all generators
~$X^+_{1,4+k}\,$, ~$X^+_{3,4+k}\,$,
~$X^+_{4,4+k}\,$ are eliminated.
These chiral irreps were
denoted ~$\chi_s\,$, $s=j_1=0,\half,1,\ldots$, in Section 3 of \DPu.
Besides the vacuum they contain only ~$N$~ states in ~$\hL_\L$~ and
 should be called ultrashort UIRs.
The bare character formula is \Dobch:
\eqna\chdbcz
$$\eqalignno{
ch~\hL_\L ~=&~
  1 ~+~ e(\a_{25}) ~+~ e(\a_{26})\,e(\a_{25}) ~+~ e(\a_{27})\, e(\a_{26})\,e(\a_{25})
  \ +\cr &+~ e(\a_{28})\,e(\a_{27})\, e(\a_{26})\,e(\a_{25})\ .
 &\chdbcz {}
}$$
Their signatures are:
\eqna\chdbczz
$$\eqalignno{
&[d\equiv 1 + j_1;j_1,0\,;\,0,0,0] \ , \qquad [d+\ha;j_1+\ha,0\,;\,0,0,1] \ ,
\qquad [d+1;j_1+1,0\,;\,0,1,0] \ , \cr
&[d+\trh;j_1+\trh,0\,;\,1,0,0] \ , \qquad [d+2;j_1+2,0\,;\,0,0,0] \ , &\chdbczz {}
}$$
All conformal entries are on the massless unitarity threshold, thus
we use formula \charthrzy{}. Then we have the explicit character formula for the
chiral massless case:
\eqna\chlessfourr
$$\eqalignno{
ch\,L_\L ~=&~ {e(\L)\over (1-t_2) (1-t_1t_2) (1-t_2t_3) (1-t_1t_2t_3)}
 \ \times &\chlessfourr  {}\cr \times \ &\Big\{\
\cp'_{n_1} ~+~ \tth_4\,\cp'_{n_1+1}\,\cs_{001} ~+~
\tth_3\tth_4\,\cp'_{n_1+2}\,\cs_{010} ~+ \cr
&~+~ \tth_2\tth_3\tth_4\,\cp'_{n_1+3}\,\cs_{100}
~+~ \tth_1\tth_2\tth_3\tth_4\,\cp'_{n_1+4}\ \Big\}
}$$

We note the mirror symmetry between character formulae \chlessfour{} and
\chlessfourr{}.

\md

\nt{\bu{\bf bd}}~~~
$d ~=~ d^2_{41} = d^4_{44} = m_1 = 1,\quad
i_0 ~=~ 0,1,2 \ , ~~~
 z = (i_0-1)/2  \ , \quad
 j_1=j_2=0 \ , \quad r_i ~=~ \d_{i,i_0+1}\ .$\nl
In Section 3 of \DPu{}
they are parametrized by $n=2,3$, and
denoted by ~$\chi'_n\,$, $n=m$, ($z= n/2-1$),
~$\chi'^+_n\,$, $n=N-m$, ($z= 1-n/2$), but there is
the coincidence for $n=2$:
~$\chi'_{2} ~=~\chi'^+_{2}\,$.
Here they are enumerated by the parameter $i_0$.
The self-conjugate case  ($i_0=1$, $z=0$) is a ~$\trq$--BPS state.
The following generators
are eliminated: the chiral ~$X^+_{1k}$, $\forall k$, ~$X^+_{2k}$, $k=5,\ldots,5+i_0$,
and the anti-chiral ~$X^+_{3k}$, $\forall k$, ~$X^+_{4,N+4-k}$, $k=0,\ldots,i_0$.

 The bare character formula is \Dobch:
\eqn\chdbdz{
ch~\hL_\L ~= ~
1 ~+~
\sum_{j=0}^{i_0}\ \prod_{i=j}^{i_0} \
e(\a_{2,8-i}) ~+~
 \sum_{k=1}^{3-i_0}\ \prod_{i=k}^{3-i_0} \
e(\a_{4,4+i}) }
We write out the signatures for the three subcases separately:
\eqna\chdbdzz
$$\eqalignno{
&[1;0,0\,;\,1,0,0] \ , \qquad [\trh;\ha,0\,;\,0,0,0] \ ,
 \qquad [\trh;0,\ha\,;\,0,1,0] \ ,&\chdbdzz{}\cr &  [2;0,1\,;\,0,0,1] \ ,
\qquad [\frh;0,1\,;\,0,0,0] \ , &{i_0=0}\cr
&[1;0,0\,;\,0,1,0] \ , \qquad [\trh;\ha,0\,;\,1,0,0] \ ,
\qquad [2;1,0\,;\,0,0,0] \ , \cr
&[\trh;0,\ha\,;\,0,0,1] \ , \qquad [2;0,1\,;\,0,0,0] \ , &{i_0=1}\cr
&[1;0,0\,;\,0,0,1] \ , \qquad [\trh;\ha,0\,;\,0,1,0] \ ,
\qquad  [2;1,0\,;\,1,0,0] \ , \cr
& [\frh;\trh,0\,;\,0,0,0]\ , \qquad [\trh;0,\ha\,;\,0,0,0]
 \ , &{i_0=2}
}$$

All conformal entries are on the massless unitarity threshold, thus
we use formulae \charthrz{} and \charthrzy{}. Then we have the explicit character formula for the
mixed chiral -- anti-chiral massless cases:
\eqna\chlessfourrr
$$\eqalignno{
ch\,L_\L^{0} ~=&~ {e(\L)\over (1-t_2) (1-t_1t_2) (1-t_2t_3) (1-t_1t_2t_3)}
 \ \times &\chlessfourrr  {a}\cr \times \ &\Big\{\
 (1-t_2t_{13})\,\cs_{100} ~+~ \tth_2\, (1+t_1-t_{12} - t_{13})
 ~+~  \bth_3\, (1+t_3-t_{23} - t_{13})  \, \cs_{010}
~+\cr &~+~ \bth_2\, \bth_3\,\cp_3 (t_3) \, \cs_{001}
~+~ \bth_1\,\bth_2\, \bth_3\,\cp_4 (t_3)   \ \Big\} \cr &\cr
ch\,L_\L^{1} ~=&~ {e(\L)\over (1-t_2) (1-t_1t_2) (1-t_2t_3) (1-t_1t_2t_3)}
 \ \times &\chlessfourrr  {b}\cr \times \ &\Big\{\
 (1-t_2t_{13})\,\cs_{010} ~+~ \tth_2\, (1+t_1-t_{12} - t_{13})\, \cs_{100}
 ~+~ \tth_1\,\tth_2\, \cp'_3 (t_1)
~+\cr &~+~ \bth_2\, (1+t_3-t_{23} - t_{13})\, \cs_{001}
~+~ \bth_1\,\bth_2\, \cp_3 (t_3)\ \Big\} \cr &\cr
ch\,L_\L^{2} ~=&~ {e(\L)\over (1-t_2) (1-t_1t_2) (1-t_2t_3) (1-t_1t_2t_3)}
 \ \times &\chlessfourrr  {c}\cr \times \ &\Big\{\
 (1-t_2t_{13})\,\cs_{001} ~+~ \tth_3\, (1+t_1-t_{12} - t_{13})\, \cs_{010}
 ~+~ \tth_2\,\tth_3\, \cp'_3 (t_1)\, \cs_{100}
~+\cr &~+~ \tth_1\,\tth_2\,\tth_3\, \cp'_4 (t_1)
~+~ \bth_2 \, (1+t_3-t_{23} - t_{13})\  \Big\}
}$$

We note the mirror symmetry between character formulae \chlessfourrr{a} and
\chlessfourrr{c},  while \chlessfourrr{b} is self-conjugate (as expected).

 \md

\nt\bu ~Next we consider the ~{\it graviton supermultiplet} \GRW.
In our classification this is a DRC case:\nl
\nt{\bu{\bf bd}}~~~
$d ~=~ d^2_{41} = d^4_{44} ~=~ d^{bd} ~ = m_1 =2$\ , \quad
~~~ $j_1=j_2= 0 \ , \quad r_2=2, r_1=r_3=0, \quad z=0$.

Here are eliminated generators ~$X^+_{1k}\,$,\ $X^+_{2k}\,$\ for $k=5,6$~ and
generators ~$X^+_{3k}\,$,\ $X^+_{4k}\,$\ for $k=7,8$~ and this is a
~$\ha$-BPS case.

The bare character formula is (3.101) from \Dobch\  taken for
the case ~$i_0=i'_0=1$~:
\eqna\chdbd
$$\eqalignno{
ch~\hL_\L ~=&~
\prod_{\a \in\D^+_\I
\atop {\a\neq \a_{j,5+N-k}
\atop {j=3,4\,, \ k=1,2
\atop {\a\neq \a_{j',4+k'}
\atop j'=1,2\,, \ k'=1,2
}}}}
\ (1 +e(\a)) ~~-~~ \car   &\chdbd{a} \cr
}$$

Explicitly, the states (fields) are:
\eqna\gravi
$$\eqalignno{
&|\L\rg \ , \qquad [2;0,0;0,2,0], & \varphi_{(1)} \cr
&X^+_{27} \, |\L\rg \ , \qquad [\frh\,;\ha,0\,;1,1,0]\ , & \l^+_{(1)}\cr
&X^+_{46} \, |\L\rg \ , \qquad [\frh\,;0,\ha\,;0,1,1]\ & \l^-_{(1)}\cr
&X^+_{17} \, X^+_{27} \, |\L\rg \ , \qquad [3\,;0,0\,;2,0,0]\ ,& \varphi_{(2)} \cr
&X^+_{36} \, X^+_{46} \, |\L\rg \ ,\qquad [3\,;0,0\,;0,0,2]\ ,& \bar{\varphi}_{(2)} \cr
&X^+_{18} \, X^+_{27} \, |\L\rg \ , \qquad [3\,;0,0\,;0,1,0]\ ,& *\varphi'_{(2)} \cr
&X^+_{35} \, X^+_{46} \, |\L\rg \ , \qquad [3\,;0,0\,;0,1,0]\ ,& *\varphi''_{(2)} \cr
&X^+_{28} \, X^+_{27} \, |\L\rg \ , \qquad [3\,;1,0\,;0,1,0]\ ,& A^+_{\mu\nu}\cr
&X^+_{45} \, X^+_{46} \, |\L\rg \ ,\qquad [3\,;0,1\,;0,1,0]\ ,& A^-_{\mu\nu} \cr
&X^+_{46} \, X^+_{27} \, |\L\rg \ , \qquad [3\,;\ha,\han\,;1,0,1]\ ,& A_{\mu}\cr
&X^+_{28} \, X^+_{17} \, X^+_{27} \, |\L\rg \ ,
\qquad [\srh\,;\ha,0\,;1,0,0]\ ,& \l^+_{2}\cr
& X^+_{35} \,X^+_{27} \, X^+_{46} \, |\L\rg \ ,
\qquad [\srh\,;\ha,0\,;1,0,0]\ ,& *\l'^+_{2}\cr
&X^+_{45} \, X^+_{36} \, X^+_{46} \, |\L\rg \ ,
\qquad [\srh\,;0,\ha\,;0,0,1]\ ,& \l^-_{2}\cr
&X^+_{18} \, X^+_{46} \, X^+_{27} \, |\L\rg \ ,
\qquad [\srh\,;0,\ha\,;0,0,1]\ ,& *\l'^-_{2} \cr
&X^+_{28} \, X^+_{46} \, X^+_{27} \, |\L\rg \ ,\qquad [\srh\,;1,\ha\,;0,0,1]\ ,& \psi^+_\mu \cr
 & X^+_{45} \, X^+_{27} \, X^+_{46} \, |\L\rg \ ,\qquad
 [\srh\,;\ha,1\,;1,0,0]\ ,&\psi^-_\mu\cr
&X^+_{18} \, X^+_{28} \, X^+_{17} \, X^+_{27} \, |\L\rg \ ,
\qquad [4;0,0;0,0,0], & \varphi_{(3)} \cr
& X^+_{35} \, X^+_{45} \, X^+_{36} \, X^+_{46} \, |\L\rg \ ,
\qquad [4;0,0;0,0,0], & \varphi_{(3)} \cr
&X^+_{18} \, X^+_{35} \, X^+_{46} \, X^+_{27} \, |\L\rg \ ,
\qquad [4;0,0;0,0,0], & *\varphi_{(3)} \cr
&X^+_{28} \, X^+_{35} \, X^+_{46} \, X^+_{27} \, |\L\rg \ ,
\qquad [4;1,0;0,0,0], & *B^+_{\mu\nu} \cr
&X^+_{18} \, X^+_{45} \, X^+_{27} \, X^+_{46} \, |\L\rg \ ,
\qquad [4;0,1;0,0,0], & *B^-_{\mu\nu} \cr
& X^+_{28} \, X^+_{45} \, X^+_{27} \, X^+_{46} \, |\L\rg \ ,
\qquad [4;1,1;0,0,0], & h_{\mu\nu} \cr
&&\gravi{}
}$$
This multiplet appeared first in \GRW\ and we use their notation for
the states, though we should note that the
fields marked with $*$ are missing there (though for some it is just a question of
multiplicity).

For the conformal entries in
the explicit character formula we shall use formula \charthr{} for the
 fields on the unitarity threshold: graviton ~$h_{\mu\nu}\,$,
gravitini ~$\psi^\pm_{\mu}\,$, vector ~$A_{\mu}\,$,  and formula \slgen{} for the rest.
Thus, we have:
\eqna\gravic
$$\eqalignno{
ch\,L_\L ~=&~ {e(\L)\over (1-t_2) (1-t_1t_2) (1-t_2t_3) (1-t_1t_2t_3)}
 \ \times &\gravic {}\cr \times \ &\Big\{\
\cs_{020} ~+~ \tth_2\, \cs_{110}\, (1+t_1) ~+~ \bth_2\, \cs_{011}\, (1+t_3) ~+\cr
&~+~\tth_2\,\bth_2\, \cs_{101}\, (1+t_1)\, (1+t_3)~+\cr
&~+~ \Big( \tth_1\, \tth_2\, (1+t_1+t_1^2)
+ \bth_1\, \bth_2\, (1+t_3+t_3^2)\Big)\,\cs_{010} ~+\cr
&~+~ (\tth_2\, \vth_1 + \bth_2\, \bvh_1) \,\cs_{010}\,  ~+ &*\cr
&~+~ \tth_2\, \vth_2\,\cs_{200} ~+~ \bth_2\, \bvh_2\,\cs_{002} ~+\cr
&~+~ \tth_2\, \Big( \tth_1\, \vth_2 + \bth_2\, \bvh_1\Big)\, (1+t_1)\,\cs_{100} ~+&*'\cr
&~+~ \bth_2\, \Big( \bth_1\, \bvh_2 + \tth_2\, \vth_1\Big)\, (1+t_3)\,\cs_{001} ~+&*'\cr
&~+~ \tth_2\, \bth_2\,\Big( \tth_1\,(1+t_1+t_1^2)\,(1+t_3)\,\cs_{001}
+ \bth_1\,(1+t_1)\,(1+t_3+t_3^2)\,\cs_{100}\Big)\cr
&~+~ \tth_1\,  \tth_2\, \vth_1\,\vth_2 ~+~ \bth_1\,  \bth_2\, \bvh_1\,\bvh_2
~+~ \tth_2\,    \vth_1\, \bth_2\,    \bvh_1 ~+&*'\cr
&~+~ \tth_2\, \bth_2\,\Big(\tth_1\,\bvh_1\,(1+t_1+t_1^2) +
\vth_1\,\bth_1\,(1+t_3+t_3^2) \Big)~+ &*\cr
&~+~ \tth_1\,\tth_2\,\bth_1\, \bth_2\,  (1+t_1+t_1^2)\,(1+t_3+t_3^2)
\Big\}}$$
where ~$*$~ designates a line that would be missing if we use \GRW,
a ~$*'$~ designates a line from which the last term would be missing.

\md

\nt\bu ~Next we consider the ~{\it Konishi supermultiplet}, \Konishi,
this the spinless $R$-symmetry scalar on the unitarity threshold, thus, ~$d=2$.
In our classification this is a DRC case ~{\bf ac}~:

\nt{\bu{\bf ac}}~~~
$d ~=~ d_{\rm max}~=~ d^1_{N1} = d^3_{NN} ~=~ d^{ac} ~ = 2$\ , \quad
~~~ $j_1=j_2= 0 \ , \quad r_i=0,\ \forall\, i \ , \quad z=0$.

Here all generators ~$X^+_{1k}\,$~ and ~$X^+_{3k}\,$~ are eliminated and this is
~$\ha$-BPS case.

The bare character formula is (3.84) from \Dobch{}:
\eqn\chaczz{\eqalign{
ch~\hL_\L ~=&~
 \sum_{k=1}^{N}\ \prod_{i=1}^{k} \
e(\a_{2,4+i}) ~+~
 \sum_{k=1}^N \prod_{i=1}^k
e(\a_{4,5+N-i}) ~+ \cr &\cr
&+~ \prod_{\a \in\D^+_\I
\atop {\ve_2> 0, ~ \ve_4> 0}}
\ (1 +e(\a))
~-~ \car ~ .}}

The states of the first line of \chaczz\ are the four chiral and anti-chiral states
of this case:
\eqn\achir{\eqalign{
& X^+_{25} \, |\L\rg \ , \quad X^+_{26} \ X^+_{25} \, |\L\rg \ , \quad
X^+_{27} \, X^+_{26} \, X^+_{25} \, |\L\rg \ , \quad
X^+_{28} \, X^+_{27} \, X^+_{26} \, X^+_{25} \, |\L\rg \ , \cr
& X^+_{48} \, |\L\rg \ , \quad X^+_{47} \, X^+_{48} \, |\L\rg \ , \quad
X^+_{46} \, X^+_{47} \, X^+_{48} \, |\L\rg \ , \quad
X^+_{45} \, X^+_{46} \, X^+_{47} \, X^+_{48} \, |\L\rg \ . }}

All other states (besides the vacuum) are of mixed chirality, and
our first task is to make explicit the second line of the above formula \chaczz.
It turns out that the counter-term has 22 states and there remain 42 states contributing
to this 2nd line, which explicitly are:
\eqna\dtacnfour
$$\eqalignno{
&|\L\rg \ , \quad X^+_{25} \, X^+_{48} \, |\L\rg \ , \qquad &* \cr
&X^+_{45} \,X^+_{25} \, |\L\rg \ , \quad X^+_{28} \,X^+_{48} \, |\L\rg \ , \cr
& X^+_{28} \,X^+_{25} \,X^+_{48} \, |\L\rg \ , \quad
X^+_{45} \,X^+_{25} \, X^+_{48} \, |\L\rg \ , \cr
&X^+_{26} \,X^+_{25} \,X^+_{48} \, |\L\rg \ ,\quad X^+_{47} \,X^+_{25} \, X^+_{48} \, |\L\rg \ , \cr
&X^+_{46} \, X^+_{26} \, X^+_{25} \, |\L\rg \ , \quad
 X^+_{27} \, X^+_{47} \, X^+_{48} \, |\L\rg \ , \cr
&X^+_{46} \, X^+_{26} \,X^+_{25} \,X^+_{48} \, |\L\rg \ ,\quad
X^+_{27} \, X^+_{47} \,X^+_{25} \, X^+_{48} \, |\L\rg \ , \cr
&X^+_{47} \, X^+_{26} \,X^+_{25} \,X^+_{48} \, |\L\rg \ ,\quad
X^+_{28} \, X^+_{45} \,X^+_{25} \, X^+_{48} \, |\L\rg \ , &*\cr
&X^+_{45} \, X^+_{46} \,X^+_{26} \,X^+_{25} \, |\L\rg \ ,\quad
X^+_{28} \, X^+_{27} \,X^+_{47} \,X^+_{48} \, |\L\rg \ ,\cr
&X^+_{27} \, X^+_{26} \,X^+_{25} \,X^+_{48} \, |\L\rg \ ,\quad
X^+_{46} \, X^+_{47} \,X^+_{25} \,X^+_{48} \, |\L\rg \ , \cr
&X^+_{26} \, X^+_{28} \,X^+_{25} \, X^+_{48} \, |\L\rg \ , \quad
X^+_{45} \, X^+_{47} \,X^+_{25} \, X^+_{48} \, |\L\rg \ , \cr
&X^+_{47} \, X^+_{27} \,X^+_{26} \,X^+_{25} \, |\L\rg \ ,\quad
X^+_{26} \, X^+_{46} \,X^+_{47} \,X^+_{48} \, |\L\rg \ ,&\dtacnfour{}\cr
&X^+_{46} \, X^+_{47} \, X^+_{27} \,X^+_{26} \,X^+_{25} \, |\L\rg \ ,\quad
X^+_{27} \,X^+_{26} \, X^+_{46} \,X^+_{47} \,X^+_{48} \, |\L\rg \ ,\cr
&X^+_{28} \, X^+_{46} \, X^+_{26} \,X^+_{25} \,X^+_{48} \, |\L\rg \ ,\quad
X^+_{45} \, X^+_{27} \, X^+_{47} \,X^+_{25} \, X^+_{48} \, |\L\rg \ , \cr
&X^+_{28} \, X^+_{27} \, X^+_{47} \,X^+_{25} \, X^+_{48} \, |\L\rg \ ,\quad
X^+_{45} \, X^+_{46} \, X^+_{26} \,X^+_{25} \,X^+_{48} \, |\L\rg \ , \cr
&X^+_{26} \, X^+_{27} \, X^+_{47} \,X^+_{25} \, X^+_{48} \, |\L\rg \ ,\quad
X^+_{47} \, X^+_{46} \, X^+_{26} \,X^+_{25} \,X^+_{48} \, |\L\rg \ , \cr
&X^+_{28} \, X^+_{27} \, X^+_{26} \,X^+_{25} \,X^+_{48} \, |\L\rg \ ,\quad
X^+_{45} \, X^+_{46} \, X^+_{47} \,X^+_{25} \, X^+_{48} \, |\L\rg \ , \cr
&X^+_{28} \,X^+_{45} \, X^+_{46} \, X^+_{26} \,X^+_{25} \,X^+_{48} \, |\L\rg \ ,\quad
 X^+_{45} \,X^+_{28} \, X^+_{27} \, X^+_{47} \,X^+_{25} \, X^+_{48} \, |\L\rg \ , \cr
 &X^+_{45} \, X^+_{46} \,X^+_{47} \, X^+_{27} \,X^+_{26} \,X^+_{25} \, |\L\rg \ ,\quad
X^+_{28} \,X^+_{27} \,X^+_{26} \, X^+_{46} \,X^+_{47} \,X^+_{48} \, |\L\rg \ ,\cr
&X^+_{47} \,X^+_{28} \, X^+_{27} \, X^+_{26} \,X^+_{25} \,X^+_{48} \, |\L\rg \ ,\quad
X^+_{26} \,X^+_{45} \, X^+_{46} \, X^+_{47} \,X^+_{25} \, X^+_{48} \, |\L\rg \ , \cr
&X^+_{46} \, X^+_{47} \,X^+_{28} \, X^+_{27} \, X^+_{26} \,X^+_{25} \,X^+_{48} \, |\L\rg \ ,\quad
X^+_{27} \, X^+_{26} \,X^+_{45} \, X^+_{46} \, X^+_{47} \,X^+_{25} \, X^+_{48} \, |\L\rg \ , \cr
&X^+_{46} \,X^+_{27} \, X^+_{47} \, X^+_{26} \,X^+_{25} \,X^+_{48} \, |\L\rg \ ,\quad
X^+_{28} \,X^+_{27} \, X^+_{26} \,X^+_{45} \, X^+_{46} \, X^+_{47} \,X^+_{25} \, X^+_{48}
 \, |\L\rg \ ,&*}$$
where ~$*$~ denotes lines in which the states are self-conjugate, while in all other cases the
two states in each line are conjugate to each other.
The corresponding signatures of both \achir\ and
\dtacnfour{} are:
\eqna\signfour
$$\eqalignno{
&[\frh;\han,0\,;\,0,0,1] \ , \quad [3;1,0\,;\,0,1,0] \ , \quad
[\srh;\trh,0\,;\,1,0,0] \ , \quad [4;2,0\,;\,0,0,0] \ , \cr
&[\frh;0,\han\,;\,1,0,0] \ , \quad [3;0,1\,;\,0,1,0] \ , \quad
[\srh;0,\trh\,;\,0,0,1] \ , \quad [4;0,2\,;\,0,0,0] \ , \cr
& [2;0,0\,;\,0,0,0] \ , \quad [3;\han,\han\,;\,1,0,1] \ , \cr
&[3;\han,\han\,;\,0,0,0] \ , \quad [3;\han,\han\,;\,0,0,0] \ , \cr
&[\srh;1,\han\,;\,0,0,1] \ ,\quad [\srh;\han,1\,;\,1,0,0] \ , \cr
&[\srh;1,\han\,;\,1,1,0] \ , \quad [\srh;\han,1\,;\,0,1,1] \ , \cr
&[\srh;1,\han\,;\,0,0,1] \ , \quad [\srh;\han,1\,;\,1,0,0] \ , \cr
&[4;1,1\,;\,1,0,1] \ , \quad [4;1,1\,;\,1,0,1] \ , \cr
&[4;1,1\,;\,0,2,0] \ , \quad [4;1,1\,;\,0,0,0] \ , \cr
&[4;1,1\,;\,0,0,0] \ , \quad [4;1,1\,;\,0,0,0] \ , \cr
&[4;\trh,\han\,;\,2,0,0] \ , \quad [4;\han,\trh\,;\,0,0,2] \ , \cr
&[4;\trh,\han\,;\,0,1,0] \ , \quad [4;\han,\trh\,;\,0,1,0] \ , \cr
&[4;\trh,\han\,;\,0,1,0] \ , \quad [4;\han,\trh\,;\,0,1,0] \ , \cr
&[\nrh;\trh,1\,;\,0,0,1] \ , \quad [\nrh;1,\trh\,;\,1,0,0] \ , \cr
&[\nrh;\trh,1\,;\,0,0,1] \ , \quad [\nrh;1,\trh\,;\,1,0,0] \ , \cr
&[\nrh;\trh,1\,;\,0,0,1] \ , \quad [\nrh;1,\trh\,;\,1,0,0] \ , \cr
&[\nrh;\trh,1\,;\,1,1,0] \ , \quad [\nrh;1,\trh\,;\,0,1,1] \ , \cr
&[\nrh;2,\han\,;\,1,0,0] \ , \quad [\nrh;\han,2\,;\,0,0,1] \ , \cr
&[5;\trh,\trh\,;\,0,0,0] \ , \quad [5;\trh,\trh\,;\,0,0,0] \ , \cr
&[5;\trh,\trh\,;\,0,0,0] \ , \quad [5;\trh,\trh\,;\,0,0,0] \ , \cr
&[5;2,1\,;\,0,1,0] \ , \quad [5;1,2\,;\,0,1,0] \ , \cr
&[\erh;2,\trh\,;\,0,0,1] \ , \quad [\erh;\trh,2\,;\,1,0,0] \ , \cr
&[5;\trh,\trh\,;\,1,0,1] \ , \quad [6;2,2\,;\,0,0,0] \ , &\signfour{}}$$
where the first two lines are the eight states in \achir\ and the rest
are from \dtacnfour{}.

Note that for all conformal signatures holds: ~$d-j_1-j_2=2$. ~For ~$j_1j_2\neq 0$~
this is ~{\it on}~ the conformal unitarity threshold and then we shall use formula \charthr{}.
For ~$j_1j_2 = 0$~
this is ~{\it above}~ the conformal unitarity threshold and then we shall use
the generic formula \slgen{}.

\eqna\chkonishi
$$\eqalignno{
ch\,L_\L ~=&~ {e(\L)\over (1-t_2) (1-t_1t_2) (1-t_2t_3)
(1-t_1t_2t_3)}
 \  \times \cr \times \ &\Big\{\
 \tth_4(1+t_1)\,\cs_{001} ~+~ \tth_3\,\tth_4(1+t_1+t_1^2)\,\cs_{010} ~+~ \cr &
 ~+~ \tth_2\,\tth_3\,\tth_4(1+t_1+t_1^2 + t_1^3) \,\cs_{100} ~+~ \cr
 & ~+~ \tth_1\,\tth_2\,\tth_3\,\tth_4 (1+t_1+t_1^2 + t_1^3 + t_1^4) ~+~ \cr &
 ~+~ \bth_4(1+t_3)\,\cs_{100} ~+~ \bth_3\,\bth_4(1+t_3+t_3^2)\,\cs_{010} ~+~ \cr
 & ~+~ \bth_2\,\bth_3\,\bth_4 (1+t_3+t_3^2 + t_3^3) \,\cs_{001}~+~ \cr
 & ~+~
 \bth_1\,\bth_2\,\bth_3\,\bth_4(1+t_3+t_3^2 + t_3^3 + t_3^4)  ~+~ \cr
 & ~+~ {\bf
1} ~+~ \Big(\tth_4\,\bth_4 \,\cs_{101}
 ~+~   \bth_1\,\tth_4 + \tth_1\,\bth_4\Big)\, \cp_{22}
+ \cr & ~+~ \Big( \big(\tth_1\,\tth_4\,\bth_4 +
            \bth_2\,\tth_3\,\tth_4\big) \,\cs_{001} ~+~
            \tth_3\,\tth_4\,\bth_4\,\cs_{110} \Big)\,\cp_{3,2}
~+~ \cr
      & ~+~ \Big( \big(\bth_1\,\tth_4\,\bth_4
        + \tth_2\,\bth_3\,\bth_4\big)\,\cs_{100}   ~+~
         \bth_3\,\tth_4\,\bth_4 \,\cs_{011}\Big)\,\cp_{2,3} ~+~ \cr
& ~+~ \Big( \big( \bth_2\,\tth_3\,\tth_4\,\bth_4 +
\tth_2\,\bth_3\, \tth_4\,\bth_4\big)\,\cs_{101} ~+~
 \bth_3\,\tth_3\,\tth_4 \, \bth_4\,\cs_{020}
 ~+~ \cr
 & ~+~ \tth_1\,\bth_1\,\tth_4\,\bth_4 ~+~ 
 \bth_1\,\bth_2\,\tth_3\,\tth_4 ~+~
   \tth_1\,\tth_2\,\bth_3\,\bth_4\Big)\, \cp_{3,3}  ~+~ \cr
 & ~+~ \Big(
\big( \tth_1\,\bth_4 +
\bth_3\,\tth_2\big)\,\cs_{010} ~+~
 \tth_2\,\bth_4\,\cs_{200}  \Big)\,\tth_3\,\tth_4\, \cp_{4,2}  ~+~ \cr
 & ~+~ \Big(\big(  \bth_1 \,\tth_4  + \tth_3\,\bth_2\big)\,\cs_{010}
 ~+~  \bth_2 \,\tth_4 \,\cs_{002}
 \Big) \,\bth_3\,\bth_4\, \cp_{2,4} ~+~ \cr
 & ~+~ \Big( \big( \tth_2\,\tth_3 \, \bth_2\,\bth_3 ~+~
 \tth_1\,\bth_2\,\tth_3\,  \bth_4~+~  \tth_1\,\tth_2\,\bth_3\,  \bth_4\big) \,\cs_{001} ~+~ \cr
 & ~+~ \tth_3\,\tth_2\,\bth_3\, \bth_4\,\cs_{110} \Big)\, \tth_4\, \cp_{4,3} ~+~ \cr
 & ~+~ \Big( \big( \tth_2\,\tth_3\,\bth_2\,\bth_3  ~+~
  \bth_1\,\tth_2\,\bth_3\, \tth_4  ~+~  \bth_1\,\bth_2\,\tth_3\, \tth_4\big) \,\cs_{100}  ~+~ \cr
 & ~+~ \bth_3\,\bth_2\,\tth_3\, \tth_4\,\cs_{011} \Big)\,\bth_4\, \cp_{3,4} ~+~ \cr
 & ~+~ \tth_1\,\tth_2\,\tth_3\, \tth_4\,\bth_4\,\cs_{100}\, \cp_{5,2}~+~
  \bth_1\,\bth_2\,\bth_3
\,\bth_4\, \tth_4\,\cs_{001}\, \cp_{2,5} ~+~ \cr
 & ~+~ \Big( \tth_1\,\bth_1\,\bth_2\,\tth_3\,\tth_4\,\bth_4~+~
  \tth_1\,\bth_1\,\tth_2\,\bth_3\,\tth_4\,\bth_4 ~+~ \cr
 & ~+~ \bth_1\,\tth_2\,\bth_2\,\tth_3\,\bth_3\,\tth_4~+~
  \tth_1\,\tth_2\,\bth_2\,\tth_3\,\bth_3\,\bth_4 ~+~ \cr
& ~+~ \tth_2\,\bth_2\,\tth_3\,\bth_3\,
\tth_4\,\bth_4\, \cs_{101}
  \Big)\,\cp_{4,4}
 ~+~ \cr & ~+~ \Big( \tth_1\,\tth_2\,\tth_3
\, \bth_3\,\tth_4\,\bth_4\, \cp_{5,3} ~+~  \bth_1\,\bth_2\,\tth_3\,\bth_3\,\tth_4\,\bth_4
 \, \cp_{3,5}\Big)\, \cs_{101} ~+~ \cr
 & ~+~ \tth_1\,\tth_2\,\tth_3\,\tth_4\,\bth_2\,\bth_3\,\bth_4\,\cp_{5,4}\,
\cs_{001}~+~
 \bth_1\,\bth_2\,\bth_3\,\bth_4 \,\tth_2\,\tth_3\,\tth_4\,\cp_{4,5}\, \cs_{100} ~+~ \cr
 & ~+~ \tth_1\,\tth_2\,\tth_3
\,\tth_4\,\bth_1\,\bth_2\,\bth_3\,\bth_4\,\cp_{5,5}
\Big\}\ . &\chkonishi {}}
$$

 \vskip 5mm

\newsec{Outlook}

\nt
With the present paper we continue the project of  construction of
the character formulae for the
positive energy unitary irreducible
representations of the $N$-extended $D=4$ conformal
superalgebras $su(2,2/N)$. In the first paper we
presented the bare characters which represent the
defining odd entries of the characters. Now we give
the full explicit character formulae for $N=1$ and
several important examples for $N=2$ and $N=4$.
In particular, for $N=4$ we give the character formulae
for the three massless cases - chiral, anti-chiral and mixed
(the latter with three subcases), also for the  graviton supemultiplet
and for the Konishi supemultiplet.

In the further development of this project we shall try to find more
compact expressions for presenting the characters which will enable us to treat
explicitly more complicated cases for arbitrary $N$.

\vskip 5mm

\noindent {\bf Acknowledgments.}  ~~ The author
thanks the Theory Division of CERN for hospitality during the course
of this work.
 This work was supported in part by the Bulgarian National Science Fund, grant DO 02-257.

\parskip=0pt \baselineskip=10pt
\listrefs
\np\end